\documentclass[fleqn,usenatbib]{mnras}

\usepackage[T1]{fontenc}
\usepackage{float}
\usepackage{graphicx}
\usepackage{amsmath}	
\usepackage[switch]{lineno}
\title[Inferring Magnetic Field Morphology and Dust Scattering Geometry from Mid-IR Polarimetry...]{Inferring Magnetic Field Morphology and Dust Scattering Geometry from Mid-IR Polarimetry: the Extended Aitken Method}
\author[F. V\'{a}rosi, C.M. Telesco, C.M. Wright and S.J. Fernández Acosta]{
Frank V\'{a}rosi$^{1}$\thanks{E-mail: varosi@astro.ufl.edu (FV)},
Charles M. Telesco$^{1},$
Christopher M. Wright$^{2},$
and Sergio José Fernández Acosta$^{3}$
\\
$^{1}$Department of Astronomy, University of Florida, 211 Bryant Space Science Center, FL 32611, USA\\
$^{2}$School of Science, UNSW Canberra, PO Box 7916, Canberra BC 2610, Australia\\
$^{3}$Gran Telescopio de Canarias, Cuesta de San José s/n, E-38712, Breña Baja, La Palma, Spain}
\date{Accepted November 5, 2025. Received September 26, 2024}
\pubyear{2025}
\begin{document}
\label{firstpage}
\pagerange{\pageref{firstpage}--\pageref{lastpage}}
\maketitle
\begin{abstract}
The Aitken method is a useful approach for decomposing mid-IR polarimetry of silicates in astronomical sources into emissive and absorptive components. Here we extend this method to include the effects of polarization caused by scattering from graphite or similar particles located along the same sightlines. To demonstrate the extended method, we apply it in the analysis of CanariCam multi-wavelength imaging polarimetry observations of the Egg Nebula, W3 IRS5, and W51 IRS2, and also spectropolarimetry of W3 IRS5. We compare these results with those obtained with the original Aitken method and show that the Egg Nebula polarimetry observations are generally fit better when this third component is incorporated into the analysis. Polarimetry observations of W3 IRS5 are fit better with the extended Aitken method, but the original method suffices to fit many sightlines. Observations of W51 IRS2 are fit well by either the original or extended Aitken method. Including scattering by dust in the decomposition of polarimetry observations of the Egg Nebula and W3 IRS5 produces better results for the emissive and absorptive components, and in particular for the position angle (PA) of those components. The distribution of the difference between absorptive and emissive PA is then found to be more peaked at a single angle, nearly perpendicular. This supports the theory that mid-IR polarization arises from elongated dust grains aligned along magnetic field lines, since then the PA of emissive and absorptive polarization would be perpendicular. When significant scattering is not present the extended Aitken method produces the same results as the original Aitken method.
\end{abstract}
\section{Introduction}

Emission and/or absorption of thermal radiation by non-spherical dust grains aligned along magnetic fields (B-fields) can account for polarized mid-IR radiation observed in the interstellar medium. Non-spherical dust grains tend to spin, which may result from their helicity and associated radiative torques from an anisotropic radiation field \citep{laz07,cho07}. A spinning and precessing particle becomes preferentially aligned with respect to a local B-field, with its elongated axis essentially perpendicular to the B-field lines. An ensemble of these aligned grains results in polarized extinction of background photons passing through the ensemble. That transmitted light is polarized parallel to the B-field, whereas thermal mid-IR emission from the ensemble is polarized perpendicular to the B-field lines; i.e., for the same grain alignment, there is a 90\degr difference in the mid-IR polarization position angle (PA) from absorption and emission by dust particles. Where both emission and absorption are relevant, we must separate the effects of both processes in order to derive the B-field morphology along each sightline. Depending on the dust composition, polarization of mid-IR radiation can also arise from scattering of photons by the dust \citep{Matsu+Seki1986,kat15,yang16}, which requires neither non-spherical grains nor a B-field. Under some circumstances, all three of these processes -- emission, absorption, and scattering -- may contribute to the observed polarization at each sightline \citep{steph17, kat16}, such as when an emitting source is embedded in an absorbing and scattering cocoon. 

Strictly speaking, radiative transfer within the entire region should be analyzed self-consistently to interpret the observations. However, using ideas from \citet{Hilde1988} and \citet{Ait1989}, \citet{Ait2004} developed a relatively simple multi-wavelength method to separate a sightline's mid-IR emissive and absorptive polarization components due to silicate dust. They did not include scattering, since the scattering albedo of silicates is near zero in the mid-IR \citep{DL1984}, with little polarization due to scattering expected.

The original Aitken method relies on the fact that the mid-IR polarization efficiency as a function of wavelength differs for the well-known 8-13 $\mu$m silicate spectral feature depending on whether the radiation is transmitted through, or emitted by, a population of mutually aligned non-spherical silicate particles. They assumed that silicates are the dominant dust component and used template spectral signatures of silicates in the Orion nebula. The method works well to infer sightline emissive and absorptive components of polarization for a variety of astronomical objects for which silicates may be the dominant dust component \citep{barnes15, Loro17, Li18, Telesco23}. The general Bayesian inference approach was presented by \cite{Loro16}. If the dust mixture also includes carbonaceous grains, such as graphite, then the mid-IR albedo of scattering and resultant polarization may need to be considered. This is evident, for example, in the outer part of the nearly face-on circumstellar disk of the Herbig star AB Aur \citep{Li2016} and in AFGL2688, the Egg Nebula \citep{Sahai1998, Weintraub2000, Kast2002}.

In this paper, we extend the Aitken method to include scattering. We illustrate its use with a selection of CanariCam observations, in particular, multi-wavelength imaging polarimetry of the Egg Nebula, W3 IRS5, and W51 IRS2, and spectropolarimetry of W3 IRS5. The Egg Nebula and W3 IRS5 data have not been previously published and will be explored in detail in future papers. These data are used here solely to compare how the extended Aitken method performs for carbon-rich (the Egg Nebula) and silicate-rich (W3 IRS5) regions. The W51 IRS2 imaging polarimetry has been previously published \citep{Telesco23} but not discussed in the context of the extended Aitken method, which we will do here briefly. Our approach to extending the Aitken method to include scattering is discussed in \hyperref[sec:ASC]{Section \ref{sec:ASC}}. Comparison of those results are presented in \hyperref[sec:ACCD]{Section \ref{sec:ACCD}}. Further analysis of the absorptive and emissive component PA differences is presented in \hyperref[sec-PAD]{Section \ref{sec-PAD}}. Results are summarized in \hyperref[sec:SC]{Section \ref{sec:SC}}.
\section{Adding Scattering Component} \label{sec:ASC}

To account for polarization due to scattering, we extend the Aitken method by using the albedo of graphite as a third spectral template. Graphite has a different spectral behavior from silicates, thereby providing a third independent basis function with which to fit the polarization spectrum. Measurements for at least three wavelengths are needed with this approach. The position angle (PA) of polarization caused by single scattering will be perpendicular to the plane containing the emitter, the scattering point, and the observer. So the PA of scattering polarization should be organized in a somewhat circular fashion around a localized emitting source, which, as we show, is found to be the case for the Egg Nebula. By including the graphite albedo template, we extend the technique of Aitken to also model scattering in sources having significant abundances of carbonaceous grains rather than solely silicates. 

A least-squares fit of the template profiles to the observed spectrum of polarization is performed at each sightline, a procedure described below. For imaging polarimetry the method derives the two-dimensional spatial distributions and spectra of the emissive, absorptive, and scattering polarization components, $p_{\text{E}}{(\lambda)}$, $p_{\text{A}}{(\lambda)}$ and $p_{\text{S}}{(\lambda)}$, with their corresponding PA distributions, PA\textsubscript{E}, PA\textsubscript{A} and PA\textsubscript{S}, thereby determining the relative contributions of the aligned grains located in the emitting and foreground absorbing regions at each sightline. For spectropolarimetry, the procedure is the same but provides a low-to-moderate resolution spectrum at each spatial position along the spectrograph slit. 

The Aitken approach relies on the fact that silicate dust emissivity exhibits a strong, fairly broad spectral feature spanning the 8-13 $\mu$m wavelength region. While there are variations in the shape of this feature from region to region, a large fraction of the YSOs studied spectrophotometrically by \citet{smith00} exhibit a silicate feature very similar to that found in the Orion nebula. The following equations \citep{Ait2004}, which are wavelength dependent, are the expressions for absorptive polarization and emissive polarization fractions as a function of $\tau_{\parallel}$, the optical depth for radiation with electric vector parallel with the dust-particle long axis, and $\tau_{\perp}$, the optical depth for radiation with electric vector perpendicular to the long axis.
Equation (\ref{Aitken-pA})
shows that the difference $\tau_{\parallel}-\tau_{\perp}$ gives rise to the observed absorptive polarization $p_A$, and since that difference is usually small the equation reduces to a simple approximation:
\begin{equation}
    p_A = \frac{e^{-\tau\parallel}-e^{-\tau\perp}}{{e^{-\tau\parallel}}+e^{-\tau\perp}} = \frac{{\text{tanh}(\tau_\perp-\tau_\parallel)}}{2} \approx  \frac{(\tau_\perp-\tau_\parallel)}{2}
\label{Aitken-pA}
\end{equation}
Using the standard equation for non-scattering radiation transfer $(1-e^{-\tau})$, the observed emissive polarization $p_E$, can also be approximated with the difference $\tau_{\parallel}-\tau_{\perp}$ as follows:
\begin{equation}
    p_E = \frac{(1-e^{-\tau\parallel})-(1-e^{-\tau\perp})}{(1-e^{-\tau\parallel})+(1-e^{-\tau\perp})}\approx\frac{\tau_\parallel-\tau_\perp}{\tau_\parallel+\tau_\perp}
\label{Aitken-pE}
\end{equation}
The approximation in Equation (\ref{Aitken-pE}) is valid to within 10\% when the total optical depth $\tau_{\parallel}+\tau_{\perp}$ of the emitting region is less than 0.4 (see Appendix), which is usually the case for mid-IR emission. On the other hand, foreground absorbing regions can be optically thick, with the magnitude of polarization being proportional to the optical depth difference, as noted. Combining the approximations in Equations (\ref{Aitken-pA}) and (\ref{Aitken-pE}) we have:
\begin{equation}
	p_E \propto  -p_A \, / \, (\tau_{\parallel} + \tau_{\perp})
\label{Aitken-pE-pA}
\end{equation}
\citet{Ait2004} assume that the observed spectro-polarimetry of the BN object in Orion is purely absorptive, providing the absorptive template profile $\emph{f}_A(\lambda)$. They then use the emissive optical depth derived from observations of the Trapezium region of Orion to derive the emissive polarization profile as:
\begin{equation}
	f_E(\lambda) = f_A(\lambda) / \tau_E(\lambda)
\label{Aitken-fE-fA}
\end{equation}
with the approximate proportionality of Eq.(\ref{Aitken-pE-pA}) key to developing independent profiles to fit spectra.

We introduce an additional profile $f_S(\lambda)$ to fit a polarization component due to scattering, which normally occurs at shorter wavelengths. The template profile $f_S(\lambda)$ is chosen to be proportional to the scattering albedo of graphite dust, since the albedo of silicates is relatively zero in the mid-IR (dust albedo curves are shown and discussed in the Appendix Figures \ref{DustAlbedo}, \ref{DustAlbedo-maxgs=1um}, and \ref{DustAlbedo-graphite-maxgs=1um+025um}). The scattering angle probability distribution by graphite in the mid-IR is essentially isotropic. Furthermore, the fraction of scattered photons that are polarized (polarization efficiency factor) has been computed using Mie theory by \citet{Matsu+Seki1986}, and found to have values greater than 0.7 at scattering angles between 50\degr and 130\degr for wavelengths that are longer than graphite grain sizes, peaking near unity at 90\degr scattering angle from the source of photons. Therefore, the functional behavior of graphite albedo is a good a template for polarization due to scattering. Photons that are polarized from scattering will have PA perpendicular to the plane formed by photons from the source, to scattering point, to observer.

\begin{figure}
\includegraphics[width=\columnwidth]{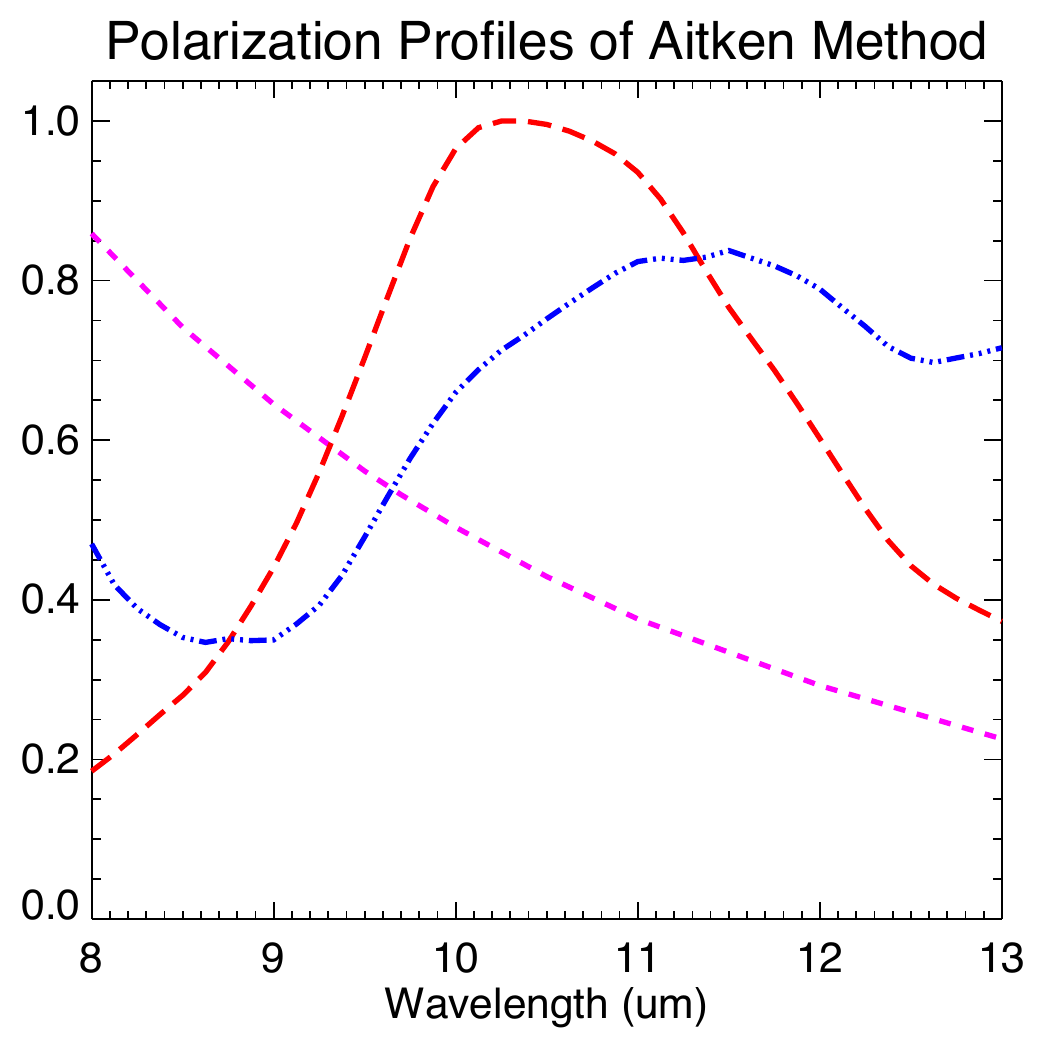}
\caption{The dashed red curve is the absorptive profile, dot-dashed blue curve is the emissive profile, and the short-dashed violet curve is the scattering polarization profile of the extended Aitken method.}
\label{Aitken-Profiles}
\end{figure}

All template profiles are shown in Figure \ref{Aitken-Profiles}: the red dashed curve is absorptive profile $f_A(\lambda)$, the blue dot-dashed curve is the emissive profile $f_E(\lambda)$, and the violet short-dashed curve is the scattering profile $f_S(\lambda)$. The silicate absorption profile is normalized to unity at it's peak, the emission template profile is normalized at 13.5 $\mu$m, and the scattering albedo profile is normalized at 7.5 $\mu$m. The exact normalization is not important since they are templates for fits. The profiles $f_A(\lambda)$ and $f_E(\lambda)$ will always be probing polarization from silicate grains because of the origin of the templates, whereas $f_S(\lambda)$ will always be modeling scattering polarization from graphite (carbonaceous) grains.

The degree of polarization $p$ and position angle PA are computed from the normalized Stokes parameters, $\emph{q}$ = $\emph{Q/I}$ and $\emph{u}$ = $\emph{U/I}$, where $\emph{I}$ is the total observed intensity \citep{tin05}, as $p=(q^2+u^2)^{0.5}$ and PA = $0.5\arctan(u/q)$. Following the derivation in \citet{Ait2004} for an emitting source of photons traveling through a dichroic absorbing layer, the observed Stokes spectra $q(\lambda)$ and $u(\lambda)$ are assumed to be linear combinations of Stokes spectra arising from dichroic absorption and emission by silicates. We propose that the polarization of photons scattered by graphite into the same sightlines can be fit with added terms proportional to $f_S(\lambda)$, giving the extended linear equations:

\begin{eqnarray}
\label{qfit}
    q(\lambda) &=&  q_A(\lambda) + q_E(\lambda)  + q_S(\lambda) \nonumber \\
			&=&  A_q f_A(\lambda) + B_q f_E(\lambda) + C_q f_S(\lambda) \\
\nonumber \\
\label{ufit}
    u(\lambda) &=&  u_A(\lambda) + u_E(\lambda)  + u_S(\lambda)  \nonumber \\
    			&=&  A_u f_A(\lambda) + B_u f_E(\lambda) + C_u f_S(\lambda) \\
\nonumber
\end{eqnarray}
where $f_A(\lambda)$, $f_E(\lambda)$, and $f_S(\lambda)$ are the absorption, emission, and scattering polarization profiles, respectively.
The coefficients of the template polarization profiles are determined by linear least-squares fits to minimize the $\chi^2$ of the difference between above equations and the Stokes $q$ and $u$ spectra independently. Values of \emph{p}$_A$, \emph{p}$_E$ or \emph{p}$_S$ and their corresponding PAs are then derived directly from the template profiles and corresponding coefficients, as for example: 

\begin{eqnarray}
\label{scatp}
	p_S(\lambda) &=& \sqrt{q_S(\lambda)^{2} + u_S(\lambda)^{2}} \,=\, f_S(\lambda)\sqrt{C_q^2 + C_u^2} \\
\nonumber \\
\label{scatPA}
	\text{PA}_S &=& \arctan( u_S / q_S )/2 \,=\, \arctan( C_u / C_q )/2 \\
\nonumber
\end{eqnarray}
Note that the PA of each component is not a function of wavelength, because the same template profiles are used to fit each Stokes parameter and so cancel out in the ratio of Equation \ref{scatPA}. This makes sense physically, since dust grains that produce polarization at a specific angle are aligned at that angle independent of the wavelength of photons they emit, absorb, or scatter. The PA of each component will be the average PA of polarization caused by dust grains along the line of sight. Of course, the polarization magnitude of each component is wavelength dependent and follows the corresponding template-profile wavelength dependence, as shown in Equation \ref{scatp}. Therefore, an observed variation of PA with wavelength would imply a mixture of polarization mechanisms along that sightline. Measuring the polarization for at least three wavelengths is required to fit the observed Stokes spectra with the above equations, thereby enabling separation into absorption, emission, and scattering components. Emissive and absorptive components of polarization are assumed to be from silicate grains, whereas the scattering polarization component is assumed to be from carbonaceous grains (since silicates have no scattering in wavelength range of Figure \ref{Aitken-Profiles}).

\begin{table*}
\caption{Polarimetry Observations with CanariCam at GTC}
\label{Table1}
\begin{tabular}{llrlrc}
\hline
Object & Filters & $\Delta\lambda$ & Date & Integration & FWHM (PSF) \\
           & ($\mu$m) & ($\mu$m) & (UT) & Time (s) & ($\arcsec$) \\
\hline
Egg Nebula  & N   \,\,(10.3) & 5.5 & 2014 June 19 & 728 & 0\farcs35 \\
\hline
Egg Nebula  & Si2  \,(8.7) & 1.1 & 2020 Oct 1 & 1165 & 0\farcs31 \\
                    & Si4(10.3) & 0.9 & 2020 Oct 1 & 1165 & 0\farcs35 \\
                    & Si6(12.5) & 0.7 & 2020 Oct 1 & 1015 & 0\farcs39 \\
\hline
W3 IRS5    & Si2 \,(8.7) & 1.1 & 2012 Oct 5 & 582 & 0\farcs26 \\
                  & Si4(10.3) & 0.9 & 2012 Oct 5 & 582 & 0\farcs33 \\
                  & Si6(12.5) & 0.7 & 2012 Oct 5 & 782 & 0\farcs36 \\
\hline
W3 IRS5 (NE) & Spectropolarimetry & 5.3 & 2014 Oct 9 & 265 & 0\farcs52 (slit-width)\\
W3 IRS5 (SW) & Spectropolarimetry & 5.3 & 2014 Oct 9 & 265 & 0\farcs52 (slit-width)\\
\hline
W51 IRS2    & Si2 \,(8.7) & 1.1 & 2013 Aug 20 & 582 & 0\farcs35 \\
                    & Si4(10.3) & 0.9 & 2013 Aug 20 & 582 & 0\farcs36 \\
                    & Si6(12.5) & 0.7 & 2013 Sep 1& 761 & 0\farcs33 \\
\hline
W51 IRS2    & Si2 \,(8.7) & 1.1 & 2014 June 18 & 509 & 0\farcs35 \\
                    & Si4(10.3) & 0.9 & 2014 June 18 & 509 & 0\farcs32 \\
                    & Si6(12.5) & 0.7 & 2014 June 9 & 761 & 0\farcs38 \\
\hline
\end{tabular}
\end{table*}
\section{Application to CanariCam Data} \label{sec:ACCD}

\subsection{Observations and Data Reduction} \label{sec:3.1}

CanariCam was used to obtain polarimetric images of the Egg Nebula, W3 IRS5, and W51 IRS2. The details of observations are listed in Table 1. CanariCam is the mid-IR multimode (imaging, spectroscopy, and polarimetry) facility camera on the 10.4 m Gran Telescopio de Canarias (GTC) on La Palma, Spain \citep{telesco03}. It has a field of view of 26\arcsec $\times$ 19\arcsec with a pixel scale of 0\farcs0794. Polarimetry is accomplished through the use of a Wollaston prism, which produces a separation between the \emph{o} and \emph{e} (ordinary and extraordinary) rays, and a half-wave plate rotated to a sequence of four orientations (0\degr, 45\degr, 22.5\degr, and 67.5\degr) repeatedly during observations. For imaging polarimetry we have used three narrow-passband filters, one centered on the central part of the silicate feature, filter Si4-10.3, and the other two on the short- and long-wavelength sides of the silicate feature, Si2-8.7 and Si6-12.5 filters, respectively, thereby providing data that can be analyzed with the Aitken method. The broad N-band filter (N-10.36) was also used for imaging polarimetry. Using the CanariCam 0\farcs52-wide slit with the Low-Resolution 10 $\mu$m grating, we also obtained 8-to-13 $\mu$m spectropolarimetry of the two stars in W3 IRS5.

Observations of the program objects were interlaced with observations of the standard star Vega for flux and point-spread-function calibration \citep{cohen99}, and the standard stars AFGL 2591 and AFGL 490 from \citet{smith00} to calibrate the polarization position angle (PA) offset. Standard chop-nod sequences were used with a chop-throw to minimize contamination from background emission. In the polarimetry mode, the actual field of view (FOV) is reduced to 26\arcsec $\times $2\farcs6 by insertion of a focal mask to avoid overlapping between \emph{o} and \emph{e} rays originating at different source locations. However, because of diffraction at edges of the masks, the reliable vertical extent of FOV is usually about 2\farcs1. When the region of interest is larger than a single field of view in the polarization mode, we split our observations of the target into multiple parts offset from each other but overlapping by a few pixel rows for registration and to avoid edge effects. 
    
\begin{figure*}
\includegraphics[width=510pt]{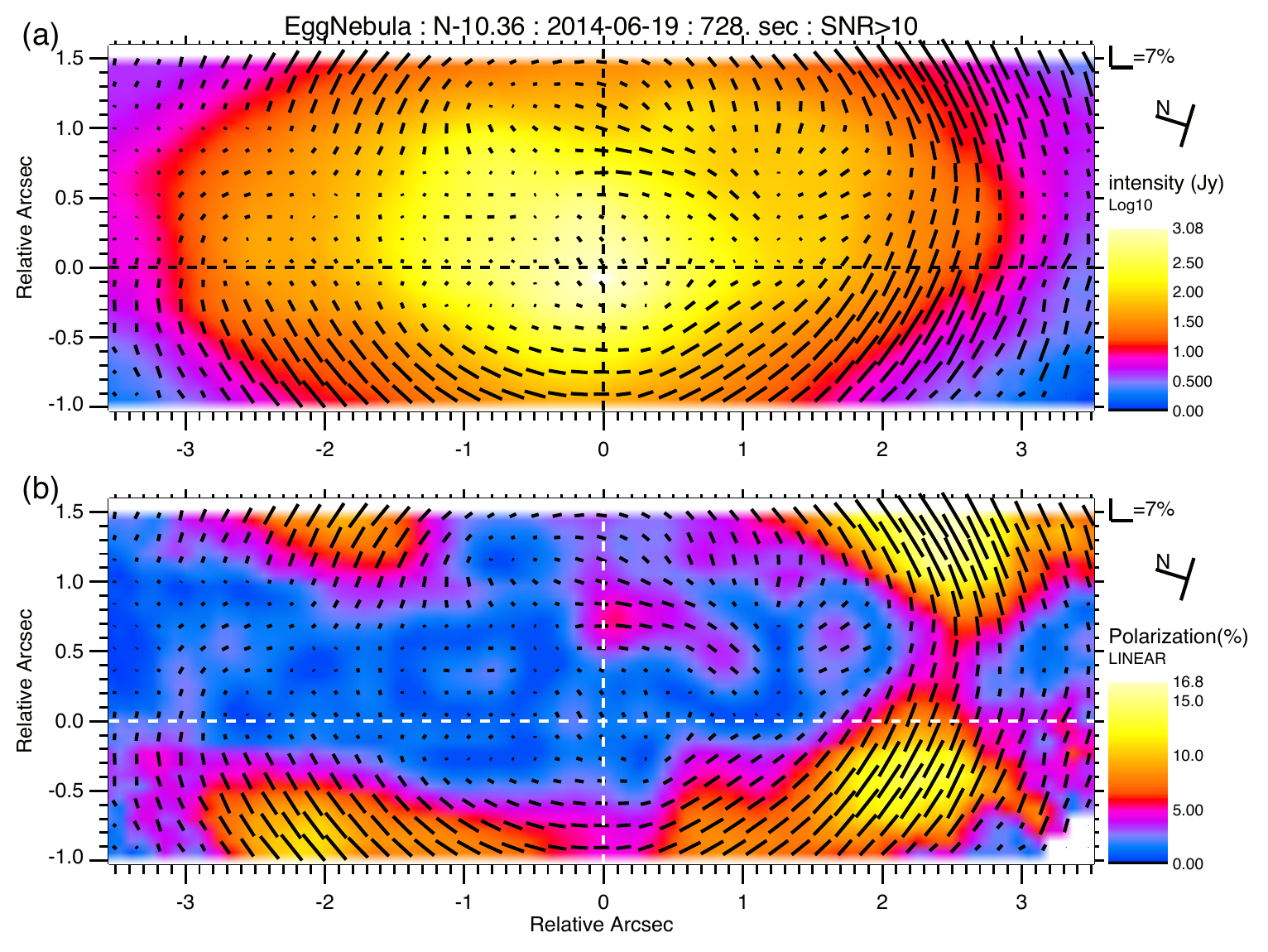}
\caption{(a) Observed intensity of the Egg Nebula with the N-band filter, in units of Jy/arcsec$^2$, indicated by the logarithmic color scale and over-plotted with N-band polarization vectors. (b) polarization image over-plotted with same vectors indicating polarization and PA, but with linear color scale indicating percent polarization. Images of Stokes parameters are first smoothed with $3 \times 3$ pixel boxcar to reduce noise before computing polarization and PA. Data is shown where SNR$>$10. Note tilt of field of view (FOV) with respect to north.}
\label{Egg-N-band}
\end{figure*}

The data were reduced and analyzed with custom IDL software (F. V\'{a}rosi, unpublished). Stokes parameters $q$ and $u$ are computed using the Ratio Method (also called the Beam-Exchange Method) from detector images at the four half-wave plate angles and the \emph{o} and \emph{e} beams created by the Wollaston prism. The Stokes parameters are corrected for half-wave plate efficiency at each wavelength. The half-wave plate is optimized for 10 $\mu$m photons, and the relative efficiency at other wavelengths was determined in the laboratory and verified with spectropolarimetry observations of Arcturus. The degree of polarization is $p=(q^2+u^2)^{0.5}$ and the position angle is PA = $0.5\arctan(u/q)$. The uncertainties $\sigma_{q}$ and $\sigma_{u}$ associated with the Stokes parameters were assumed to equal 1/SNR, where SNR is the signal-to-noise ratio at each corresponding pixel, found to be a conservatively good estimate from studies of many data sets.
The uncertainties were then propagated through the computation to obtain $\sigma_p$ 
and the PA uncertainty $\sigma\textsubscript{PA}=\sigma_p/2$.
The degree of polarization is debiased using the modified asymptotic estimator developed by \citet{pla14}, a new method of debiasing that more correctly handles the case of $\sigma_p \geq p$:
\begin{equation}
	p_\text{debias}\, =\, p \,-\, \frac{\sigma_{p}^{2}}{2p} \, \left(\,1- \emph{exp}\left[-\left(\frac{p}{\sigma_{p}}\right)^2\right]\,\right)
\label{debias}
\end{equation}
When $\sigma_p \ll p$ the above equation tends toward the usual debiasing technique $p_\text{debias} = \sqrt(p^2 - \sigma_{p}^2)$,
but as $\sigma_p \gg p$ the equation tends smoothly toward $p_\text{debias} = p/2$.

Instrumental polarization was determined during the commissioning of CanariCam and found to be $\sim$0.6\% with PA that is a linear function of telescope and instrument angles. Many subsequent observations of non-polarized stars indicate that the instrumental polarization is stable and accurately characterized at both the Nasmyth and Folded-Cassegrain focal stations of the telescope. Stokes parameters are corrected for instrumental polarization in the $(q,u)$ plane of the detector frame. 

\subsection{The Egg Nebula} \label{sec:Egg}

The Egg Nebula (AFGL 2688) is a post-Asymptotic Giant Branch (post-AGB) star in the process of shedding its outer layers and transitioning between the AGB and planetary nebula phases. Illuminated by the hot core, the nebula's unique bipolar structure and striking appearance are due to the complex interactions of stellar winds and expelled material. The Egg Nebula is a well-known carbon-rich evolved object
\citep{goto2002, Wesson2010} and is a useful end-point in the carbon-silicate sequence.

\subsubsection{N-Band Imaging Polarimetry of the Egg} \label{sec:Egg-N}

Imaging polarimetry observations of the Egg Nebula using the N-band filter were performed with CanariCam in 2014 (Table 1). The N-band filter covers the full range of 7.5 to 13 $\mu$m, centered at 10.36 $\mu$m, so it cannot be used in the Aitken method of decomposition but does provide a useful perspective. The resulting intensity and polarization images are shown in Figure \ref{Egg-N-band} for pixels that have SNR $> 10$, where signal S is always the total intensity. Orientations and lengths of the over-plotted line segments indicate the PAs and amounts of polarization, respectively. Line segments are two-sided vectors, since PA is identical for 0\degr and 180\degr. Here, we refer to them simply as vectors. These are plotted at two pixel intervals on the images.

Colors in Figure \ref{Egg-N-band}a show the total N-band intensity in units of Jy/$arcsec^2$ with values indicated by the adjacent logarithmic color scale. Colors in Figure \ref{Egg-N-band}b show the percent polarization as indicated by corresponding linear color bar scale, with peak polarization of 16.8\%. Observations at two telescope pointings were performed with enough offset to image most of the Egg nebula region while also overlapping to facilitate registration of the data.
The origin of the images is set at the intensity peak and is indicated by horizontal and vertical dashed lines. The images have been smoothed with $3 \times 3$ pixel boxcar moving average to reduce noise, which decreases the resolution slightly from 0\farcs35 to 0\farcs42 full-width half-max (FWHM) of the PSF (point spread function).

The result in Figure \ref{Egg-N-band} shows for the first time the presence of strong scattering polarization in the mid-IR for the Egg Nebula. The circular arrangement of PA vectors around the mid-IR peak are consistent with polarimetry of the Egg at wavelengths from visible to near-IR \citep{Sahai1998, Weintraub2000, Kast2002}, measuring up to 70\% polarization at 2 $\mu$m. However, because the filter is so broad, N-band imaging polarimetry by itself is not sufficient to assess the relative contributions to polarization from silicates, however small, and carbonaceous dust. To do that we must utilize multi-wavelength polarimetry and the extended Aitken method to "unfold" the sightline components. 

\begin{figure*}
\centering\includegraphics[width=505pt]{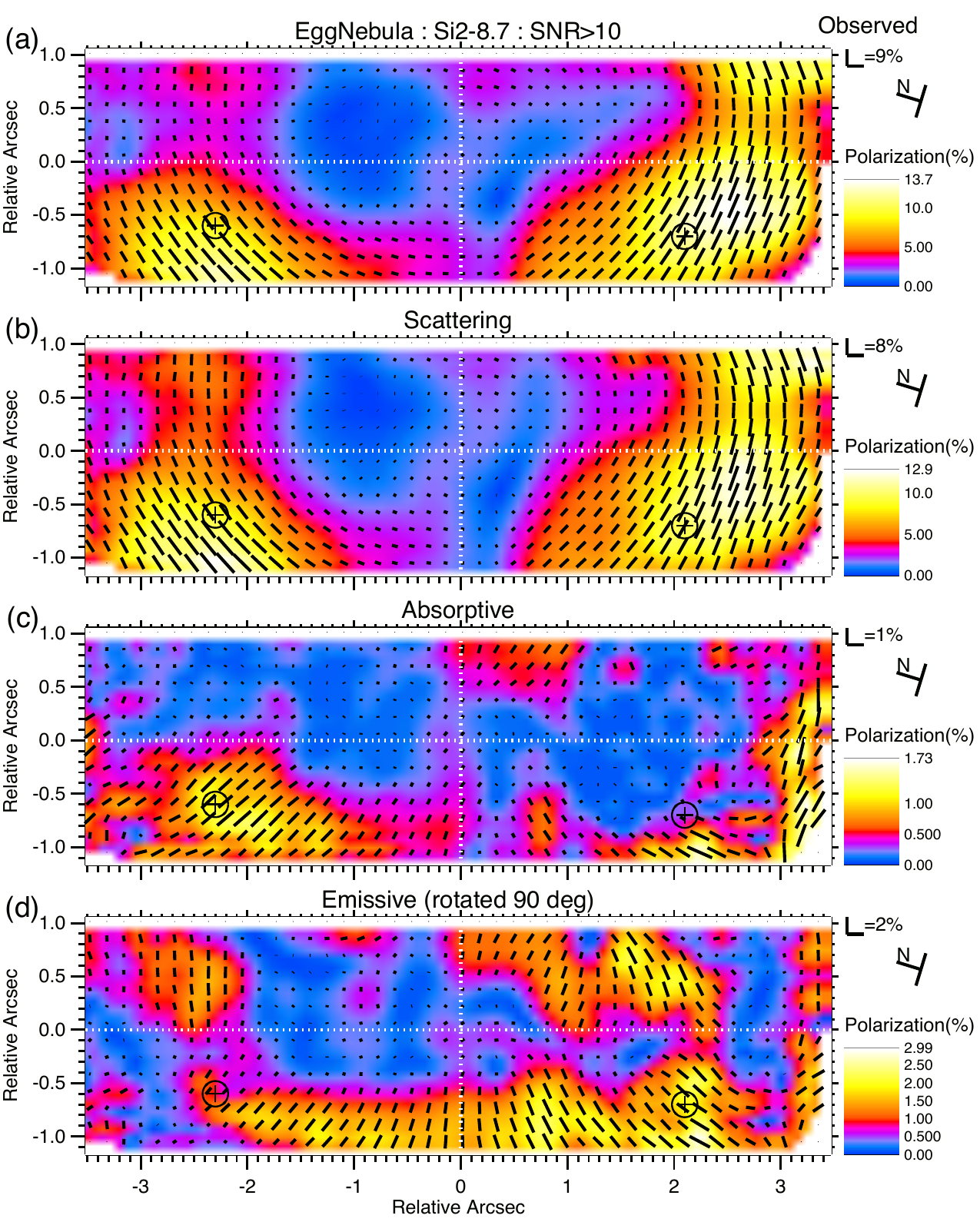}
\caption{Results of the extended Aitken method fits at 8.7 $\mu$m for the Egg Nebula: (a) observed polarization; (b) scattering component; (c) absorptive component; (d) emissive component rotated by 90\degr to show alignment with possible B-field. Linear color scales indicate the polarization magnitudes \emph{p}(\%), and over-plotted vectors show the PA and polarization. Images of Stokes parameters are first smoothed with $3 \times 3$-pixel boxcar, followed by computation of the polarization (\%) and PA. Data are shown where SNR$>$10 for all wavelengths. Note the tilt of FOV with respect to north. Circled cross symbols are sightlines at which details of fits are presented in Figures \ref{Egg-AE+EAS-PeakPolo} and \ref{Egg-AE+EAS-SPL}.}
\label{Egg-imgPolVecs-EAS}
\end{figure*}

\begin{figure*}
\centering\includegraphics[width=505pt]{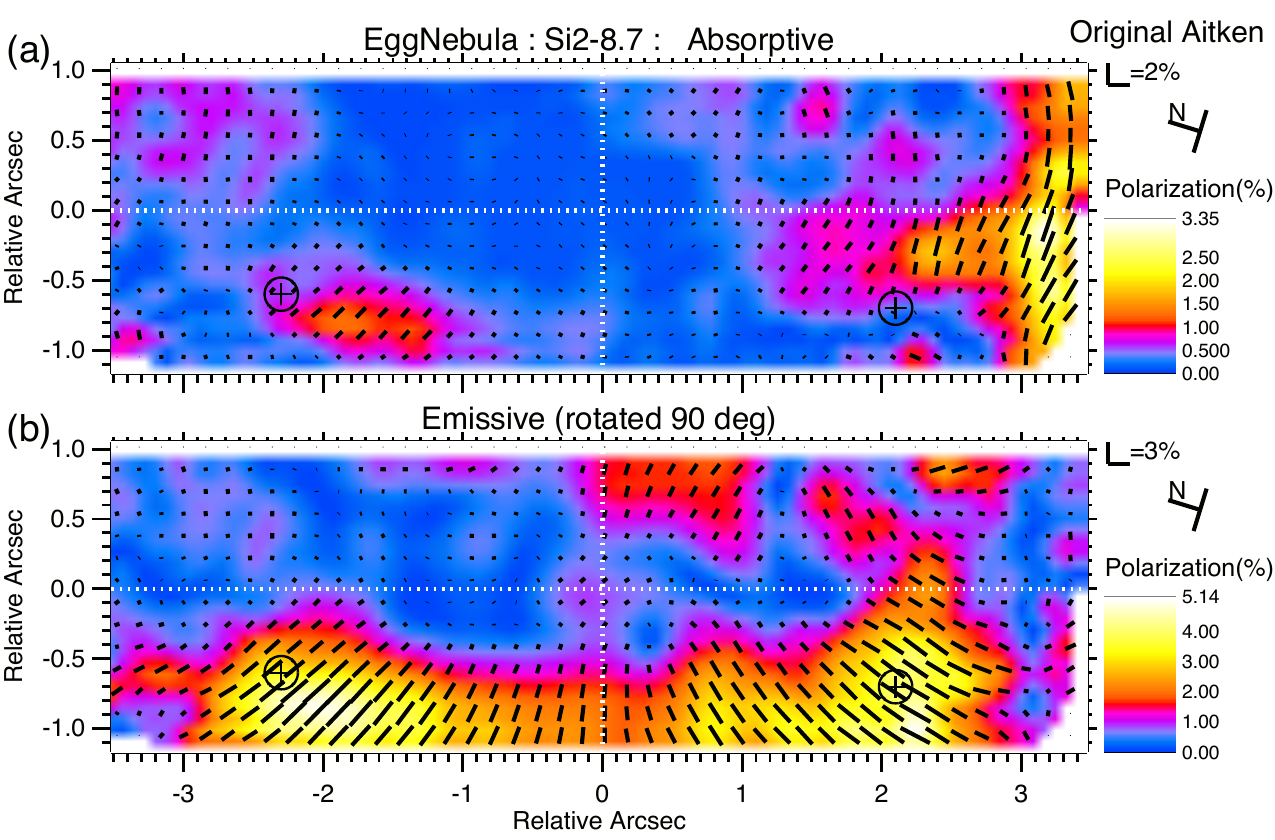}
\caption{Results of the original Aitken method fits at 8.7 $\mu$m for the Egg Nebula: (a) absorptive component; (b) emissive component rotated by 90\degr to show alignment with possible B-field. Linear color scales indicate the polarization magnitudes \emph{p}(\%), and over-plotted vectors show the PA and polarization. Images of Stokes parameters are first smoothed with $3 \times 3$-pixel boxcar, followed by computation of the polarization (\%) and PA. Data are shown where SNR$>$10 for all wavelengths. Note the tilt of FOV with respect to north. Circled cross symbols are sightlines for which details of fits and comparison with the extended method are presented in Figures \ref{Egg-AE+EAS-PeakPolo} and \ref{Egg-AE+EAS-SPL}.}
\label{Egg-imgPolVecs-EA}
\end{figure*}

\subsubsection{Narrow-Band Imaging Polarimetry of the Egg} \label{sec:Egg-NB}

Multi-wavelength imaging polarimetry observations of the Egg Nebula were performed in October 2020, after CanariCam was moved from the Nasmyth platform to the Folded-Cass focal station. Unfortunately, even though observations were performed at two telescope pointing positions (for each wavelength), the offset between positions was such that the frames overlapped significantly in such a way that the vertical extent of final (combined) image is smaller than the field of view of the final N-band image. The imaging polarimetry observations taken at three wavelengths (8.7 $\mu$m, 10.3 $\mu$m, 12.5 $\mu$m) were reduced into Stokes \textit{q} and \textit{u} images, that were each smoothed with $3 \times 3$ pixel box moving average and then fit with the original and extended Aitken methods at each pixel sightline.

The observed narrow-band polarization at 8.7 $\mu$m is shown in Figure \ref{Egg-imgPolVecs-EAS}a, with the other panels in Figure \ref{Egg-imgPolVecs-EAS} showing the polarization due to (b) scattering, (c) absorption, and (d) emission, as computed by the extended Aitken method. The origin of the images corresponds to the observed mid-IR intensity peak at the intersection of the horizontal and vertical dotted white lines. Over-plotted vectors show the PA, and the lengths indicate the amount of polarization, which is also shown by the adjacent linear color bar scales. Only pixels with SNR $>10$ for the intensity at all wavelengths are shown, and the cutoff is evident at the right edges of images. The emissive component's PA vectors are rotated by 90\degr\, to correspond to the orientation of the projected magnetic field inferred to align the silicate dust grains in the emitting region, which would be spinning with longer axis perpendicular to the magnetic field lines. We see that the emissive polarization's PA rotated by 90\degr tends to agree with the PA of the absorptive polarization. From this we can conclude that the magnetic field morphology changes slowly enough with distance along the various sightlines that in effect the same magnetic field threads the emitting and foreground absorbing regions. More analysis of the absorptive and emissive component PA distribution is presented in Section \ref{sec-PAD}.

The polarization attributed to scattering by carbonaceous grains, shown in Figure \ref{Egg-imgPolVecs-EAS}b, is organized with PA perpendicular to rays from the central source, similar to Figure \ref{Egg-N-band}, indicating that the scattering component is successfully separated from the emissive and absorptive components of polarization due to silicates, using the extended Aitken method. This data set provided the motivation for developing the extended Aitken method. The scattering component is mostly detected by the 8.7 $\mu$m polarization, whereas absorptive and emissive polarization components are connected with the longer wavelengths. Keep in mind that the spatial distributions of polarization and PA for all components is exactly the same at all wavelengths because of the fitting method presented by Equations \ref{qfit} and \ref{ufit}, in which the solution coefficients are independent of wavelength. Only the magnitude of polarization for each component changes according to the template profiles variation with wavelength, as shown by Equation \ref{scatp}, whereas the PA of each component is exactly the same at all wavelengths as shown by Equation \ref{scatPA}. The observed PA of polarization may vary with wavelength, which indicates the existence of multiple components of polarization. Table \ref{Table-PeakPol-Egg} shows the peak polarization at each wavelength for observations and each component of the extended Aitken method.

\begin{table}
\caption{Peak Polarization of Egg Observations and Components}
\label{Table-PeakPol-Egg}
\centering
\begin{tabular}{lrrr}
\hline
 Component & 8.7 $\mu$m & 10.3 $\mu$m & 12.5 $\mu$m \\
\hline
 Observed & 13.7 \% & 11.4 \% & 8.3 \% \\
\hline
 Scattering & 12.9 \% & 8.2 \% & 4.3 \% \\
\hline
 Absorptive & 1.7 \% & 7.1 \%  & 2.8 \% \\
\hline
 Emissive & 3.0 \% & 7.7 \%  & 7.5 \% \\
\hline
\end{tabular}
\end{table}

An alternative explanation for the circular arrangement of polarization PA could be due to radiative external alignment of elongated graphite grains, as presented by \cite{And2022}. Their observations of IRC+10216 at far-IR wavelengths ($> 50 \mu$m) found emissive polarization with centrosymmetric radial PA pattern, pointing toward the source. Radially aligned graphite grains that produce emissive polarization in the far-IR could be creating absorptive polarization at mid-IR wavelengths that would be perpendicular to rays from the source. Since the absorption cross-section of graphite has similar variation as the albedo versus wavelength in the mid-IR, the third component of the extended Aitken method would then be fitting the absorptive polarization.

However, since the Egg Nebula exhibits about 50\% polarization at near-IR wavelengths, which is attributed mostly to scattering \citep{Weintraub2000, Kast2002}, it seems logical that such scattering polarization would also occur at mid-IR wavelengths. The observed polarization and inferred scattering polarization given in Table \ref{Table-PeakPol-Egg} approximately follow the albedo of graphite. Since the albedo increases with shorter wavelengths, that is consistent with scattering being the main cause of polarization in the Egg Nebula at both mid-IR and near-IR wavelengths. Given the large amount of scattering polarization, the albedo curve of Figure \ref{DustAlbedo-maxgs=1um}, which includes grain sizes up to 1 $\mu$m, could be more relevant to the Egg Nebula than the albedo from maximum grain size 0.25 $\mu$m in Figure \ref{DustAlbedo}. Figure \ref{DustAlbedo-graphite-maxgs=1um+025um} shows that the two cases of maximum grain size produce albedo curves with almost the same variation in the mid-IR, but have much different variations in the near-IR. Estimation of the maximum grain size and discussion of the consequences of large grains is an interesting topic for a paper specifically about the Egg Nebula.

Figure \ref{Egg-imgPolVecs-EA} shows the results of fitting the observed Stokes spectra with the original Aitken method having just two components, emissive and absorptive silicates. Again, emissive component's PA vectors, in panel (b), are rotated by 90\degr\, to see if the vectors agree with the absorptive component in panel (a). Agreement of absorptive PA and rotated emissive PA components is evident to the left of the origin in the image, whereas to the right of the origin the PA vectors mostly do not agree. Note that the magnitudes of polarization inferred by the original Aitken method are higher overall than in Figure \ref{Egg-imgPolVecs-EAS}, and are actually not good fits to the data, as we discuss next.

\begin{figure*}
\centering
\includegraphics[width=240pt, keepaspectratio=true]{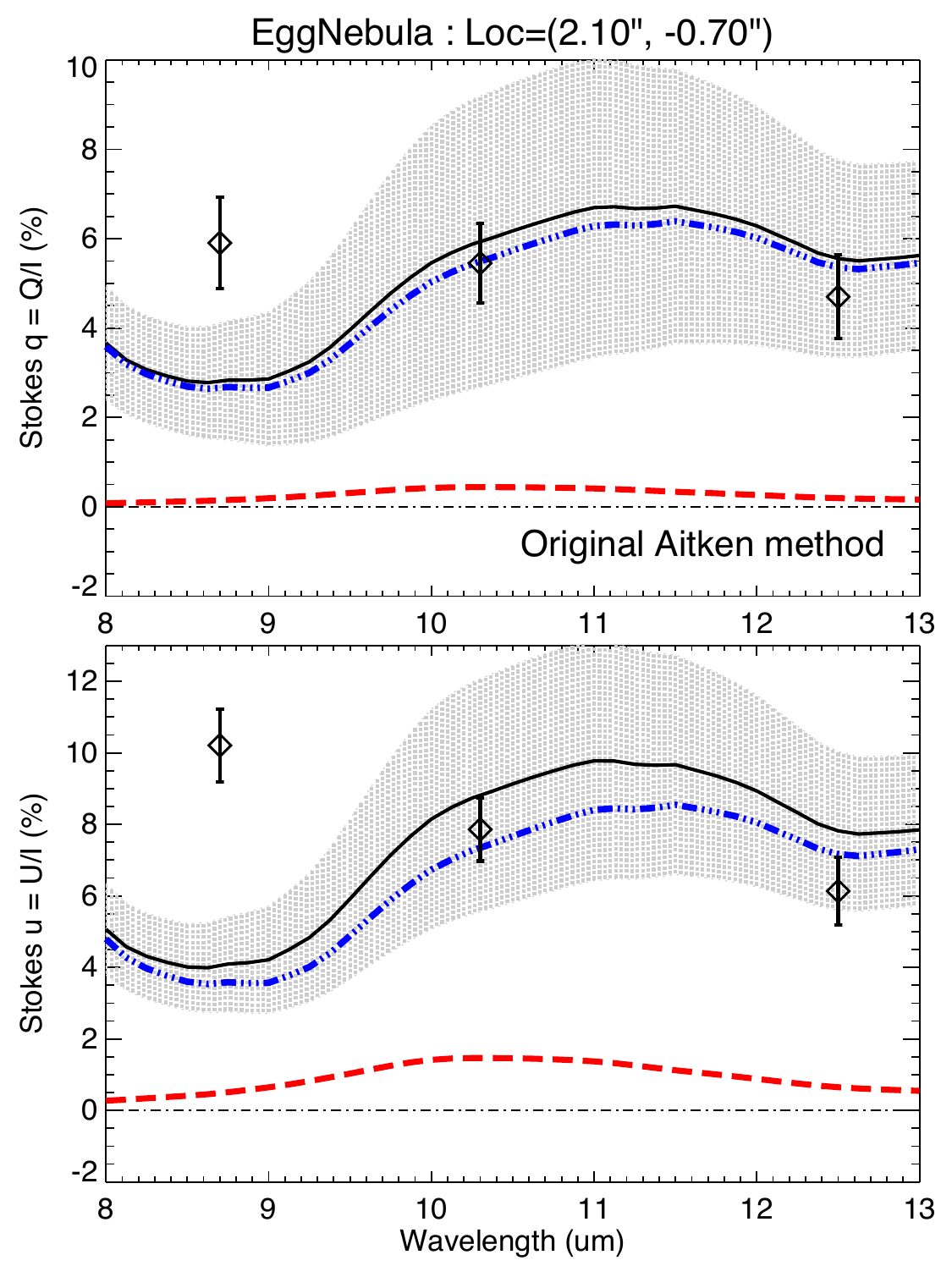}
\includegraphics[width=240pt, keepaspectratio=true]{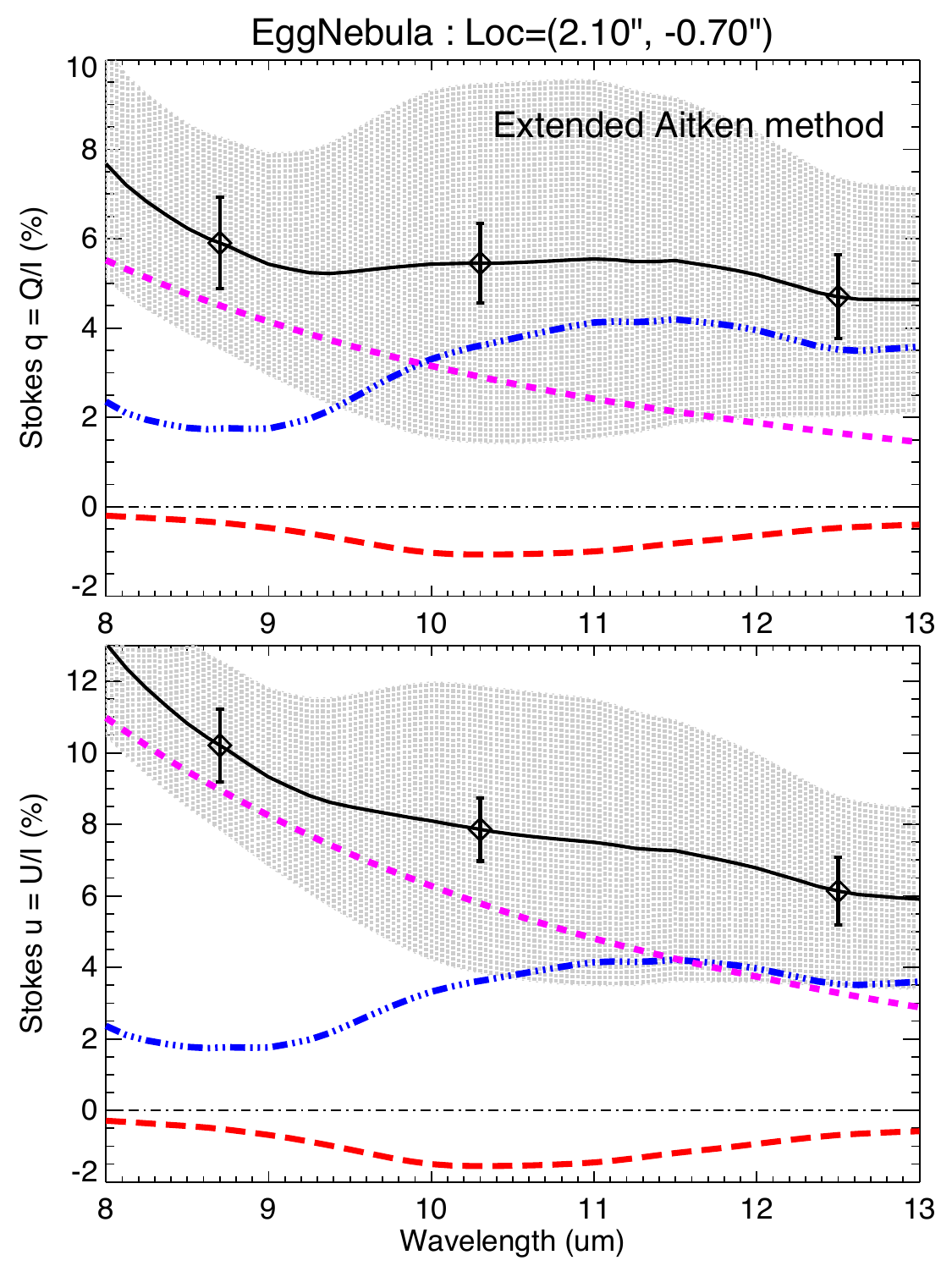}
\includegraphics[width=240pt, keepaspectratio=true]{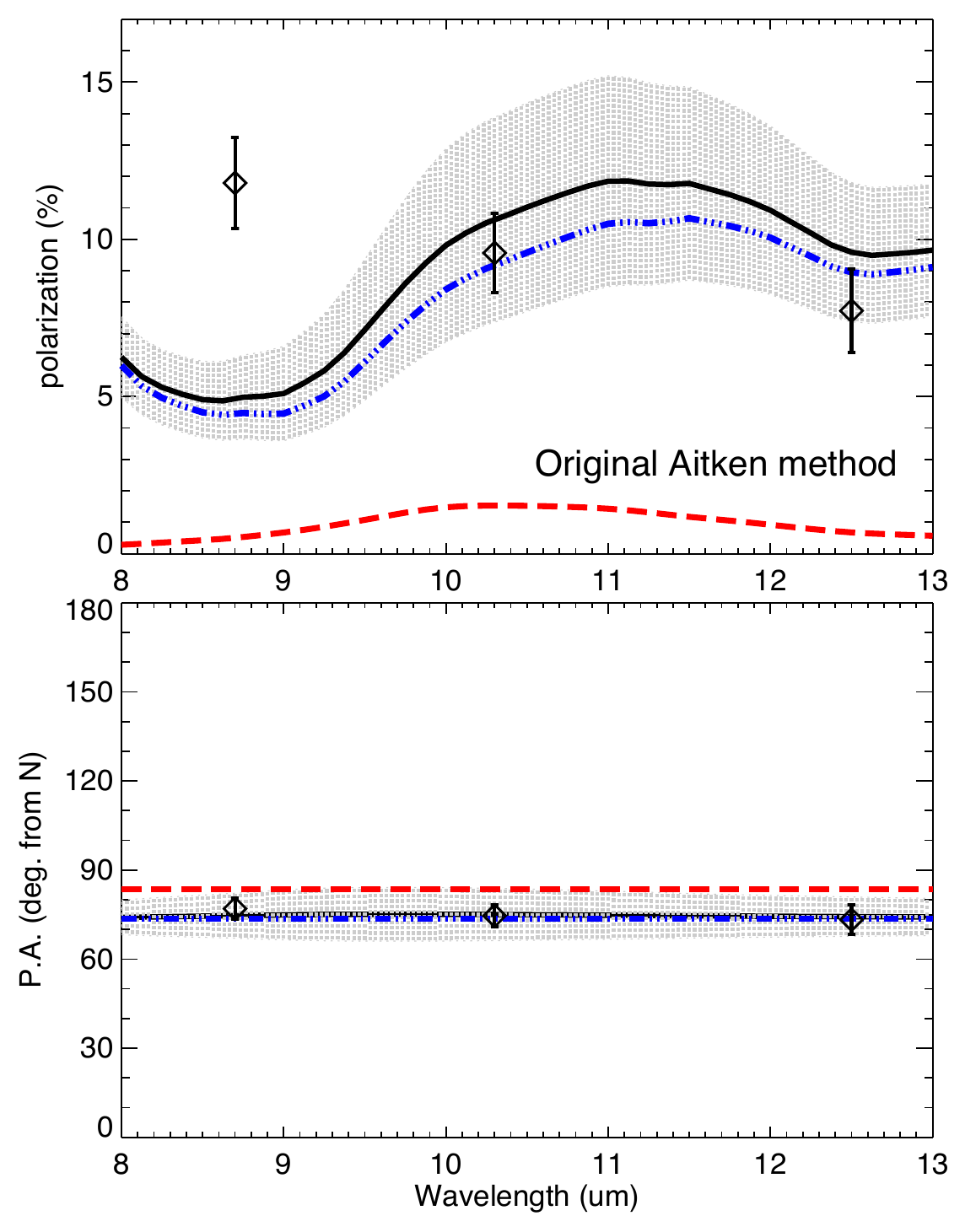}
\includegraphics[width=240pt, keepaspectratio=true]{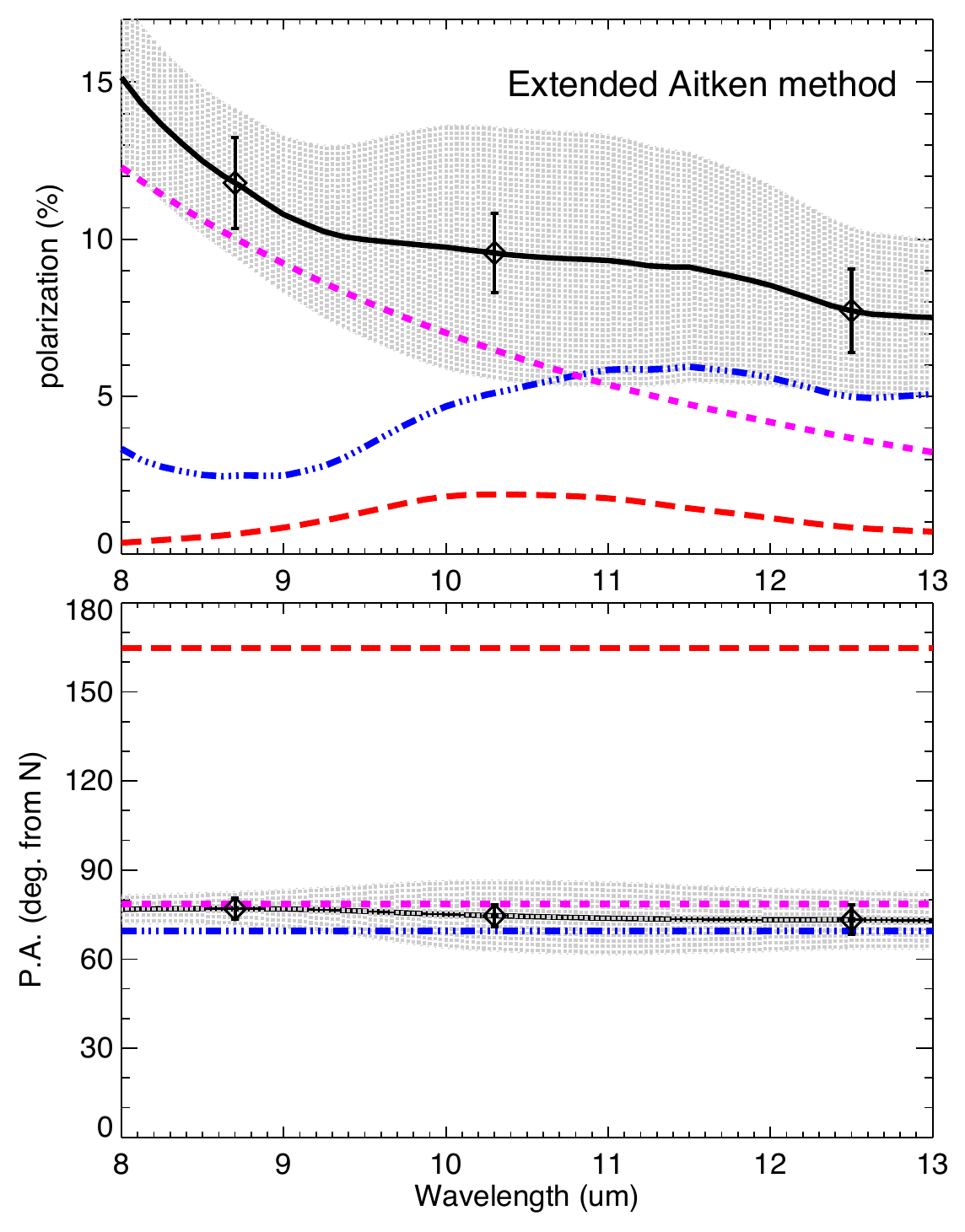}
\caption{Original and extended Aitken method fits to multi-wavelength imaging polarimetry of Egg Nebula at the sightline marked with circled cross symbol to the right of origin in images, with offset
(+2\farcs1, -0\farcs7). Observed Stokes $q$ and $u$ are plotted as diamonds with error bars (averages and uncertainty in 0\farcs24$\times$0\farcs24 apertures). Bottom row of plots show the \% polarization and PA computed from the Stokes spectra. Left column of plots show the fits with two components of the original Aitken method: absorptive (red dashed) and emissive (blue dot-dashed). Right column of plots show the extended Aitken method fits that also include scattering (violet short-dashed curves). Solid black curves are the sum of components, with gray shading indicating the range of 1-sigma uncertainty.}
\label{Egg-AE+EAS-PeakPolo}
\end{figure*}

\begin{figure*}
\centering
\includegraphics[width=240pt, keepaspectratio=true]{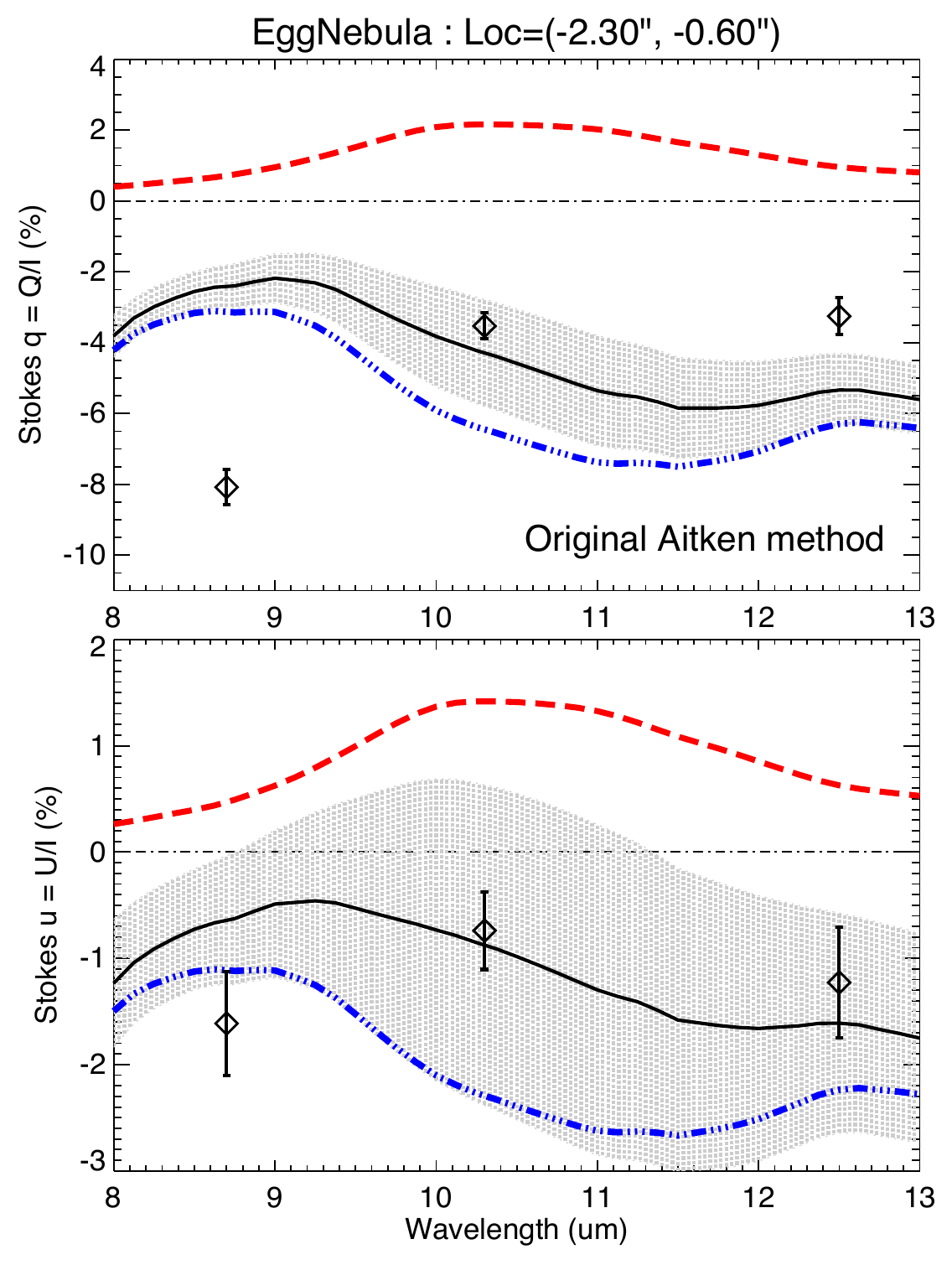}
\includegraphics[width=240pt, keepaspectratio=true]{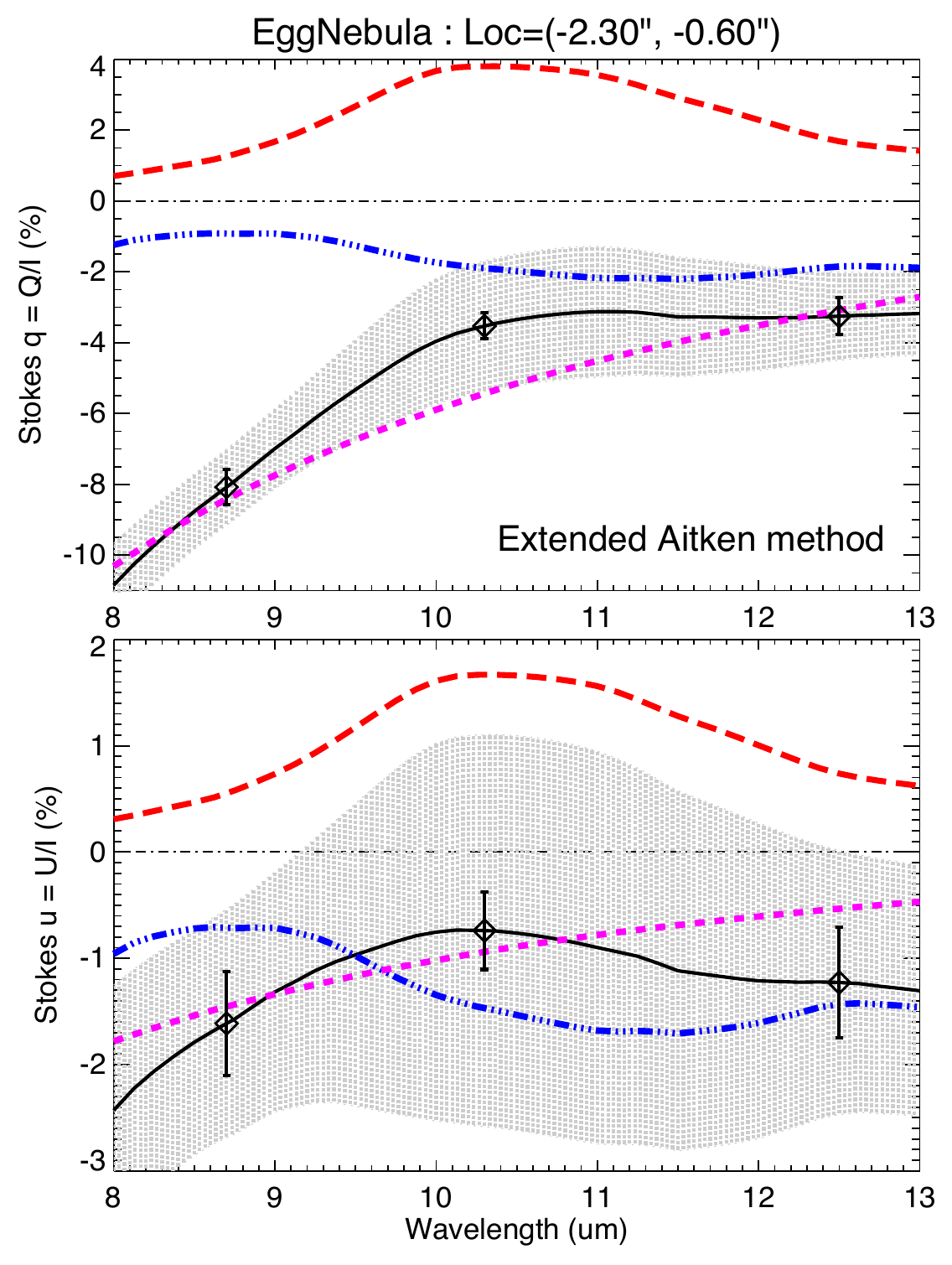}
\includegraphics[width=240pt, keepaspectratio=true]{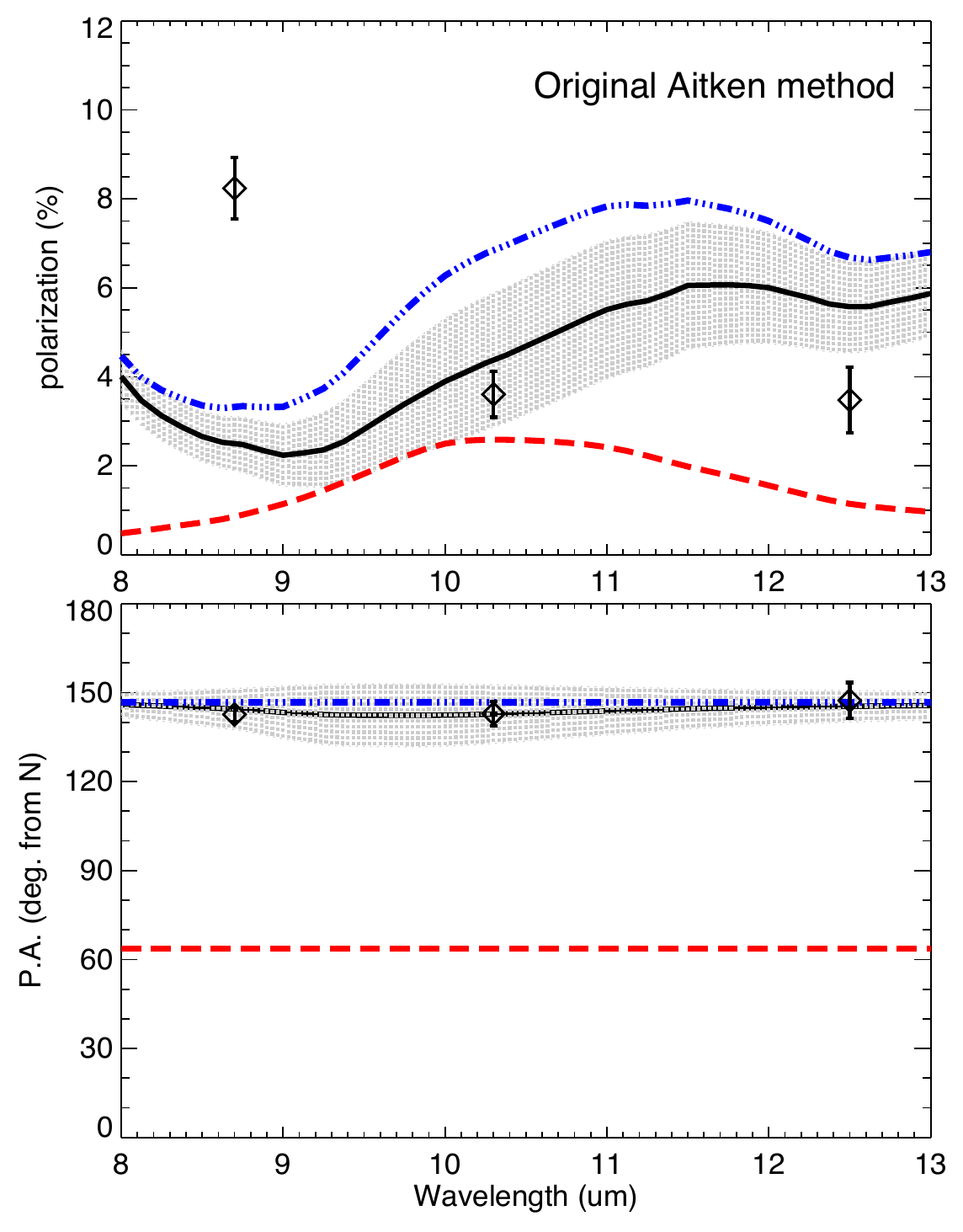}
\includegraphics[width=240pt, keepaspectratio=true]{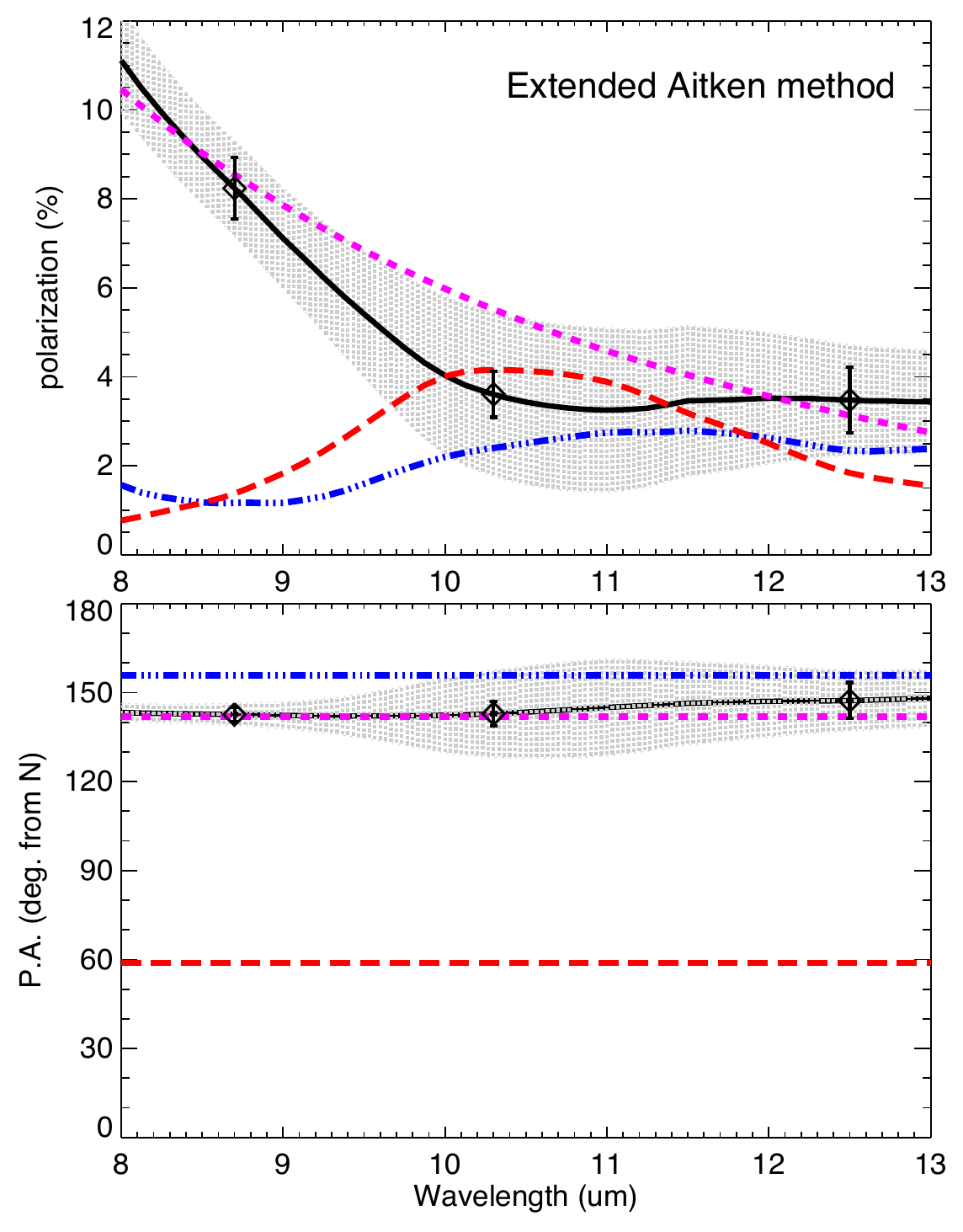}
\caption{Aitken method fits to multi-wavelength imaging polarimetry of the Egg Nebula at sight-line with offset (-2\farcs3, -0\farcs6) to the left of origin in Figure \ref{Egg-AE+EAS-PeakPolo}, marked with circled cross. Observations of Stokes $q$ and $u$ are are plotted as diamonds with error bars in top rows (averages and uncertainty in 0\farcs24$\times$0\farcs24 apertures). Corresponding \% polarization and PA in degrees are plotted in bottom rows. Left column of plots show the fits to Stokes spectra with the original Aitken method: absorptive (red dashed curves) and emissive (blue dot-dashed curves). Right column of plots show the extended Aitken method fits that also include scattering (violet short-dashed curves). Solid black curves are the sum of components, with gray shading indicating the range of 1-sigma uncertainty.}
\label{Egg-AE+EAS-SPL}
\end{figure*}

To compare results obtained with the original and extended Aitken methods, we consider fits to the data using both methods at specific sightlines marked on the images by the crosses inside circles. In plots of Stokes spectra and fits below, plots on the left side will show results from the original Aitken method, and plots on the right side will show results from the extended Aitken method. For each sightline, plots of both the fits to Stokes spectra and the resulting percent polarization and PA calculated from the Stokes spectra are presented.

Figure \ref{Egg-AE+EAS-PeakPolo} shows details of fits to Stokes spectra at the sight-line with offset (+2\farcs1, -0\farcs7) from the origin. This is at the circled cross mark to the right of origin in the images of Figure \ref{Egg-imgPolVecs-EAS}, near the peak of the observed polarization. The average Stokes $q(\lambda)$ and $u(\lambda)$ in $3\times3$ pixel box apertures (0\farcs24$\times$0\farcs24) at each of the three wavelengths are plotted as diamonds with error bars. The error bars show estimates of the uncertainty estimate at the center pixel of each aperture. The SNR of intensity in the apertures at the three wavelengths is between 50 and 110, which gives estimates of the absolute uncertainty of the polarization to be about +/-2.8\% to +/-1.3\%. As indicated, the plots on the left show the original Aitken method fits to the observed Stokes spectra using just two components, and the plots on the right side show the fits to Stokes spectra using three components of the extended Aitken method. The dashed-dotted blue curves are the emissive components, long-dashed red curves are the absorptive components, and short-dashed violet curves show the scattering component (right side only). The solid curves are the sum of components, which aim to fit to the observed Stokes spectra. The ranges of 1-sigma uncertainty of the fits are indicated with gray shading. Using just two components cannot fit the Stokes values at all wavelengths, but including the scattering component (dashed violet curves) achieves an exact fit to the observed Stokes spectra.

The bottom row of plots in Figure \ref{Egg-AE+EAS-PeakPolo} shows the percent polarization and PA computed from the Stokes spectra. Polarization and PA from observations are plotted with diamond symbols, and the polarization components from Aitken method fits are blue dot-dashed curves for emissive, red long-dashed curves for absorptive, and violet short-dashed curves for the scattering component. At this sightline near maximum observed polarization, the scattering polarization is the dominant component, with $p_S(8\mu\text{m}) = 15\%$ and $p_S(13\mu\text{m}) = 7\%$. This was expected from the fact that the PA of observed polarization has mostly a circular arrangement around the peak of mid-IR intensity, and the typical polarization detected at 2 $\mu$m and 4.5 $\mu$m is around 50\% \citep{Weintraub2000, Kast2002}, which is mostly from scattering.

Note that the difference between PA of emissive and absorptive components is closer to being perpendicular when three components are used to fit the Stokes spectra, as would be expected if silicate dust grains are aligned by magnetic fields. However, since the actual emission and absorption occurs in a three dimensional environment, the emissive grains could be in different regions of the magnetic field than the absorptive grains along the sightline, so then the observed emissive and absorptive polarization PA may not be exactly orthogonal. More about the distribution of differences in PA between components is presented in Section \ref{sec-PAD}. 

Figure \ref{Egg-AE+EAS-SPL} shows fits to the Stokes spectra observed at the second sightline, marked with a circled cross at offset (-2\farcs3, -0\farcs5) to the left of the origin, near another peak of polarization in the Egg nebula. Observations are plotted as diamonds with error bars. The SNR of intensity in 0\farcs24$\times$0\farcs24 apertures is between 120 and 230 so that the uncertainty of Stokes values is $<1\%$ (on percentage scale). Again, plots on the left side show the results from fits to the observed Stokes spectra using just two components of the original Aitken method: red-dashed curve is absorptive component, blue dot-dashed curve is the emissive component, and the solid black curve is the sum of components.
The ranges of 1-sigma uncertainty are indicated with gray shading. The plots on the right side show the results using the extended Aitken method with scattering component that is plotted with the violet short-dashed curve. The percent polarization and PA computed from Stokes spectra are plotted in bottom rows of Figure \ref{Egg-AE+EAS-SPL}. The original Aitken method appears to fit the Stokes $u(\lambda)$ reasonably well, but fails to fit the observed $q(\lambda)$ spectrum, thereby failing to fit the overall polarization and overestimating the emissive component due to silicates. The extended Aitken method finds that this sightline is also dominated by scattering polarization, with $p_S(8\mu\text{m})=10\%$ and $p_S(13\mu\text{m})=3\%$, and so again the scattering component is needed to fit the Stokes spectra and provide more accurate values for the emissive and absorptive components.

We find that most sightlines in the Egg Nebula imaging polarimetry data cube require three components to correctly fit the Stokes spectra, and this is accomplished with the extended Aitken method. Thus, for objects that have scattering in the mid-IR, it is essential to include the scattering polarization component with the Aitken method to correctly model the emissive and absorptive components of polarization by silicates.
\subsection{W3 IRS5} \label{sec:W3}

W3 IRS5 is a high-mass star-forming region with two bright infrared sources, a bipolar outflow, diffuse emission, and multiple radio sources \citep{Tak2005}. CanariCam observed W3 IRS5 in both imaging polarimetry (2012) and spectropolarimetry (2014) modes (Table \ref{Table1}). We have fit the Stokes spectra with both the original and extended Aitken methods and show that including the third component improves the fit. The third component could be due to scattering, emission from dust other than silicates, or possibly silicates different than those in Orion. Spectropolarimetry observations support the results from imaging polarimetry.
\newline

\begin{figure*}
\includegraphics[height=467pt, keepaspectratio=true]{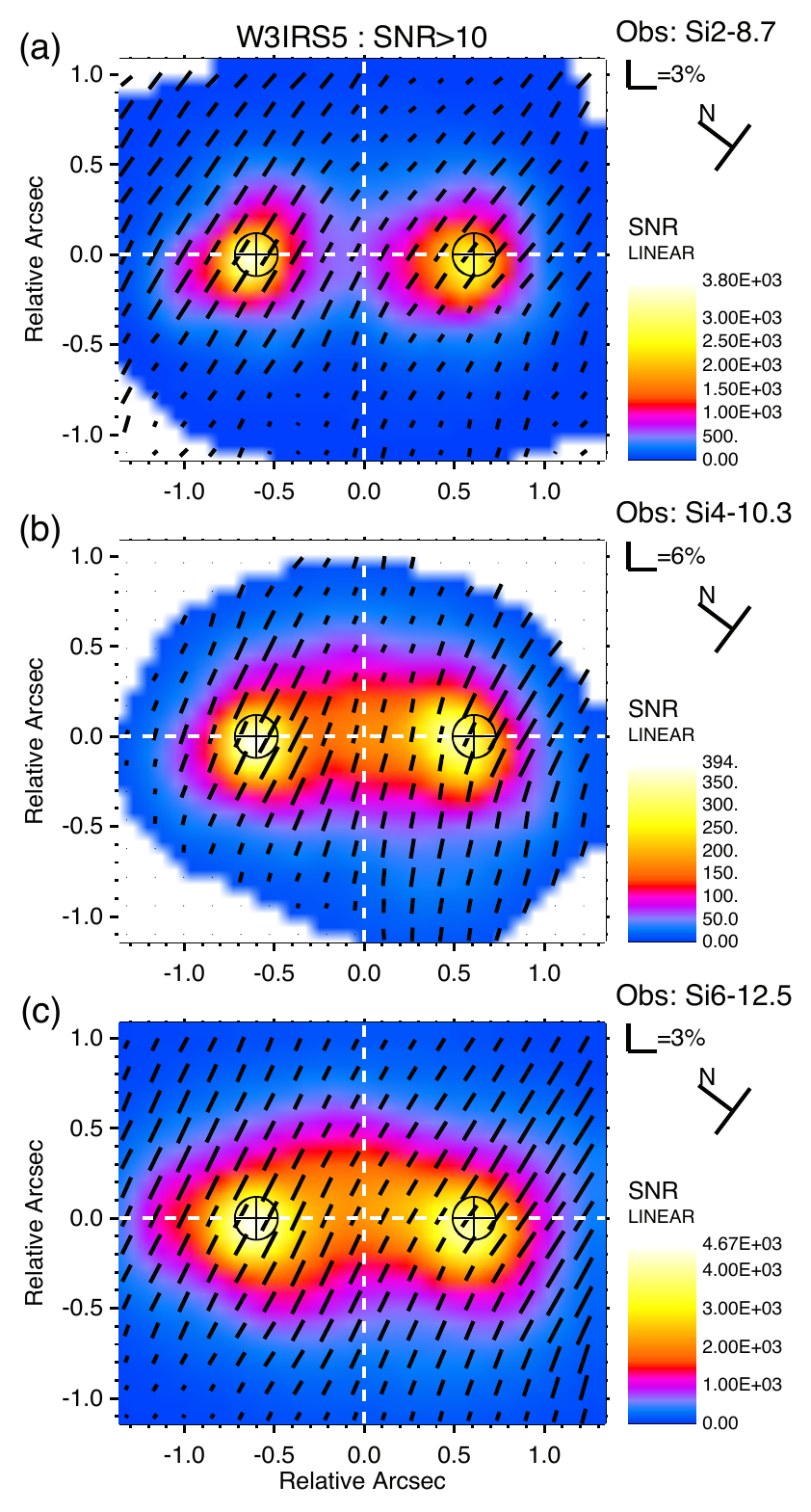}
\includegraphics[height=467pt, keepaspectratio=true]{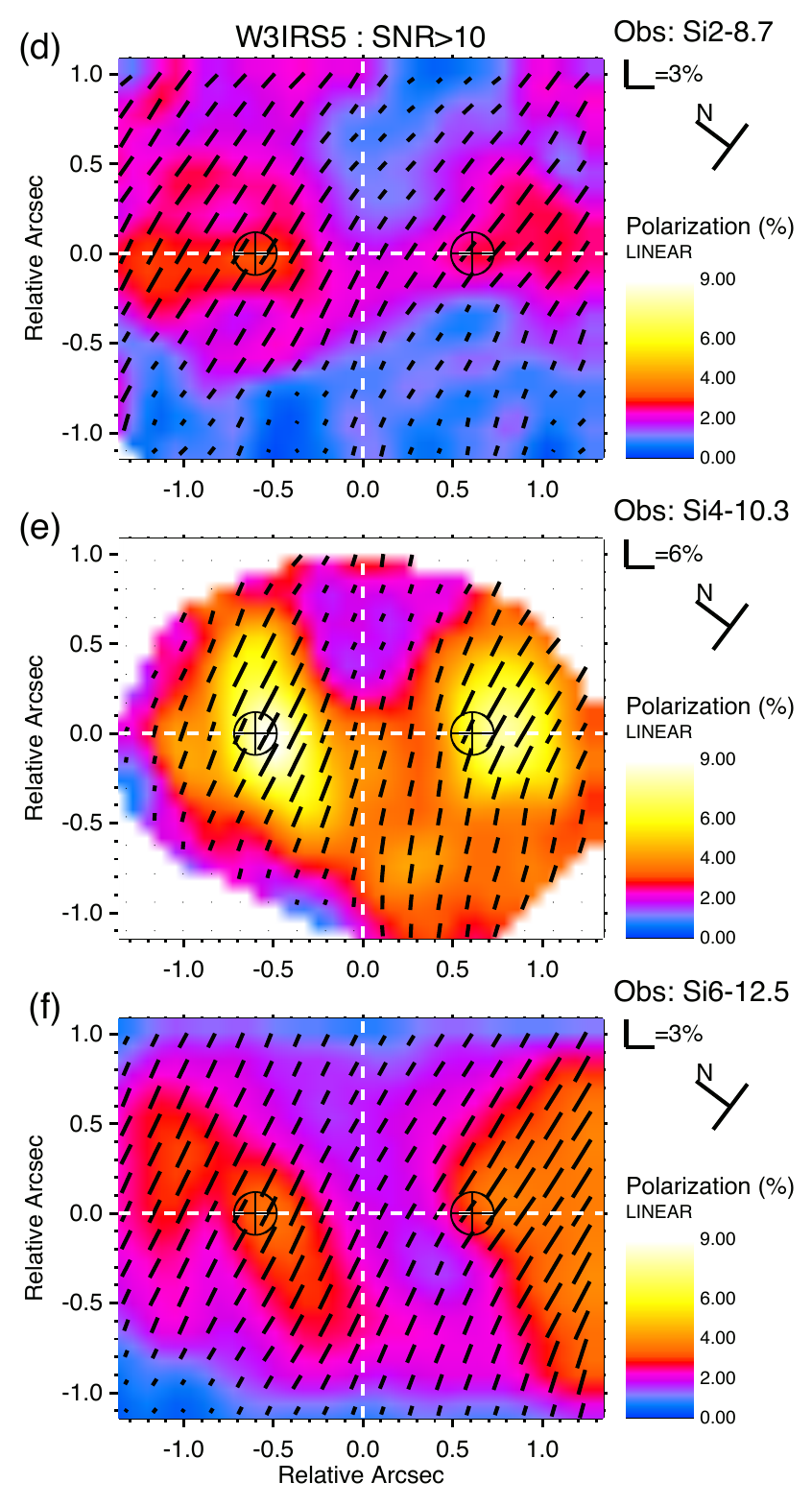}
\caption{Multi-wavelength imaging polarimetry of W3 IRS5. The left column of images show in colors the signal to noise ratio (SNR) of the intensity at the three wavelengths:
(a) 8.7 $\mu$m, (b) 10.3 $\mu$m, and (c) 12.5 $\mu$m. Linear color bar scales indicate the SNR corresponding to colors, and polarization vectors are plotted over the images. Only data for pixels at which SNR$>$10 are displayed. The right column of images show the percent polarization, as indicated by the color bar scales, with the same polarization vectors, at: (d) 8.7 $\mu$m, (e) 10.3 $\mu$m, and (f) 12.5 $\mu$m. Note that all the color bar scales of polarization have the same range, and the polarization is larger for the 10.3 $\mu$m images. The centroids of the two sources at 8.7 $\mu$m in (a) are marked with circled cross symbols on all images. The origin of images is set at the average of the two centroids, marked by horizontal and vertical white dashed lines. Note the tilt of the FOV with respect to north.}
\label{W3-pvecsPA-Obs-SNR+p}
\end{figure*}

\begin{figure*}
\includegraphics[height=467pt, keepaspectratio=true]{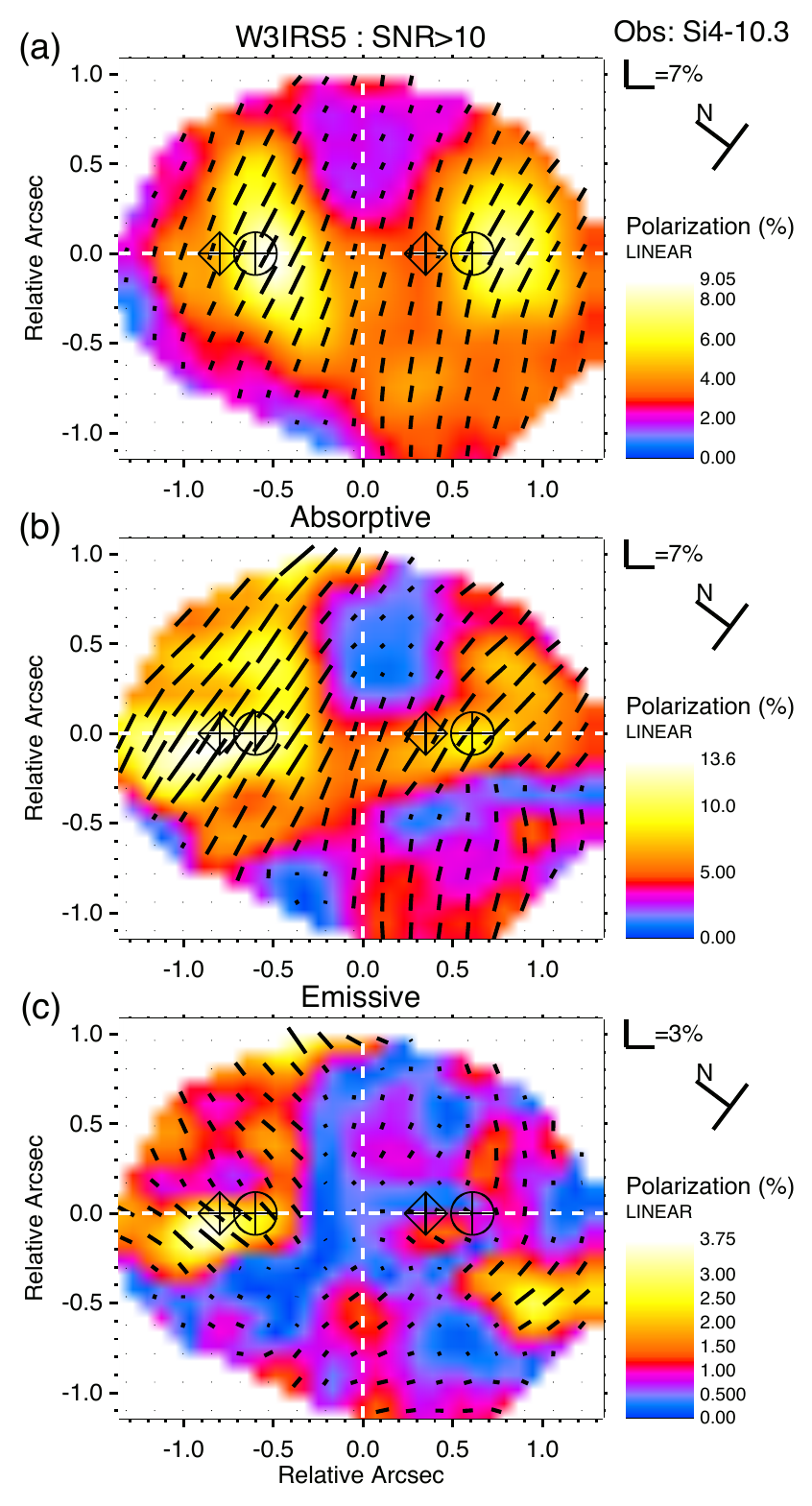}
\includegraphics[height=467pt, keepaspectratio=true]{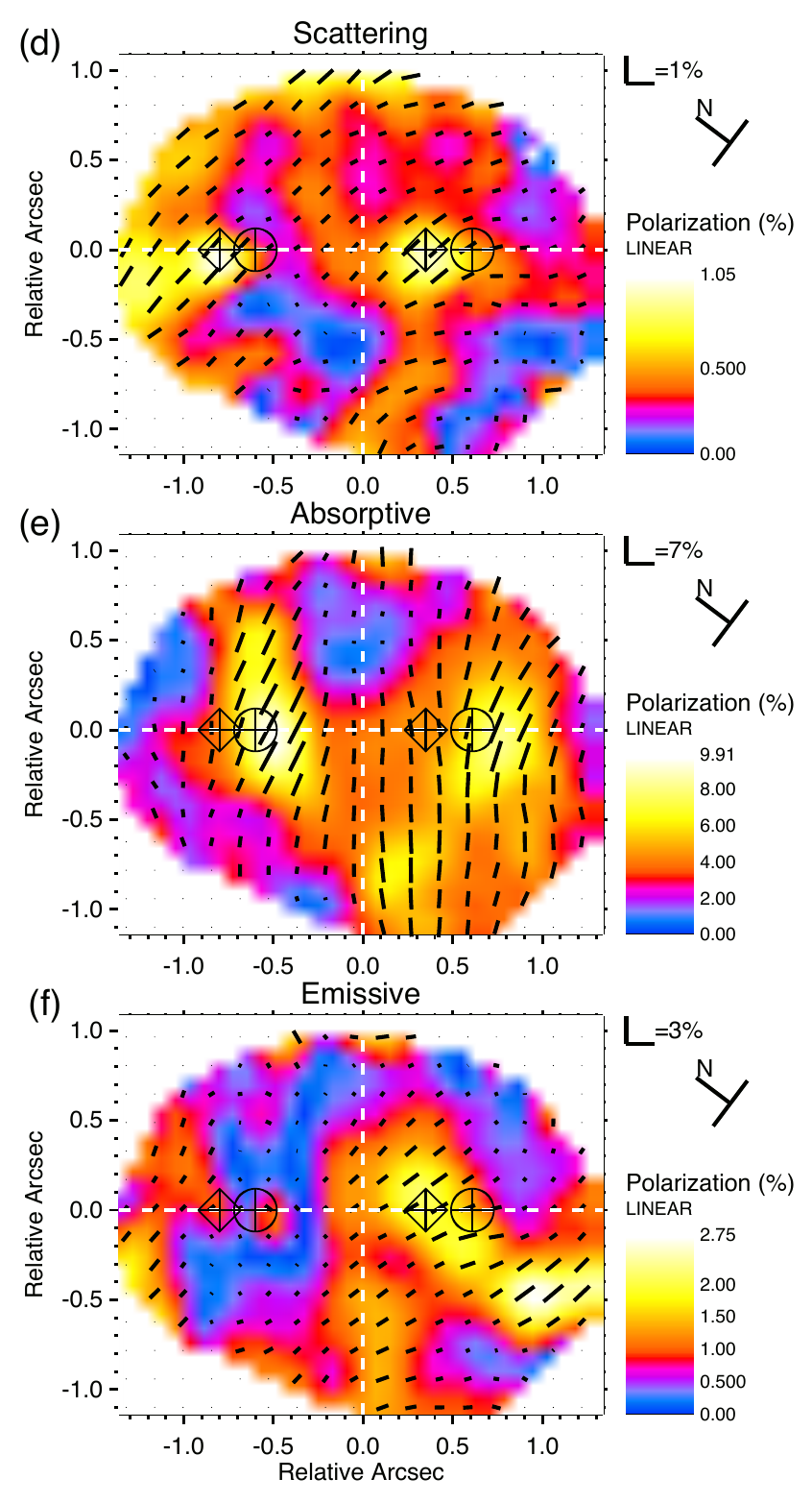}
\caption{Original and extended Aitken method fits to multi-wavelength imaging polarimetry observations of W3 IRS5. Left top panel (a) shows the observed polarization image of W3 IRS5 at 10.3 $\mu$m for pixels with SNR$>$10. Images (b) and (c) show the original Aitken method decomposition of the polarimetry into components of (b) absorptive and (c) emissive polarization. The right column of images show the extended Aitken method decomposition of the same data cube into (d) scattering, (e) absorptive, and (f) emissive polarizations, all at 10.3 $\mu$m. The centroids of the two sources at 8.7 $\mu$m are marked with circled cross symbols on all of the images. The origin of images is set at the average of the two centroids, marked by horizontal and vertical white dashed lines. The diamond symbols around crosses mark sightlines of scattering polarization peaks for which Aitken method fits of Stokes spectra are compared in Figures \ref{W3-AE+EAS-SPL} and \ref{W3-AE+EAS-SPR}.}
\label{W3-pvecsPA-EAO+EAS}
\end{figure*}

\subsubsection{Imaging Polarimetry of W3 IRS5} \label{sec:W3-imp}

Our multi-wavelength imaging polarimetry observations of W3 IRS5 were reduced into Stokes $q$ and $u$ images for each wavelength, and each image was smoothed with $3 \times 3$ pixel box moving average. Figure \ref{W3-pvecsPA-Obs-SNR+p} shows, in the left column, images of the signal-to-noise ratio (SNR) of the observations (proportional to intensity) at wavelengths (a) 8.7 $\mu$m, (b) 10.3 $\mu$m, and (c) 12.5 $\mu$m. Colors correspond to SNR values specified by the linear color-bar scale adjacent to each image; note that the 10.3 $\mu$m image has a factor of 10 lower SNR because of absorption by silicates. Vectors indicating observed polarization and PA are plotted over the SNR images at 2 pixel separations. The right column of images in Figure \ref{W3-pvecsPA-Obs-SNR+p} shows the percent polarization for the same wavelengths (d) 8.7 $\mu$m, (e) 10.3 $\mu$m, and (f) 12.5 $\mu$m, as indicated by linear color-bar scales, with the same polarization vectors plotted over the images. All the color-bar scales have the same maximum value of 9\% polarization in order to show that the polarization at 10.3 $\mu$m in image (e) is more than a factor of two larger than polarization at the other two wavelengths in images (d) and (f). However the polarization vector length scale is set a factor of two larger for the 10.3 $\mu$m in images. The regions shown in the images have pixels with SNR $>$10 for each wavelength, and so the 10.3 $\mu$m region is smaller than at other wavelengths because of the extreme absorption by silicates. The centroids of the two sources at 8.7 $\mu$m in (a) are marked with circled crosses on all images, and the average of the two centroids defines the origin of the images (horizontal and vertical white dashed lines). We refer to the two stars as the NE peak and SW peak, relative to the origin and rotated coordinate system. The peaks of sources at 10.3 $\mu$m in image (b) are slightly closer together than at other wavelengths, with about 1\farcs05 separation, whereas at 8.7 and 12.5 $\mu$m the separation of the two source peaks is about 1\farcs2. The change in separation at 10.3 $\mu$m is most likely due to the large amount of absorption by silicates, which is clearly found in spectropolarimetry observations.

We applied both the original Aitken method and extended Aitken method to fit the three wavelengths of observed Stokes spectra at each pixel sightline of the images, but only for pixels which have SNR$>$10 in the 10.3 $\mu$m data. Figure \ref{W3-pvecsPA-EAO+EAS} shows the image of the observed polarization at 10.3 $\mu$m and results of the original Aitken method decomposition into (b) absorptive polarization and (c) emissive polarization. For the same region, the right column in Figure \ref{W3-pvecsPA-EAO+EAS} shows the decomposition resulting from fits of Stokes spectra with the extended Aitken method, giving (d) scattering polarization, (e) absorptive polarization, and (f) emissive polarization. Over-plotted vectors indicate PA orientation, and the vector lengths are proportional to polarization, with colors also indicating the percent polarization. Since W3 IRS5 has very strong silicate absorption, the constraint of SNR$>$10 at 10.3 $\mu$m reduces the field of view to 2\farcs7$\times$ 2\farcs1. Other wavelengths have more SNR, but all three wavelengths are needed to apply the Aitken method, therefore restricting the region that can be analyzed. The maximum of the inferred scattering polarization (d) at 10.3 $\mu$m is about 1.1\%, and two peaks occur at locations of about 0\farcs3 to the NE of each star. Those relative peaks of scattering polarization are marked by crosses within diamond symbols on all the images, and are discussed in more detail in following graphs of the fits to Stokes spectra.

We see that the absorptive component inferred from the extended Aitken method (Figure \ref{W3-pvecsPA-EAO+EAS}e) has almost the same morphology as the observed polarization at 10.3 $\mu$m, shown in image (a), whereas the absorptive component's distribution resulting from the original Aitken method (Figure \ref{W3-pvecsPA-EAO+EAS}b), is noticeably different. This tells us that most of the polarization at 10.3 $\mu$m is due to absorption by non-spherical aligned silicate grains in cooler foreground material, possibly an outer nebular shell. This conclusion would not be so obvious if one compares image (a) to image (b), the absorptive component inferred from the original method, and it provides clear evidence that fits with three components produce a better result than using just two components.

However, the PA of the scattering polarization component does not appear to be organized around the sources, so either the origin of scattering photons coincides with a source not in our images, or it is not polarization due to scattering but instead dichroism by some other dust component or mechanism. Since the absorption cross-section of graphite has similar behavior versus wavelength as the albedo, we speculate that we could be detecting polarization caused by absorption or emission from elongated graphite particles instead of scattering. Elongated graphite grains could be subject to radiative external alignment, as proposed and supported by observations in \cite{And2022}, and then the extended Aitken method would be fitting the absorptive or emissive polarization from such radiatively aligned graphite grains.

\begin{figure*}
\centering
\includegraphics[width=239pt, keepaspectratio=true]{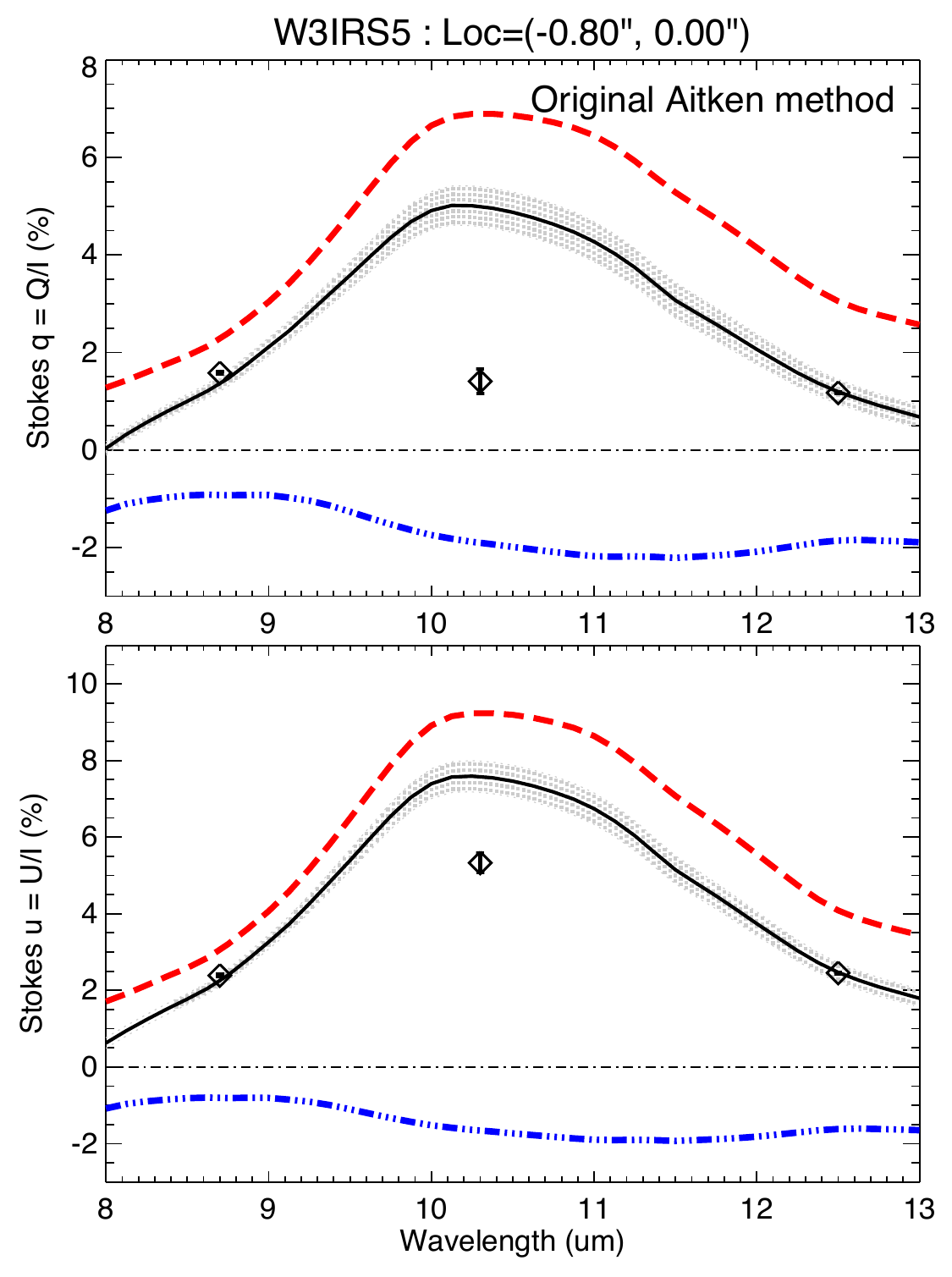}
\includegraphics[width=239pt, keepaspectratio=true]{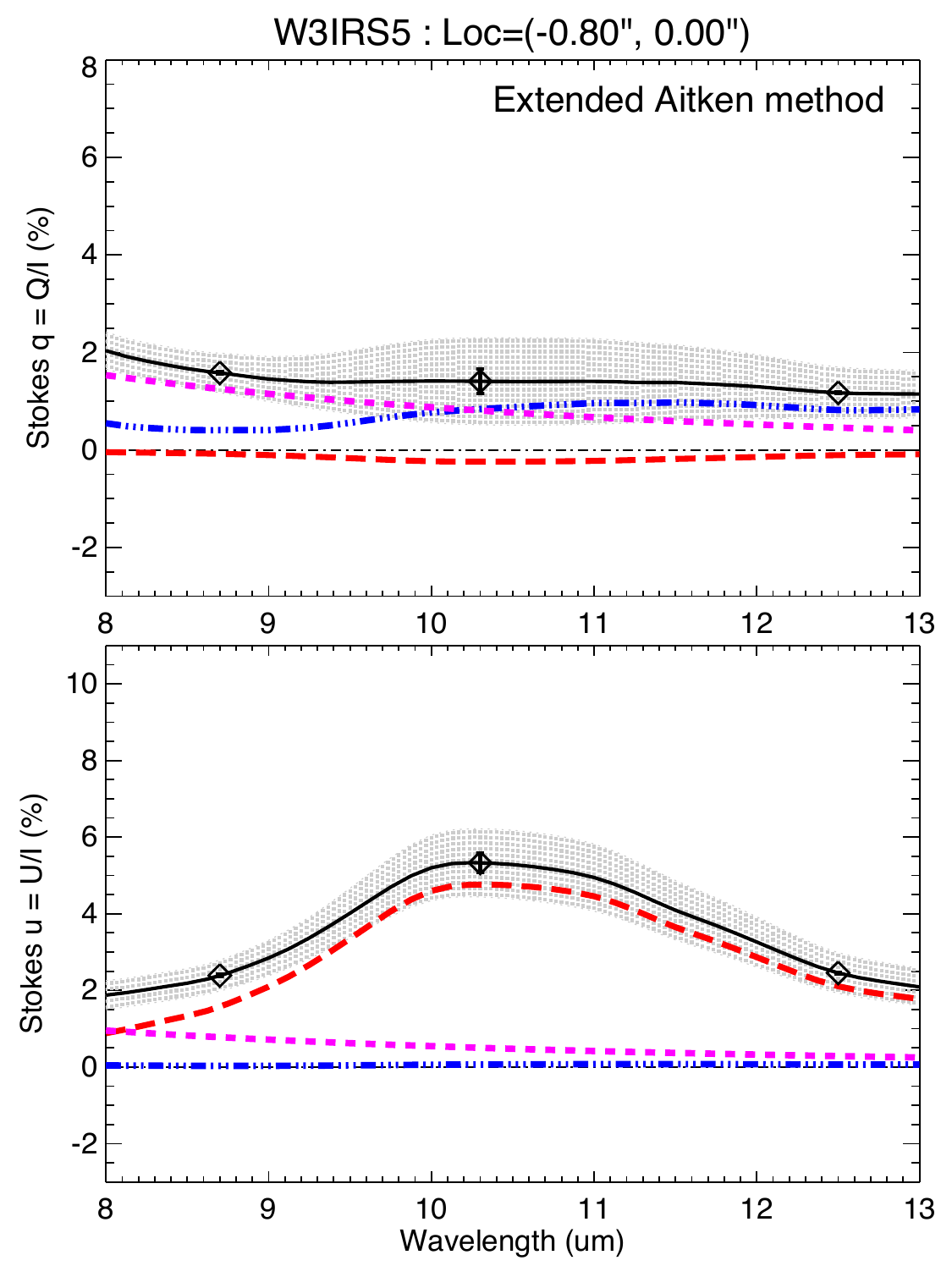}
\includegraphics[width=239pt, keepaspectratio=true]{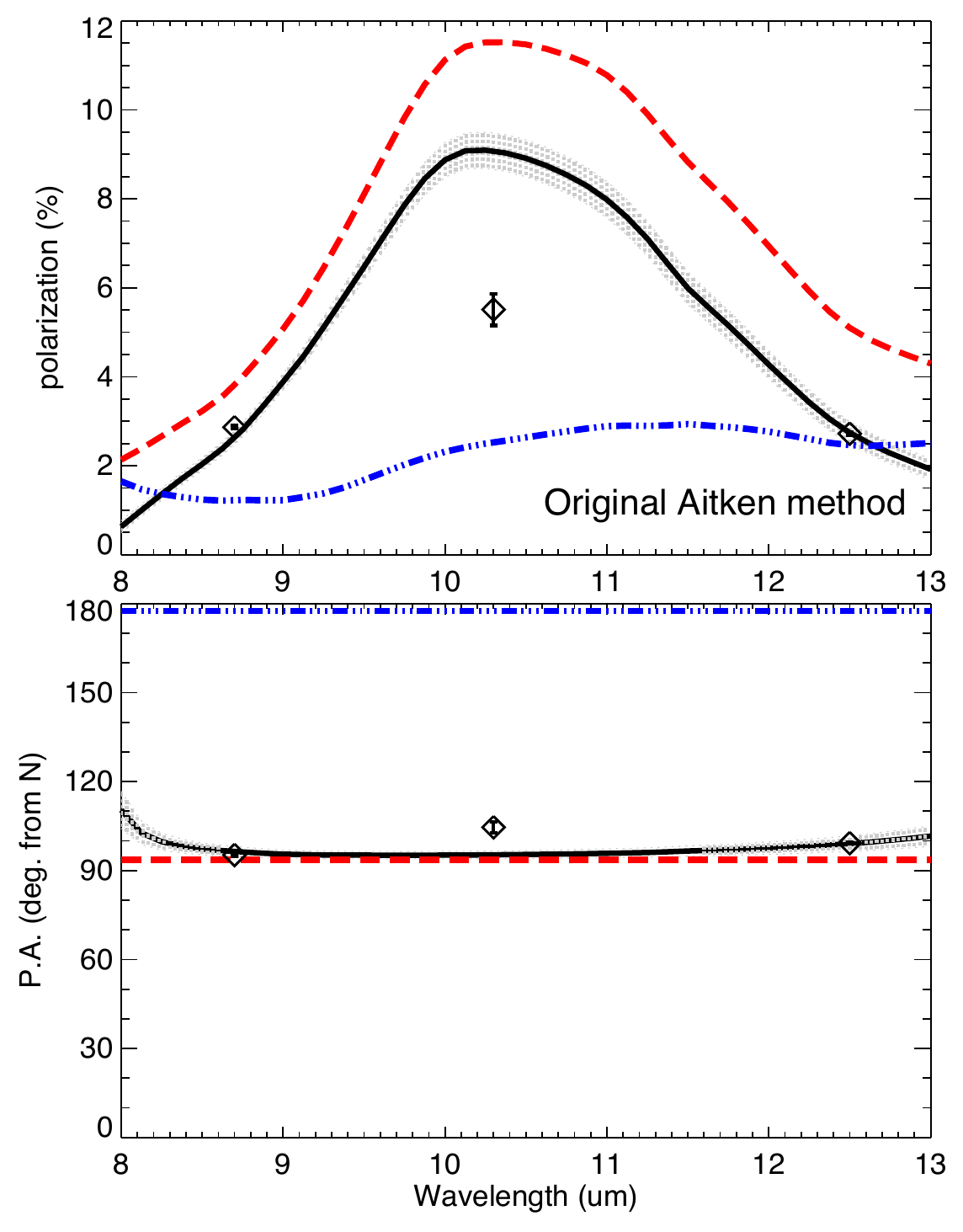}
\includegraphics[width=239pt, keepaspectratio=true]{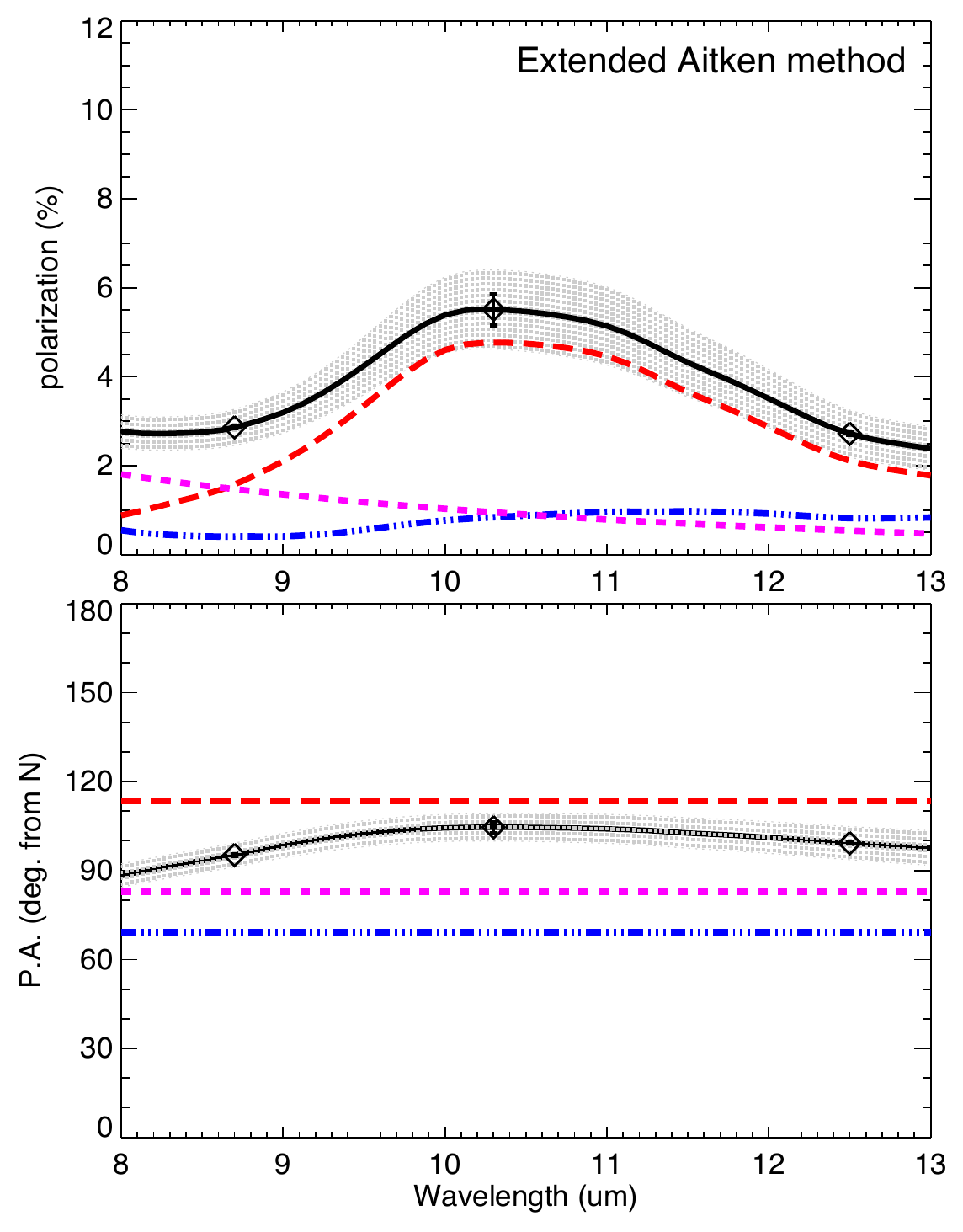}
\caption{Original and extended Aitken method fits to imaging polarimetry of W3 IRS5 at sightline with offset (-0\farcs8, 0\farcs0) from the origin, which is a peak in scattering polarization (marked with diamond and cross on images in Figure \ref{W3-pvecsPA-EAO+EAS}). This sightline is offset (-0\farcs3, 0\farcs0) from the source NE of the origin. Stokes $q$ and $u$ from observations are plotted as diamonds with error bars, in the top rows. Bottom rows show \% polarization and PA computed from Stokes spectra. Left column of plots show the fits with two components: absorptive (red long-dashed curves) and emissive (blue dot-dashed curves). Right column shows the extended Aitken method fits with scattering (violet short-dashed curves). Solid black curves are the sum of components, with gray shading indicating the range of 1-sigma uncertainties.}
\label{W3-AE+EAS-SPL}
\end{figure*}

\begin{figure*}
\centering
\includegraphics[width=240pt, keepaspectratio=true]{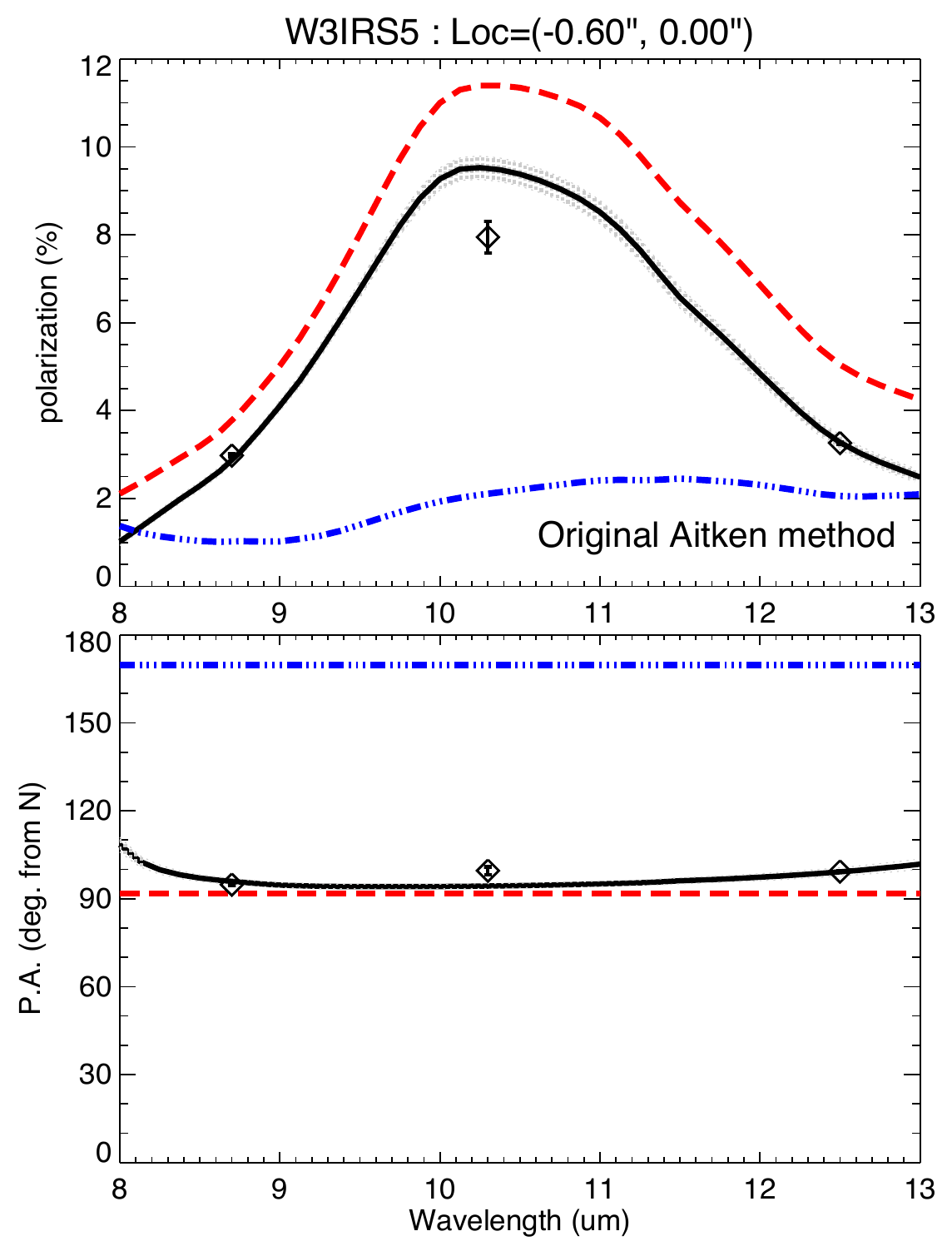}
\includegraphics[width=240pt, keepaspectratio=true]{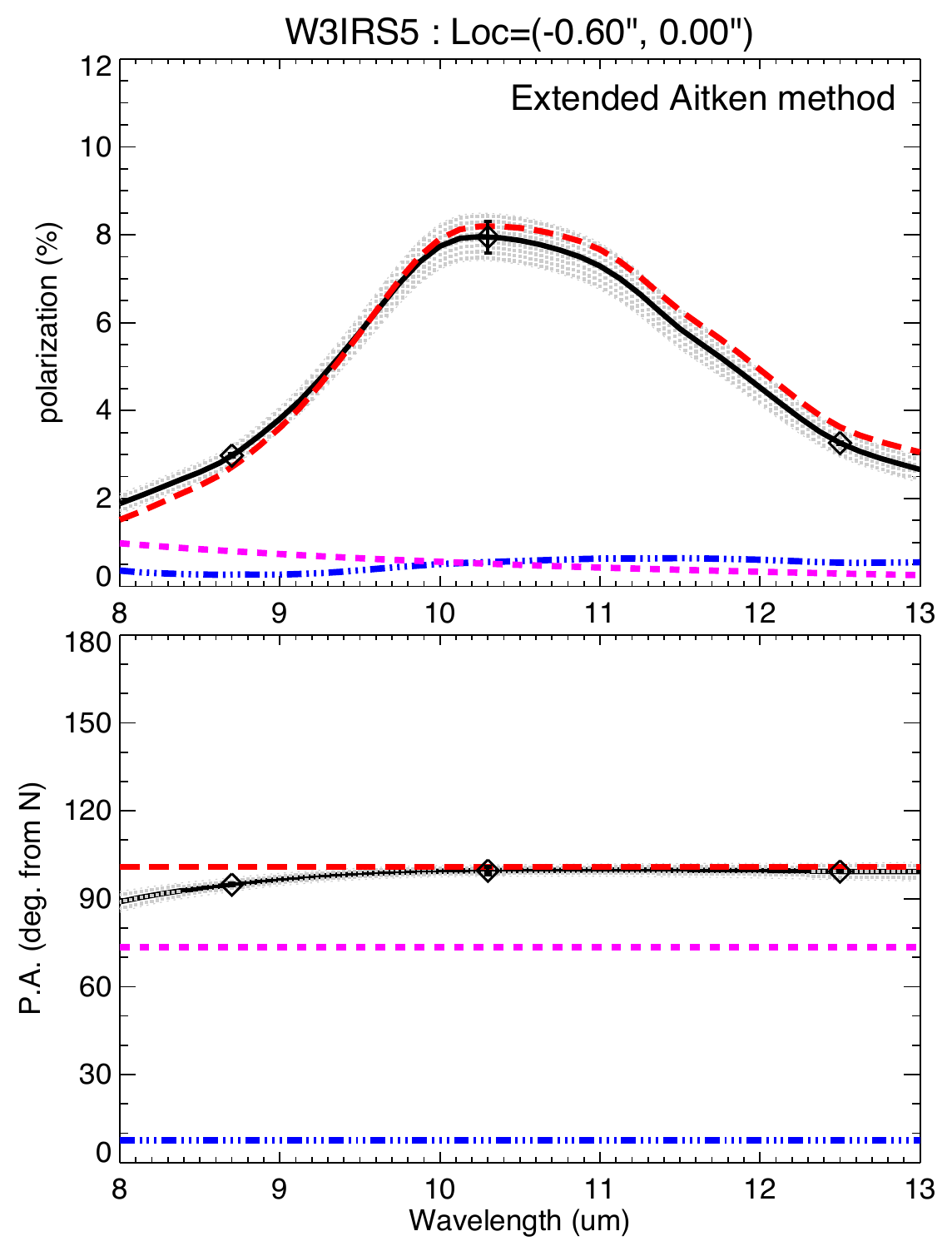}
\caption{Original and extended Aitken method fits to multi-wavelength imaging polarimetry at the brighter source sightline in W3 IRS5, marked with a circled cross NE of the origin on images in Figure \ref{W3-pvecsPA-Obs-SNR+p}. Plots show polarization and PA in degrees east from north that result from fits to observed Stokes $q$ and $u$ spectra, which are plotted as diamonds with error bars. Left plot shows the fit with just two components: absorptive (red long-dashed curves) and emissive (blue dot-dashed curves). Right plot shows the extended Aitken method fit with three components: absorptive, emissive, and scattering (short-dashed violet curves). Solid black curves are the sum of components, with gray shading indicating the range of 1-sigma uncertainty.}
\label{W3-AE+EAS-Main-NE}
\end{figure*}

\begin{figure*}
\includegraphics[width=240pt, keepaspectratio=true]{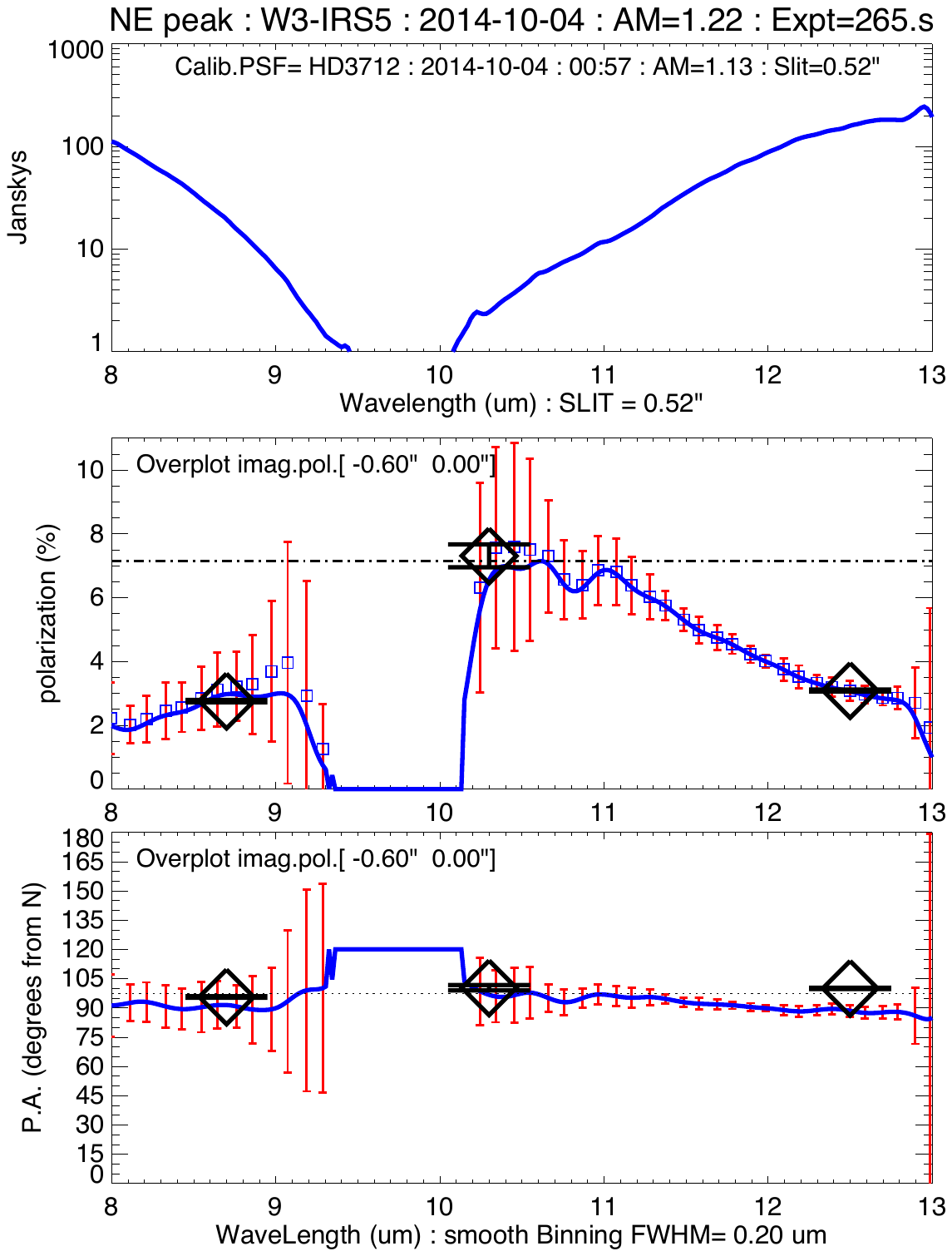}
\includegraphics[width=240pt, keepaspectratio=true]{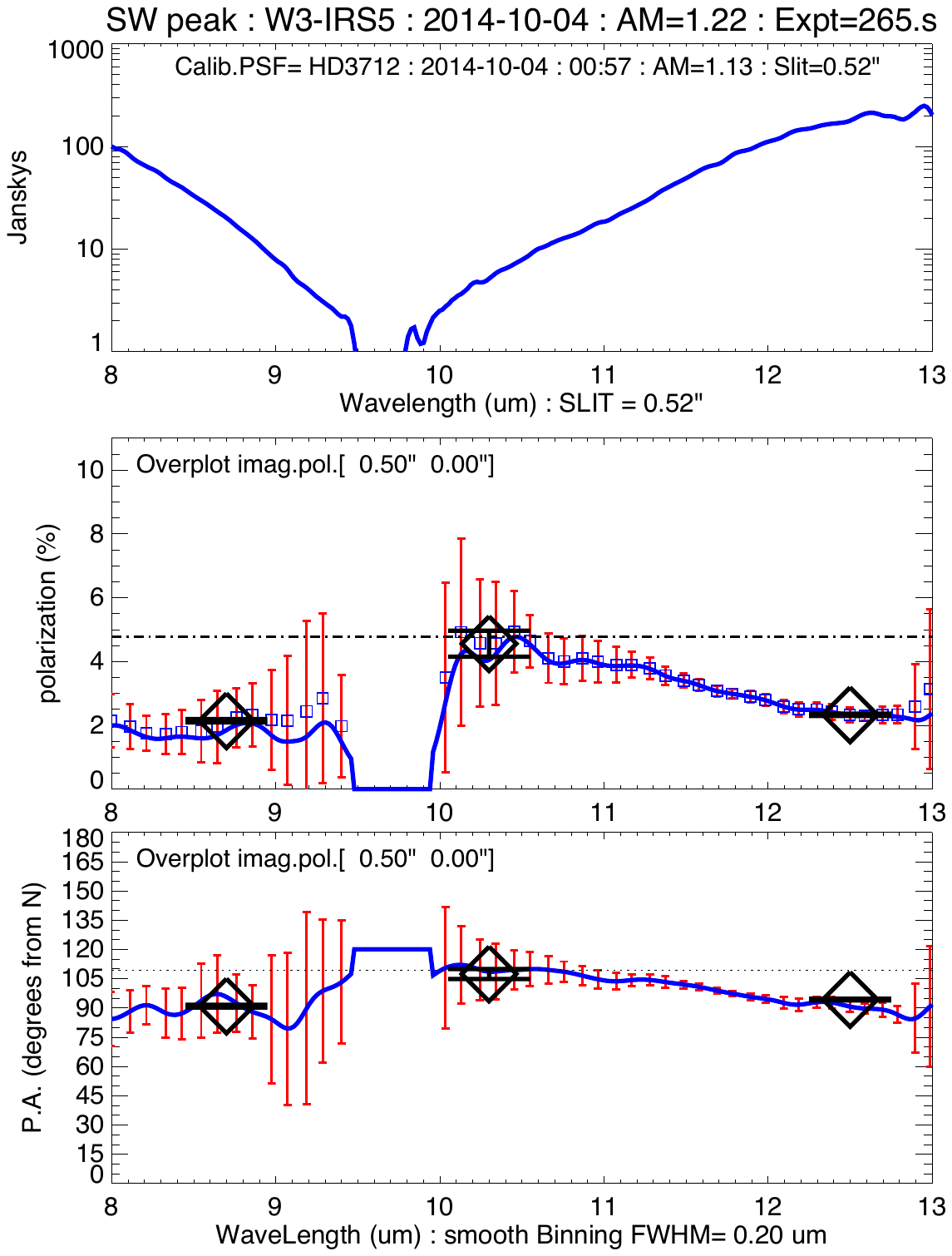}
\caption{Spectropolarimetry observations of the mid-IR peaks in W3 IRS5 using slit of width 0\farcs52. Left column of plots show the polarization spectrum of the brighter peak in W3 IRS5, NE of origin in images, whereas plots on the right side show the secondary peak, which is SW of the origin. Top panels are flux spectrum in Janskys (logarithmic scale), calibrated with HD3712. Middle and bottom panels show polarization (\%) and PA spectrum, respectively, binned at 0.2 $\mu$m resolution, with bin averages shown as squares and error bars showing the standard deviation in bins. The solid curve is the polarization debiased with the $\sigma_p$ of bins. Average values of polarization and PA from imaging polarimetry, in 0\farcs56 diameter apertures at corresponding sightlines, are plotted with large diamond symbols.}
\label{W3-SpecPol-NE+SW}
\end{figure*}

To further explore the extended Aitken method, we consider fits of Stokes spectra at specific sightlines in more detail. Not surprisingly, we find that including the scattering polarization component generally improves the fits to the Stokes spectra. For example, we consider the sightline at (-0\farcs8, 0\farcs0), i.e., NE from the origin and marked with a cross-diamond symbol, which is 0\farcs3 NE of the brighter mid-IR source in W3 IRS5 (circled cross). Figure \ref{W3-AE+EAS-SPL} shows the average Stokes $q$ and $u$ spectra observed in 0\farcs24$\times$0\farcs24 box apertures centered on the sightline, plotted as diamonds with error bars in the top rows of plots. The plots in the bottom row show polarization and PA computed from the Stokes spectra. As indicated, the left column of plots present the original Aitken method fits to Stokes spectra (two components), and the right column of plots present the extended Aitken method fits, using three components. The resulting absorptive components are plotted as long-dashed red curves, the emissive components are plotted as dot-dashed blue curves, and the scattering component is short-dashed violet curves (right column of plots only). The solid black curves are the sums of the components, and the gray shading indicates the range of 1-sigma uncertainty of the fits. Note that these spectra are not debiased, and so the magnitude of the polarization shown there is slightly higher than presented in the images. Debiasing is not performed, because, given that the uncertainties of the Stokes values between observed wavelengths cannot be accurately estimated, debiasing produces artifacts in the curves.

The original Aitken method with just two components cannot fit the Stokes spectra shown in
Figure \ref{W3-AE+EAS-SPL}, causing the inferred absorptive polarization to be overestimated. The extended Aitken method with three components does fit the Stokes spectra exactly, resulting in a reasonable amount of absorptive polarization by including a small amount of the third component of polarization. The inferred scattering component of polarization at 10.3 $\mu$m at this sightline has value of about 1.1\%, a relative maximum in the image. Instead of scattering, the third component of polarization could be due to absorption by or emission from graphite, or from a different type of dust.

Next we focus on the brighter infrared source marked with a circled cross at an offset of (-0\farcs6, 0\farcs0), i.e., to the left (NE) of the origin of images. Figure \ref{W3-AE+EAS-Main-NE} shows the average percent polarization and PA observed in 0\farcs24$\times$0\farcs24 box apertures plotted as diamonds with error bars, computed from the Stokes spectra (plot not shown). The left-hand plot shows the polarization and PA derived using the original Aitken method: red dashed curves are the absorptive components, blue dot-dashed curves are the emissive components. The right-hand plot shows the polarization and PA resulting from the extended Aitken method, with the scattering polarization plotted as short-dashed violet curves. The solid black curves are the sums of the components, and the gray shading indicates the range of 1-sigma uncertainty of the fits. Again, the extended Aitken method fits the data better by including a small amount of scattering polarization component. The original method fit is somewhat acceptable but overestimates the polarization. The secondary star of W3 IRS5, with offset (+0\farcs6, 0\farcs0) to the right (SW) of the origin, is also marked with a circled cross. Plots of Aitken method fits at the sightline of the secondary source of W3 IRS5 are shown in the Appendix Figure \ref{W3-Decomp-AE+EAS-SW}. We revisit these two sources in the next section with spectropolarimetry observations, and more detailed analysis is presented in the Appendix.

Another sightline having a relative peak in scattering polarization is at offset (+0\farcs35, 0\farcs0), i.e., to the right (SW) of the origin and is also marked with a cross-diamond symbol in the images of Figure \ref{W3-pvecsPA-EAO+EAS}. Plots comparing the original and extended Aitken method fits to Stokes spectra at this scattering peak sightline are shown in Figure \ref{W3-AE+EAS-SPR} of the Appendix. The plots present a similar result: the Stokes spectra cannot be fit with just two components, whereas including the scattering polarization component produces an exact fit. This scattering peak sightline is offset 0\farcs3 NE from the secondary mid-IR source in W3 IRS5 (marked by a circled cross), which is the same distance and direction offset of the other scattering peak sightline from the brighter source NE of the origin.

The difficulty of fitting Stokes spectra with the original Aitken method is mainly due to the elevated polarization often observed at 8.7 $\mu$m that cannot be fit with the Orion silicate template profiles, thus requiring an additional component that the scattering template profile provides. As the plots of Stokes spectra in Figures \ref{W3-AE+EAS-SPL} and \ref{W3-AE+EAS-SPR} show, in many cases even the Stokes parameters at 10.3 $\mu$m cannot be fit with just the two components based on silicate profiles of the original Aitken method. Not surprisingly, the extended Aitken method is able to always fit three wavelengths of observations exactly. The three wavelengths used for imaging polarimetry, taken together, are pivotal to probing mid-IR polarization: the 10 $\mu$m band clearly probes the silicate absorption feature; the 12 $\mu$m band mostly probes emission from silicates; and the 8 $\mu$m band probes scattering or additional emission/absorption, as also suggested by the Bayesian approach in \cite{Loro16}. As Figure \ref{W3-pvecsPA-EAO+EAS}e shows, the extended Aitken method fit produces a decomposition with the absorptive component having the same morphology as the polarization at 10.3 $\mu$m, which is expected from the extreme silicate feature absorption. This results in PA distributions for absorptive and emissive silicate components having less deviation from a fixed angle difference, as shown in Figure \ref{PA-Diff-W3} of Section \ref{sec-PAD}. In the next section, we compare the results from imaging polarimetry with spectropolarimetry and show that the latter observations are also fit better with the three component model of the extended Aitken method.

\begin{figure*}
\centering
\includegraphics[width=240pt, keepaspectratio=true]{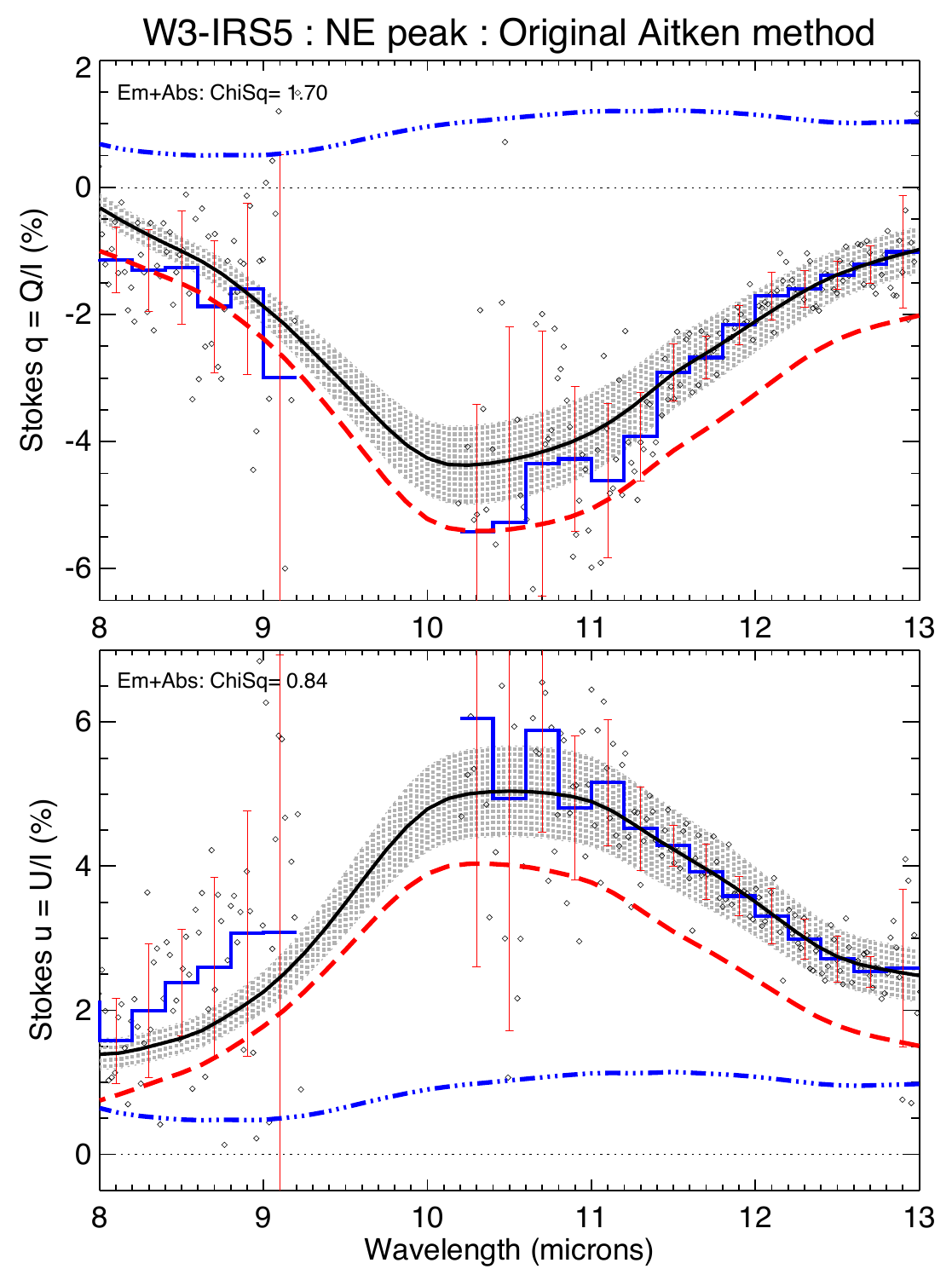}
\includegraphics[width=240pt, keepaspectratio=true]{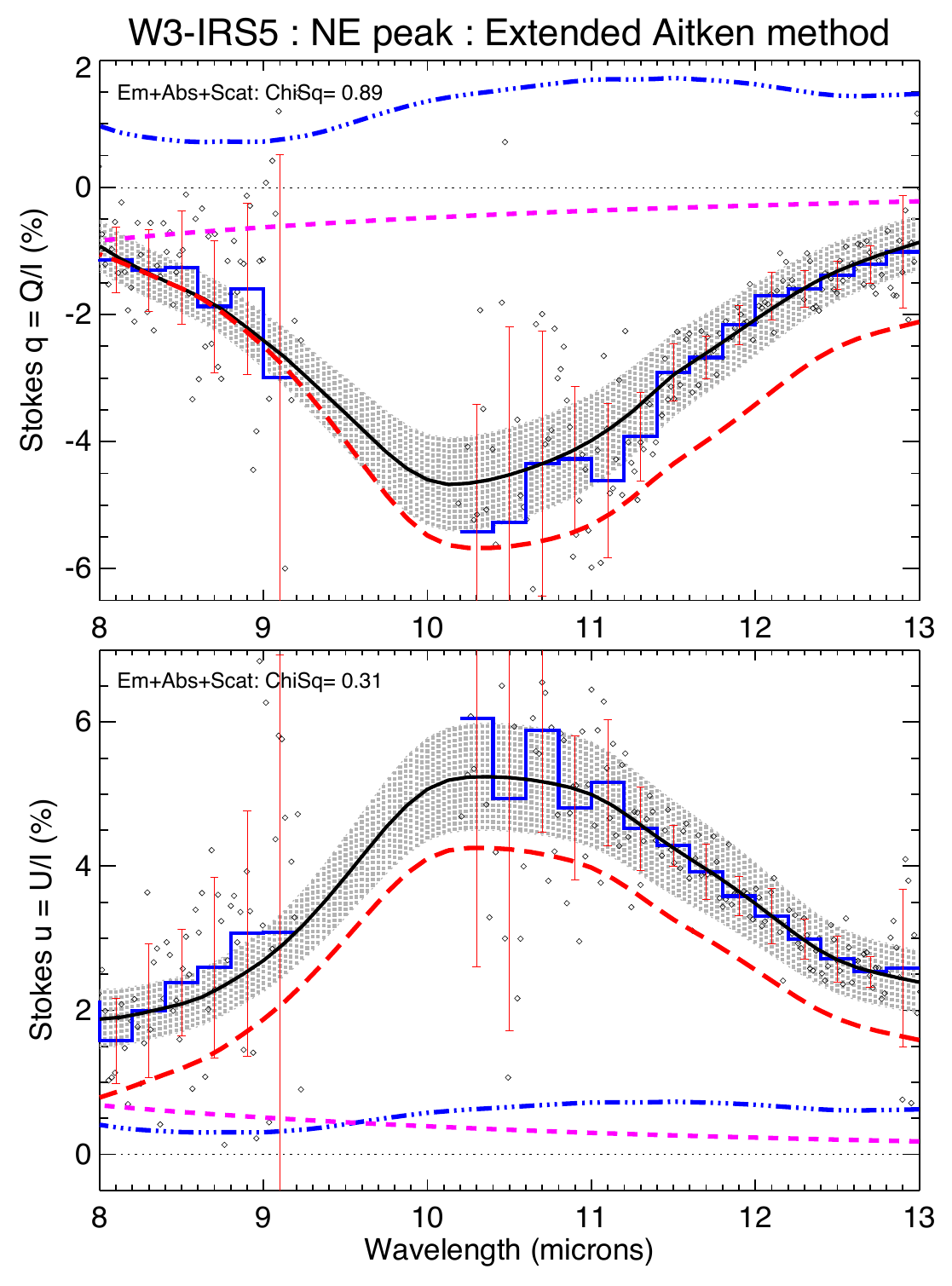}
\includegraphics[width=243pt, keepaspectratio=true]{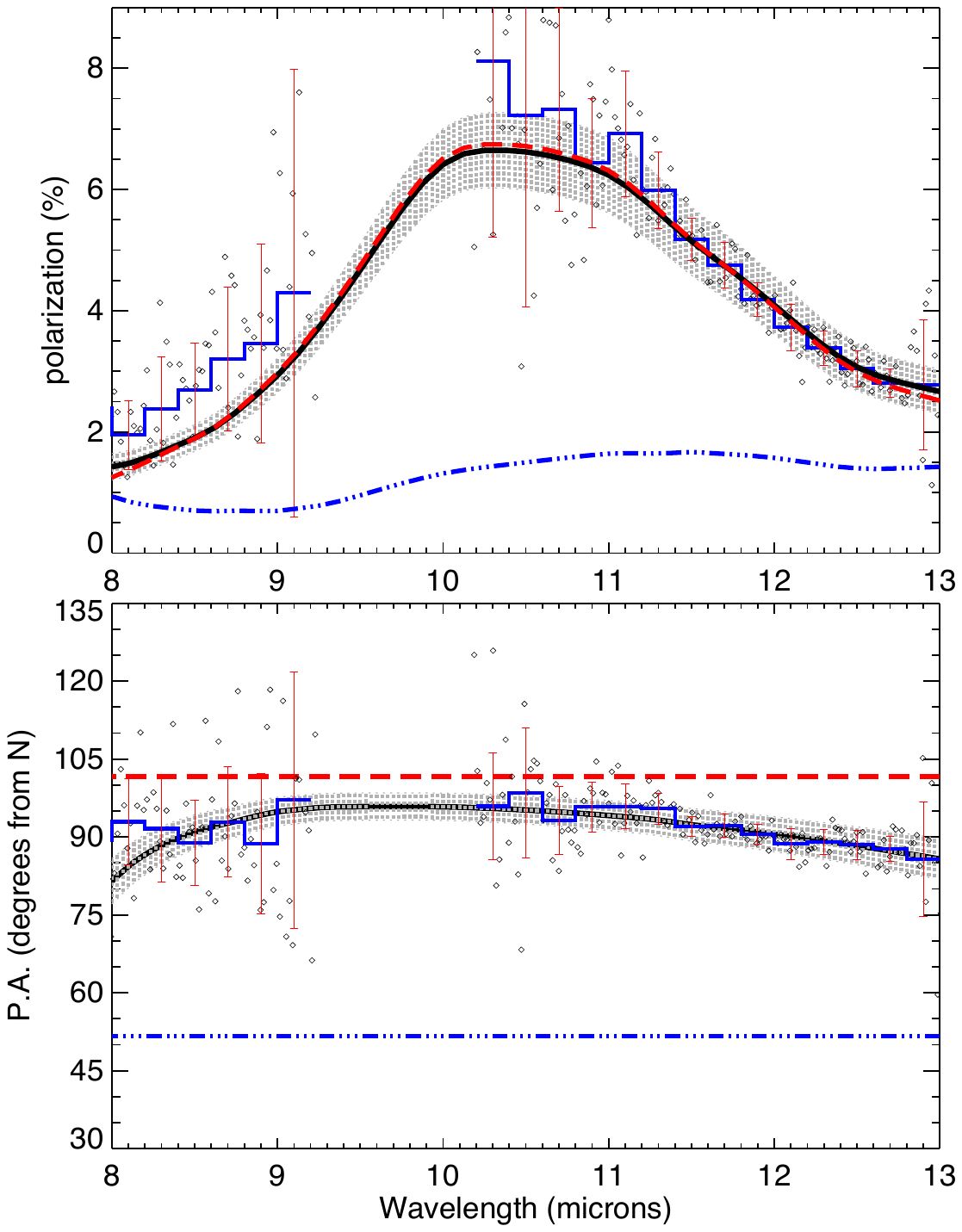}
\includegraphics[width=243pt, keepaspectratio=true]{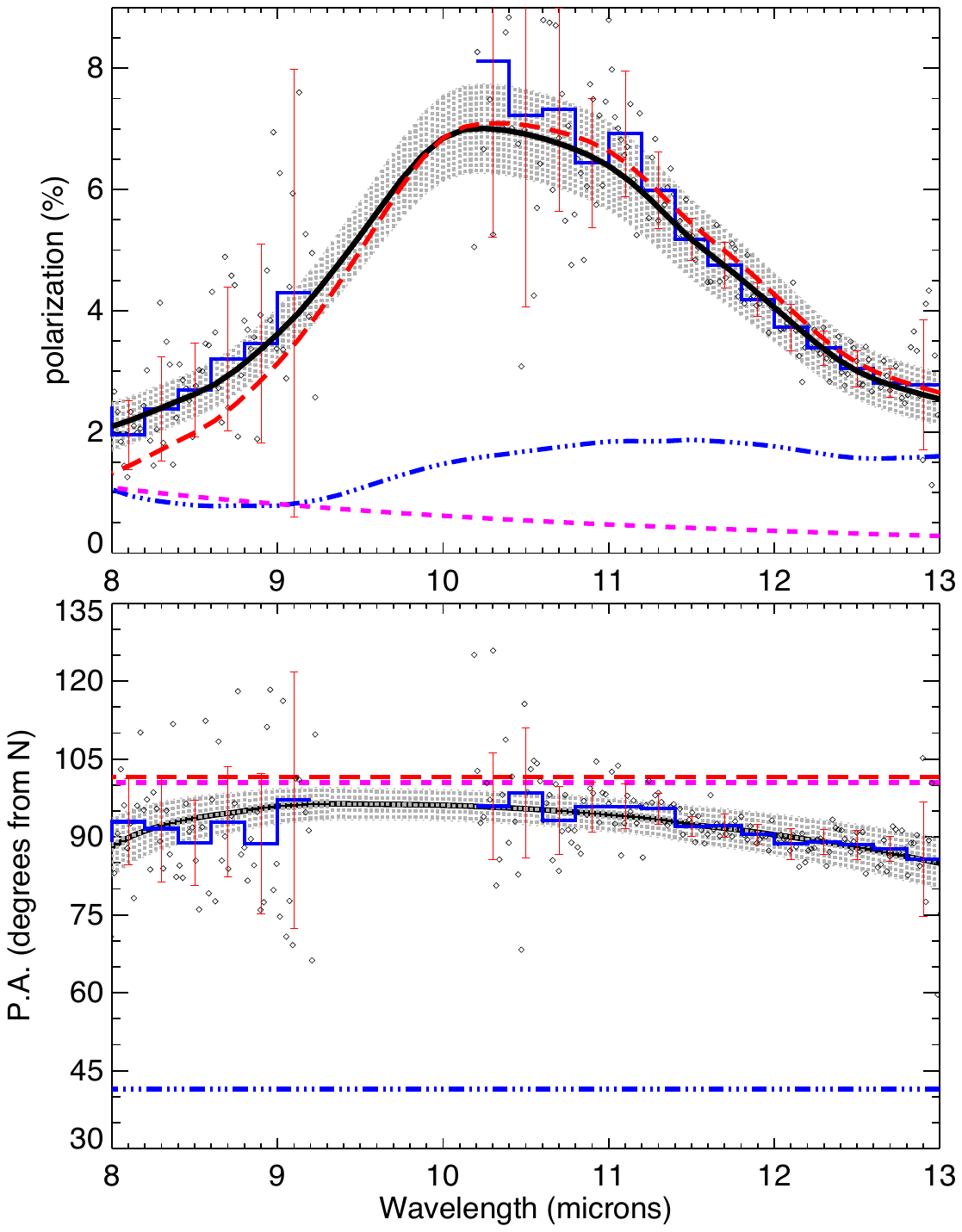}
\caption{Stokes spectra of the brighter peak in W3 IRS5 (NE of origin of images) are plotted with small diamonds, and shown in top row of plots. Values are binned at 0.2 $\mu$m resolution and bin averages are shown with blue histogram lines. Bottom row of plots show the polarization (\%) and PA computed from the Stokes spectra. Left column of plots show the original Aitken method fits to Stokes spectra: absorptive components are red dashed curves, emissive components are blue dot-dash curves. Right column of plots show the extended Aitken method fits, with scattering component plotted as violet short-dashed curves. Solid black curves are the sum of components, with gray shading indicating the 3-sigma range of uncertainties.}
\label{W3-SpecPol-AE+EAS-NEpeak}
\end{figure*}

\subsubsection{Spectropolarimetry of W3 IRS5} \label{sec:W3-SP}

Spectropolarimetry of the two main mid-IR sources in W3 IRS5 was carried out with CanariCam (Table 1), and the results, with comparison to imaging polarimetry, are presented in Figure \ref{W3-SpecPol-NE+SW}. The telescope pointing was alternated between the two source sightlines for four iterations, thus obtaining four exposures with the slit centered on the brighter source, NE of the origin, and four exposures with the slit centered on the secondary source, SW of the origin (see Figure \ref{W3-pvecsPA-Obs-SNR+p}). The $0\farcs52$ slit width ensured that each target source is isolated, since they are separated by $\sim1\farcs1$ and the sources are $<0\farcs5$ in size (full-width-at-half-maximum intensity). Each exposure in the interleaved sequence was reduced into Stokes spectra, each corrected for specific instrumental polarization, and the four Stokes spectra for each source were averaged. Figure \ref{W3-SpecPol-NE+SW} shows the flux and polarization spectra computed from the Stokes spectra at the brighter main source (NE of origin) in the left column of plots, and the secondary mid-IR source (SW of origin) in the right column of plots. Stokes spectra derived from observations are binned at 0.2 $\mu$m resolution (11 wavelength values per bin). The bin averages of Stokes $q$ and $u$ are used to compute polarization and PA spectra, which are plotted with squares, and red error bars show the uncertainty derived from the standard deviations of Stokes values in each bin. Polarization spectra are plotted in middle panels, with corresponding PA values (degrees east from north) in bottom panels. Top panels show the flux in logarithmic scale of Janskys, calibrated with HD3712. The solid blue curves show the polarization debiased with Equation (\ref{debias}) using the standard deviation in each bin, and so have lower values than the squares when the uncertainty is larger. 

The absorption by silicates is extreme, as shown by the flux plots in Figure \ref{W3-SpecPol-NE+SW}, and we only display the data for which SNR$>$20. The average polarization values in 0\farcs56-diameter disk apertures from imaging polarimetry observations (Figure \ref{W3-pvecsPA-Obs-SNR+p}d,e,f) are plotted over the spectra with large diamonds and error bars that also show the approximate filter bandpasses. That aperture is chosen to approximately match the slit mask width of 0\farcs52. The horizontal dash-dotted lines show the maximum polarization and the PA at that wavelength, which is at 10.5 $\mu$m. Clearly, the spectropolarimetry and imaging polarimetry results are in agreement.

To further test the extended Aitken method we focus on the brighter mid-IR peak at offset (-0\farcs6, 0\farcs0), which is NE from the origin. Figure \ref{W3-SpecPol-AE+EAS-NEpeak} shows plots comparing the fits of Stokes spectra using the original Aitken method in the left column of plots and extended Aitken method in right column of plots as indicated in titles. The Stokes spectra at all wavelengths are plotted with small diamonds, and the blue histogram style lines are the average of values in 0.2 $\mu$m wavelength bins, with error bars showing the standard deviation of the 11 values in each bin. Only data with SNR $>$ 30 is analyzed, thus leaving a gap in data analysis between 9.2 and 10.2 $\mu$m of the plots. All fits are to the blue histogram values of binned Stokes spectra, since less noisy. The derived absorptive components are plotted as red dashed curves, emissive components are blue dot-dashed curves, and the scattering component is plotted with violet short-dashed curves. The black curves are the sum of components, with shading indicating the range of 3-sigma uncertainty of the fits. It is evident that the original Aitken method has some difficulty fitting the Stokes spectra at shorter wavelengths, even accepting 3-sigma uncertainty, so including the scattering component (violet dotted curves) of the extended Aitken method produces a better fit to the Stokes spectra. The $\chi^2$ goodness of fits (square root of differences between data and fit) is improved by about 50\%. Table \ref{Table-SP-W3-NE} in the Appendix presents the coefficients, uncertainties, and $\chi^2$ of the fits.

The bottom two rows of plots in Figure \ref{W3-SpecPol-AE+EAS-NEpeak} show the polarization and PA computed from the Stokes spectra. The plots in the bottom-right quadrant show that the extended Aitken method can fit the polarization spectrum better by including a small amount of scattering component, having about 1\% polarization at 8 $\mu$m, and about 0.5\% at 10.3 $\mu$m, in approximate agreement with the polarization images in Figure \ref{W3-pvecsPA-EAO+EAS}. The PA of the scattering component is found to be the same as the PA of the absorptive component, with a value of 100\degr, which could indicate they are similar components. The plots of $p$ are in reasonable agreement with Figure \ref{W3-AE+EAS-Main-NE} for imaging polarimetry. However, average values of $p$ in Figure \ref{W3-SpecPol-AE+EAS-NEpeak} are lower because of the larger aperture of spectropolarimetry (0\farcs52 $\times$ 2\farcs1), whereas the smaller 0\farcs24 $\times$ 0\farcs24 aperture used for imaging polarimetry gives higher average values.

Figure \ref{W3-SpecPol-AE+EAS-SWpeak} in the Appendix compares Aitken method fits to the spectropolarimetry observation of the secondary source of W3 IRS5, at offset (+0\farcs61, 0\farcs0), or SW from the origin, also marked with circled cross in the images of Figure \ref{W3-pvecsPA-Obs-SNR+p}. Again, the extended Aitken method fits are slightly better at the SW peak sightline. Table \ref{Table-SP-W3-SW} presents the coefficients and uncertainties of the fits. The third component may not always be due to scattering. Since the absorption cross-section of graphite has similar functional behavior as the albedo, the extended Aitken method could be fitting absorption by or emission from graphite grains. Another possibility is that the emissive and absorptive silicate profiles need to be adjusted for this source, but including the third component provides that adjustment.

\begin{figure*}
\centering
\includegraphics[width=240pt, keepaspectratio=true]{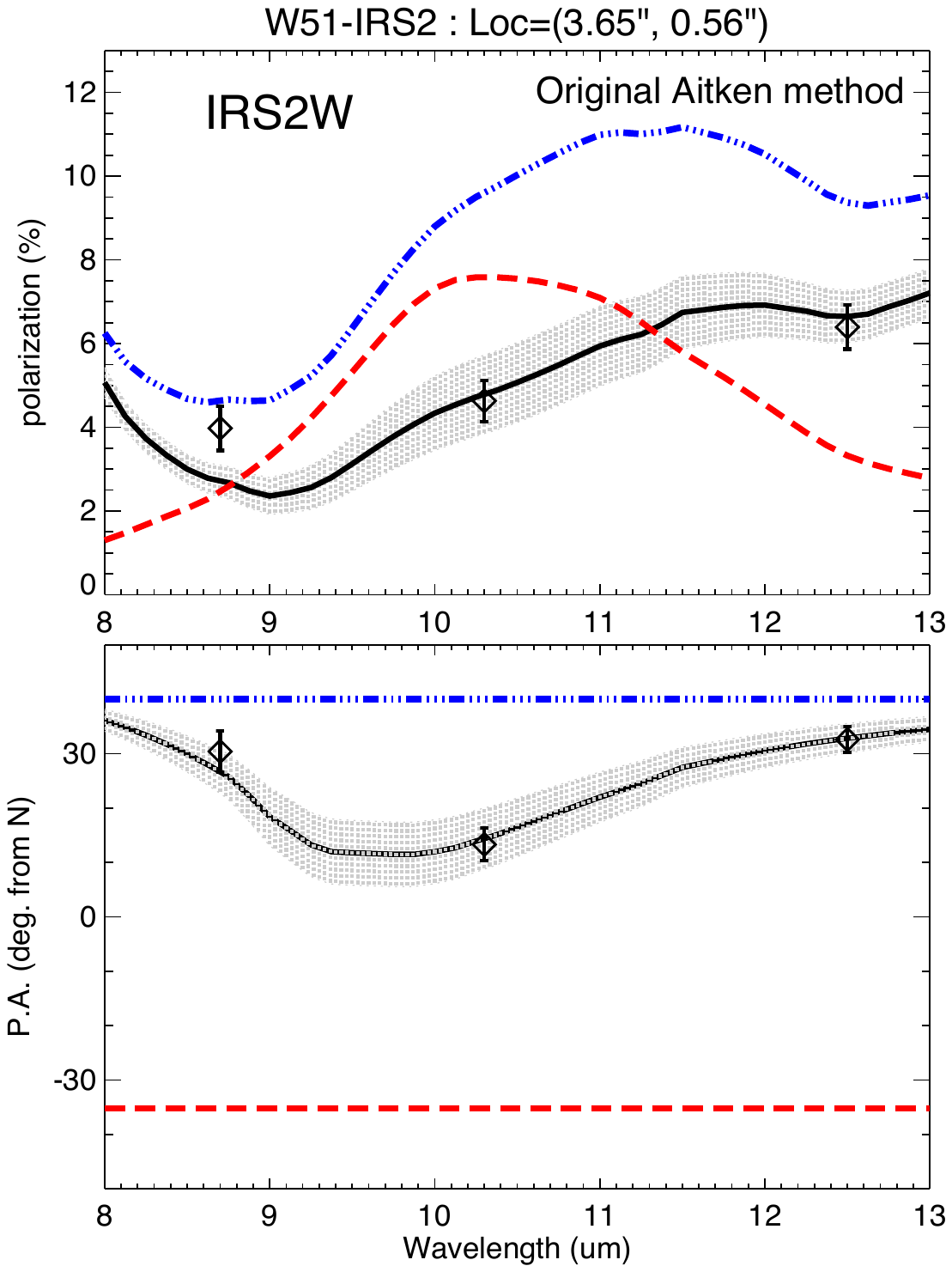}
\includegraphics[width=240pt, keepaspectratio=true]{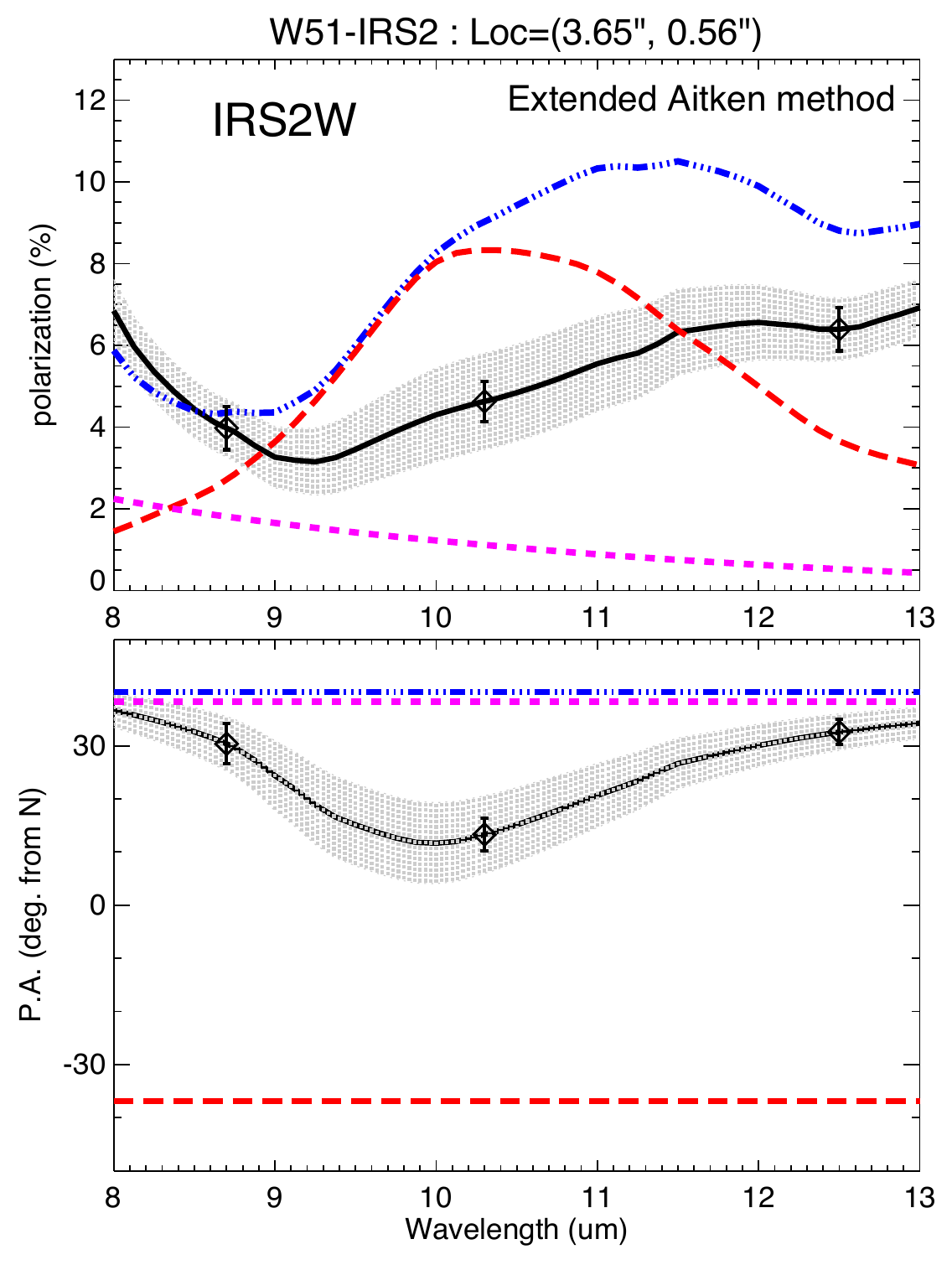}
\caption{Fits of multi-wavelength imaging polarimetry at the source IRS2W of W51 IRS2 (offset +3\farcs65 and +0\farcs56 from IRS2E). Observations are plotted as diamonds with error bars, with percent polarization in top panels and PA in lower panels. Left plot shows the original Aitken method fit, and plot on right shows the extended Aitken method fit with three components. Absorptive components are the red dashed curves, emissive components are the blue dot-dashed curves, and the scattering component is the violet short-dashed curve of extended Aitken method. The solid black curves are the sum of components, with gray shading indicating the 1-sigma range of uncertainties.}
\label{W51-AE+EAS-IRS2W}
\end{figure*}

\subsection{Imaging Polarimetry of W51 IRS2} \label{sec:W51}

Extensive analysis of W51 IRS2 imaging polarimetry was presented in \citet{Telesco23}. Applying the Aitken method to Stokes spectra finds emissive and absorptive polarization magnitudes across W51 IRS2 that are among the highest ever detected in the mid-IR. Using the extended Aitken method to fit the data does not significantly change the results shown in that publication. Figures \ref{W51-AE+EAS-IRS2W} and \ref{W51-AE+EAS-LacyJet} show two sightlines at which the elevated 8.7 $\mu$m polarization can be fit better with the extended Aitken method (right side plots), but the original Aitken method still produces an adequate fit with just two components. The spatial distribution of polarization is essentially the same when the extended Aitken method is applied. There is a small amount of scattering polarization component detected, having maximum $p_S(10.3\mu\text{m}) = 2.8\%$, but mostly $p_S(10.3\mu\text{m}) < 2\%$. This reduces the maximum polarization of the absorptive component from 19.7\% to 18\%, and maximum of emissive component from 16.5\% to 16\%, at 10.3 $\mu$m. The PA of the scattering polarization component is essentially the same as the emissive component. Most sightlines are fit well with the original Aitken method, and the extended method produces mostly the same results.

\section{PA of Absorptive and Emissive Dust} \label{sec-PAD}

Disentangling the PA values of polarization from emissive and absorptive silicate components at any given sightline is critical for understanding the variations in the magnetic field morphology along that sightline and therefore for piecing together a picture of the global magnetic field structure. As described in the Introduction, for a given ensemble of elongated dust particles threaded by the same magnetic field, the PA of the emissive polarization component should be perpendicular to the PA of the absorptive component of polarization.  However, variations of the local 3D structure of magnetic fields and the dust distribution along an ensemble of sightlines are expected to result in distributions of emissive PA and absorptive PA which may not be exactly perpendicular, but would still have a peak in the distribution of the PA differences.

At least for some compact, somewhat isolated sources, with relatively simple geometries in which the same magnetic field threads both emitting and foreground absorbing regions we would expect the derived values of the absorptive and emissive PA to be nearly perpendicular to each other. For such cases the distribution of PA differences would be peaked at a specific angle near 90\degr. In fact, we find this to be the case by applying the Aitken method to polarimetry observations of various sources with mid-IR emission \citep{CMW2007}, as presented most recently in \citet{Telesco23}. By comparing in more detail the distribution of polarization PA differences that are obtained from the original and extended Aitken method fits to Stokes spectra from imaging polarimetry of the Egg Nebula, W3 IRS5, and W51 IRS2, we find that the extended Aitken method provides a more refined picture of magnetic field morphology.

The following figures present histograms of $\text{PA}_A$ of the absorptive component minus $\text{PA}_E$ of the emissive component, computed at all sightlines (pixels) of the imaging polarimetry that have SNR $>20$, and also have $p_A$ and $p_E \geq 1$\%. The histogram bins of the $\text{PA}_A-\text{PA}_E$ distributions are all 10\degr. The red histograms are results from the original Aitken method and blue histograms are results from the extended Aitken method fits to Stokes spectra. Note that the histograms of PA differences are the same at all wavelengths because the emissive, absorptive, and scattering components each have constant PA versus wavelength, as seen in all the previous plots. This is simply because the dust grain alignment angle of each component is independent of wavelength.

\begin{figure}
\includegraphics[width=\columnwidth]{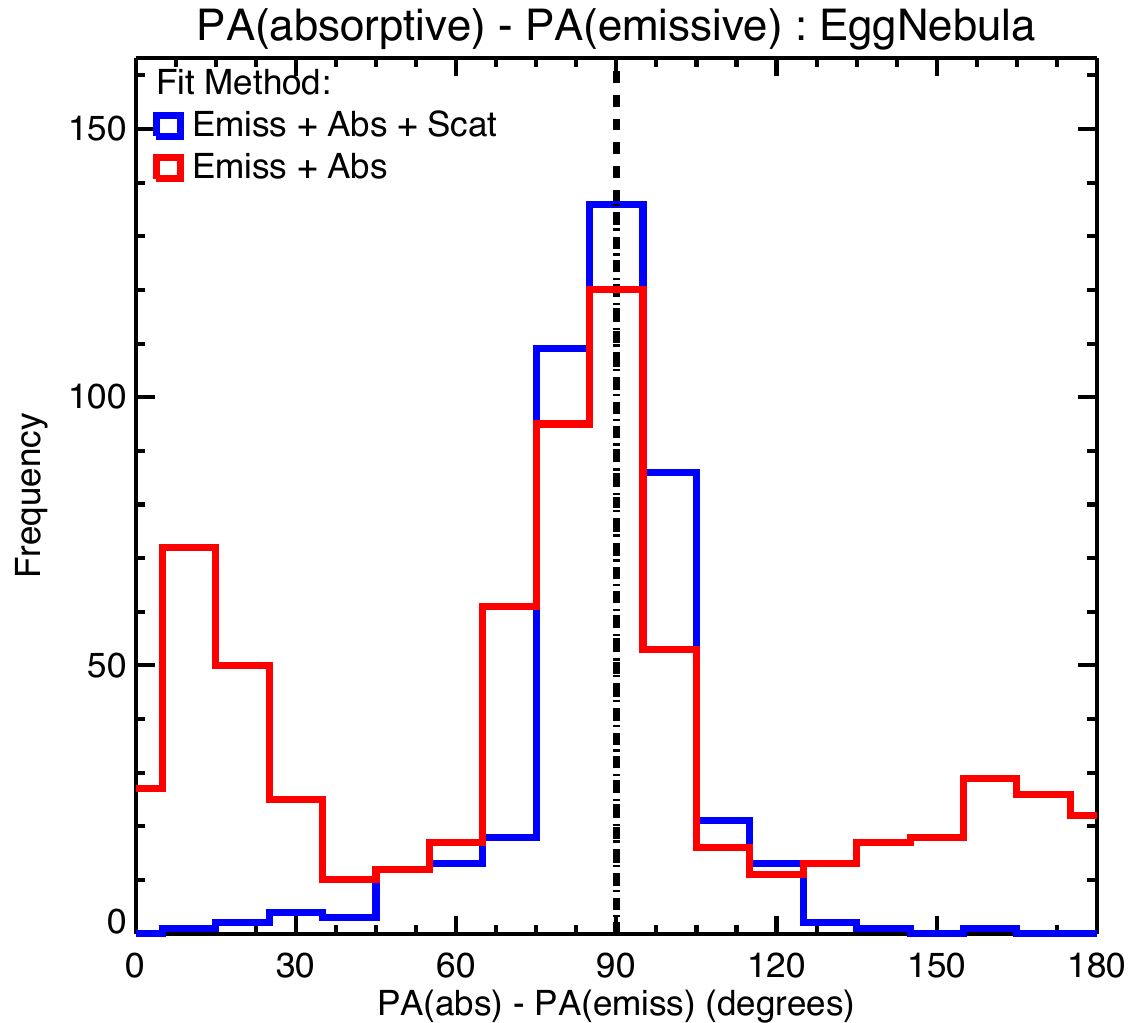}
\caption{Egg nebula: histograms showing the distribution of $\text{PA}_A - \text{PA}_E$ from fits to Stokes spectra at sightlines of imaging polarimetry: red = original, and blue = extended Aitken method.}
\label{PA-Diff-EggNebula}
\end{figure}

Figure \ref{PA-Diff-EggNebula} shows histograms of the PA difference distributions from imaging polarimetry observations of the Egg Nebula. When including the scattering polarization component the extended Aiken method fits (in blue) produce a distribution of $\text{PA}_A-\text{PA}_E$ that is more peaked at 90\degr, whereas the original Aiken method fits (in red) produce a broader distribution with multiple peaks. The more peaked distribution around 90\degr obtained from the extended Aitken method supports the idea that polarized emission and absorption arises from elongated dust grains aligned by the same or similar magnetic fields threading both regions. Since it was shown in Section \ref{sec:Egg} that the Egg Nebula clearly exhibits polarization due to scattering, we found that having three components of polarization is necessary to correctly fit the Stokes spectra. The result of a single peaked distribution of $\text{PA}_A-\text{PA}_E$ provides more evidence that including the third component of scattering polarization is a better than just two components.

\begin{figure}
\includegraphics[width=\columnwidth]{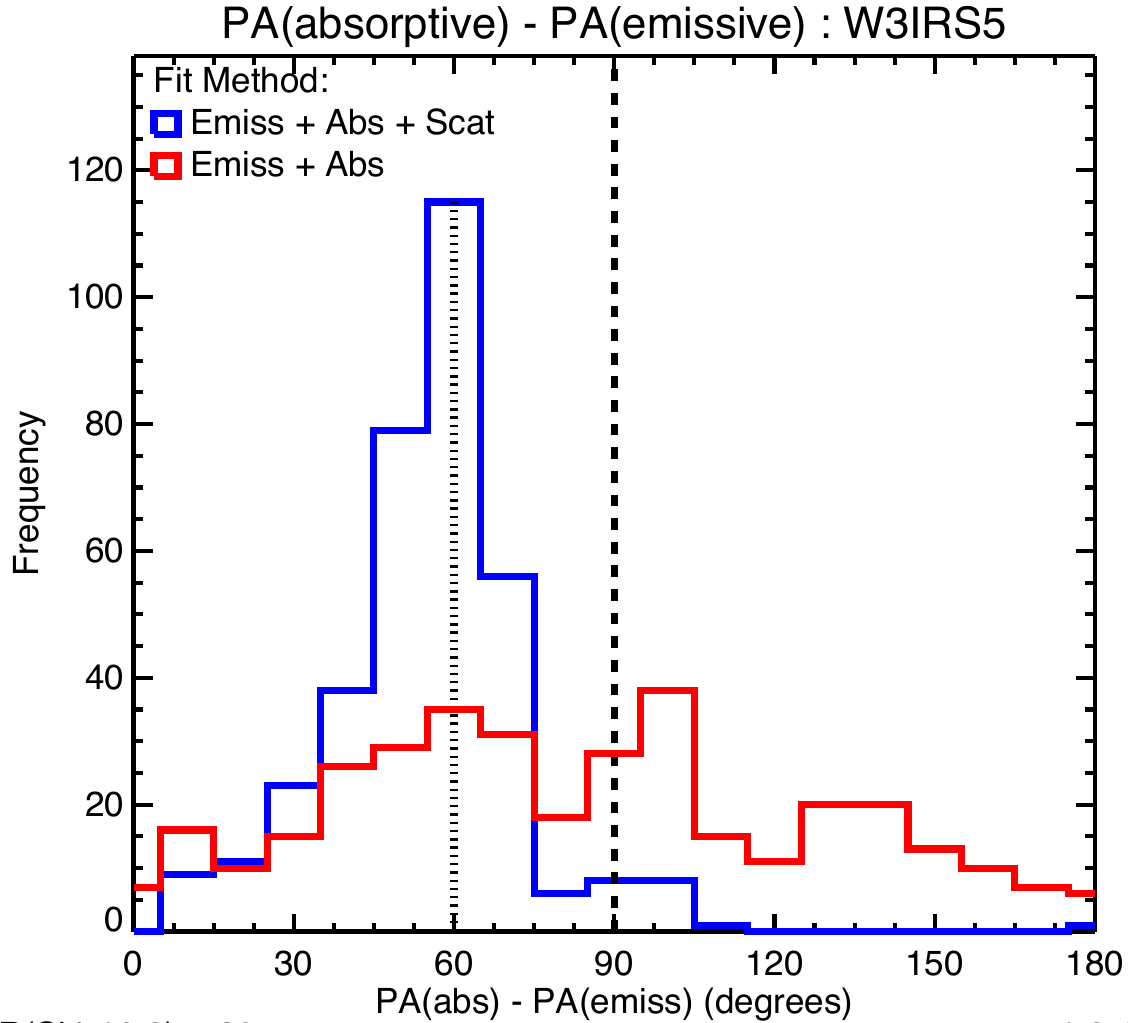}
\caption{W3 IRS5: histograms showing the distribution of $\text{PA}_A - \text{PA}_E$ from fits to Stokes spectra at sightlines of imaging polarimetry: red = original, and blue = extended Aitken method.}
\label{PA-Diff-W3}
\end{figure}

\begin{figure}
\includegraphics[width=\columnwidth]{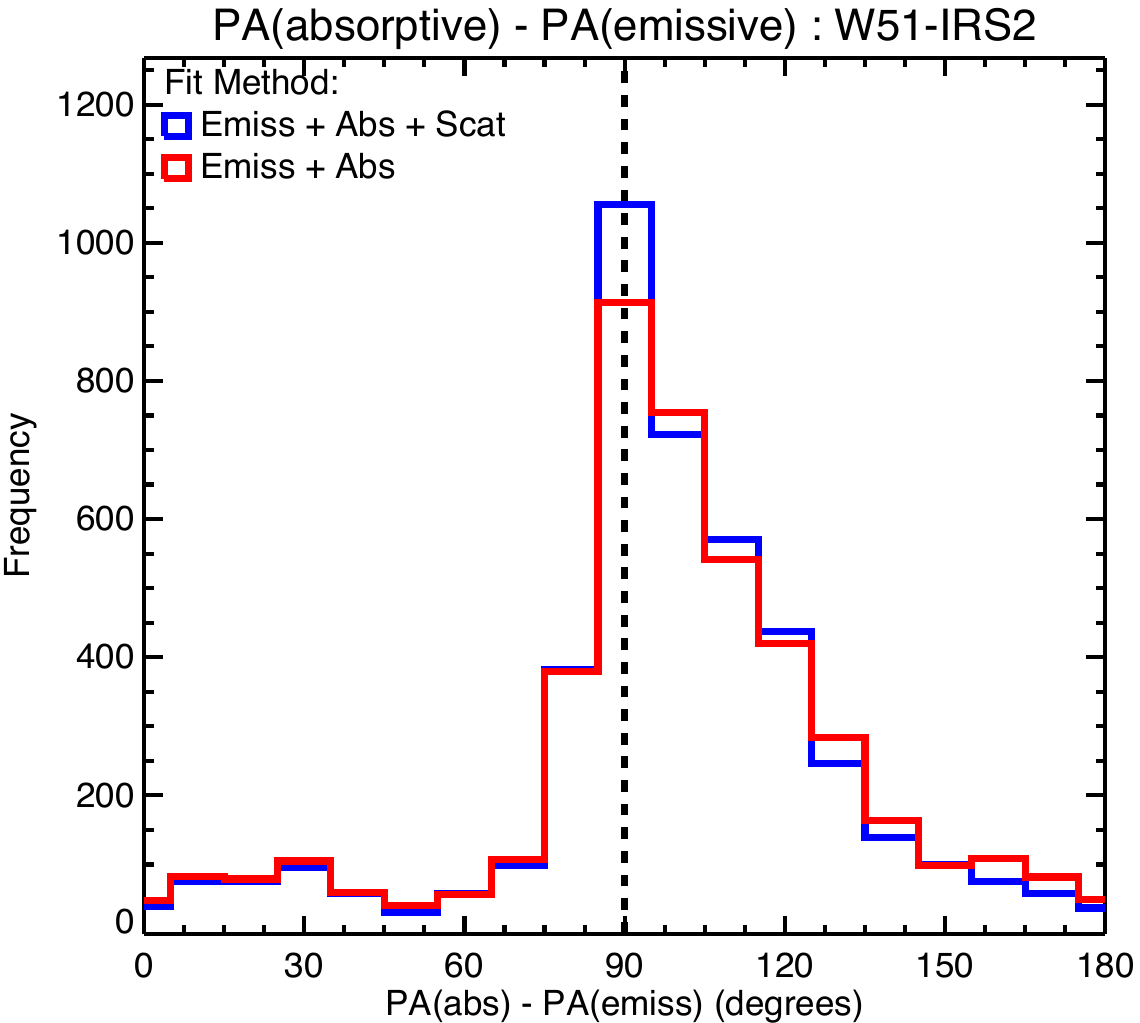}
\caption{W51 IRS2: histograms showing the distribution of $\text{PA}_A - \text{PA}_E$ from fits to Stokes spectra at sightlines of imaging polarimetry: red = original, and blue = extended Aitken method.}
\label{PA-Diff-W51}
\end{figure}

Figure \ref{PA-Diff-W3} shows histograms of the PA difference distributions found in fits of W3 IRS5 imaging polarimetry observations. The original Aitken method fits (in red) produce a somewhat random distribution of $\text{PA}_A-\text{PA}_E$, whereas the extended Aitken method fits (in blue) produce a more peaked distribution. However, the peak of the distribution is at 60\degr instead of 90\degr. This non-perpendicular angle is almost certainly due to a more complicated 3D structure along sightlines, since the optically thin warm emitting dust grains could be aligned along a B-field having a different direction from the B-field aligning the cold absorbing dust component. The fact that the extended Aitken method produces a more peaked distribution of the PA differences is consistent with the theory that the emissive and absorptive components of polarization are directly related to dust grains aligned along magnetic field lines, but likely in a more complex geometry \citep{CMW2007}.

Figure \ref{PA-Diff-W51} shows histograms of $\text{PA}_A-\text{PA}_E$ distributions from fits of W51 IRS2 imaging polarimetry observations. Fits of Stokes spectra with just two components of the original Aitken method (red) produce a peaked distribution of the PA differences around 90\degr. Using the extended Aitken method with three components (blue) produces essentially the same result, but with a slightly more peaked distribution of PA differences. Obtaining essentially the same result by fitting with two or three components occurs probably because W51 IRS2 is dominated by silicates and has very little scattering in the observed field of view. Since the PA of emissive and absorptive components are mostly perpendicular it is likely that the magnetic field direction (and hence silicate grains alignment) has not changed much for the sources of photons along the lines of sight between warm and cold dust regions.

\section{Summary and Conclusions} \label{sec:SC}

We have extended the technique developed by \citet{Ait2004} to also account for polarization due to scattering by dust grains. We show that observed mid-IR polarization spectra can be fit with three components: one due to warm emitting silicate dust grains, the second due to absorption by colder foreground silicate grains, and the (new) third component of polarization due to scattering by graphite, or similar, grains. The original Aitken method can be used to infer properties of the respective magnetic fields, $\text{B}_E$ and $\text{B}_A$, in the plane of the sky at each sightline, in effect, tomography of the magnetic field. With the extended method we can now also analyze nebulae that include scattering by other dust components. The motivation for extending the Aitken method is the discovery of mid-IR scattering polarization in the Egg Nebula, from observations by CanariCam. Applying the extended Aitken method to multi-wavelength imaging polarimetry of the Egg Nebula reveals that three components of polarization are likely present.

The circular arrangement of polarization PA in the Egg Nebula could be due to radiative external alignment of elongated graphite grains, as presented by \cite{And2022}. Centrosymmetric radially aligned graphite grains could create absorptive polarization at mid-IR wavelengths that would have PA perpendicular to rays from the source. Since the absorption cross-section of graphite has similar variation as the albedo versus wavelength, the third component of the extended Aitken method would then be fitting absorptive polarization. This may be the case for some sources, but we believe that scattering is the main source of polarization in the Egg Nebula.

From the application of the extended Aitken method to three different mid-IR sources we conclude the following.

(1) If scattering by dust is present, the extended Aitken method determines the polarization component due to scattering and improves the results for emissive and absorptive components, as compared to the original Aitken method.

(2) If scattering is not present or minimal, the extended Aitken method produces essentially the same result as the original Aitken method.

(3) If polarization is elevated at wavelengths shorter than 10 $\mu$m, the extended Aitken method can fit those type of spectra better and thus enhance the science that can be obtained from mid-IR multi-wavelength imaging polarimetry or spectropolarimetry.

We propose that the extended Aitken method can be used to derive the three components of mid-IR polarization from observations that measure at least three wavelengths which bracket and sample the 9.7 $\mu$m silicate feature.
\section*{Acknowledgements} \label{sec:acknowledgements}
We thank the referee for suggestions that improved the paper. This research is based on observations made with the Gran Telescopio CANARIAS (GTC), installed at the Spanish Observatorio del Roque de los Muchachos of the Instituto de Astrofísica de Canarias, on the island of La Palma. We thank Laurence Sabin and Romano Corradi for observations of the Egg Nebula, which were made with the GTC under programme 105-GTC53/14A to RC and with Directors Discretionary Time to LS. We acknowledge the outstanding support of the GTC science and engineering staff who made these observations possible. This research was supported in part by the National Science Foundation under grant AST-1908625 to CMT. CMW acknowledges financial support during the period 2012-2017 from an Australian Research Council Future Fellowship FT100100495. The upgrade of CanariCam was co-financed by the European Regional Development Fund (ERDF), within the framework of the ``Programa Operativo de Crecimiento Inteligente 2014-2020'', project ``Mejora de la ICTS GTC (2016-2020).''

\section*{Data Availability} \label{sec:DA}
CanariCam data is available from the GTC archive using the search form at https://gtc.sdc.cab.inta-csic.es/gtc. Reduced data and results of analysis are available upon request to the authors at University of Florida, Department of Astronomy.

\appendix
\section{Additional Graphs} \label{sec:appendix}

\subsection{Dust Albedos}

\begin{figure}
\includegraphics[width=\columnwidth]{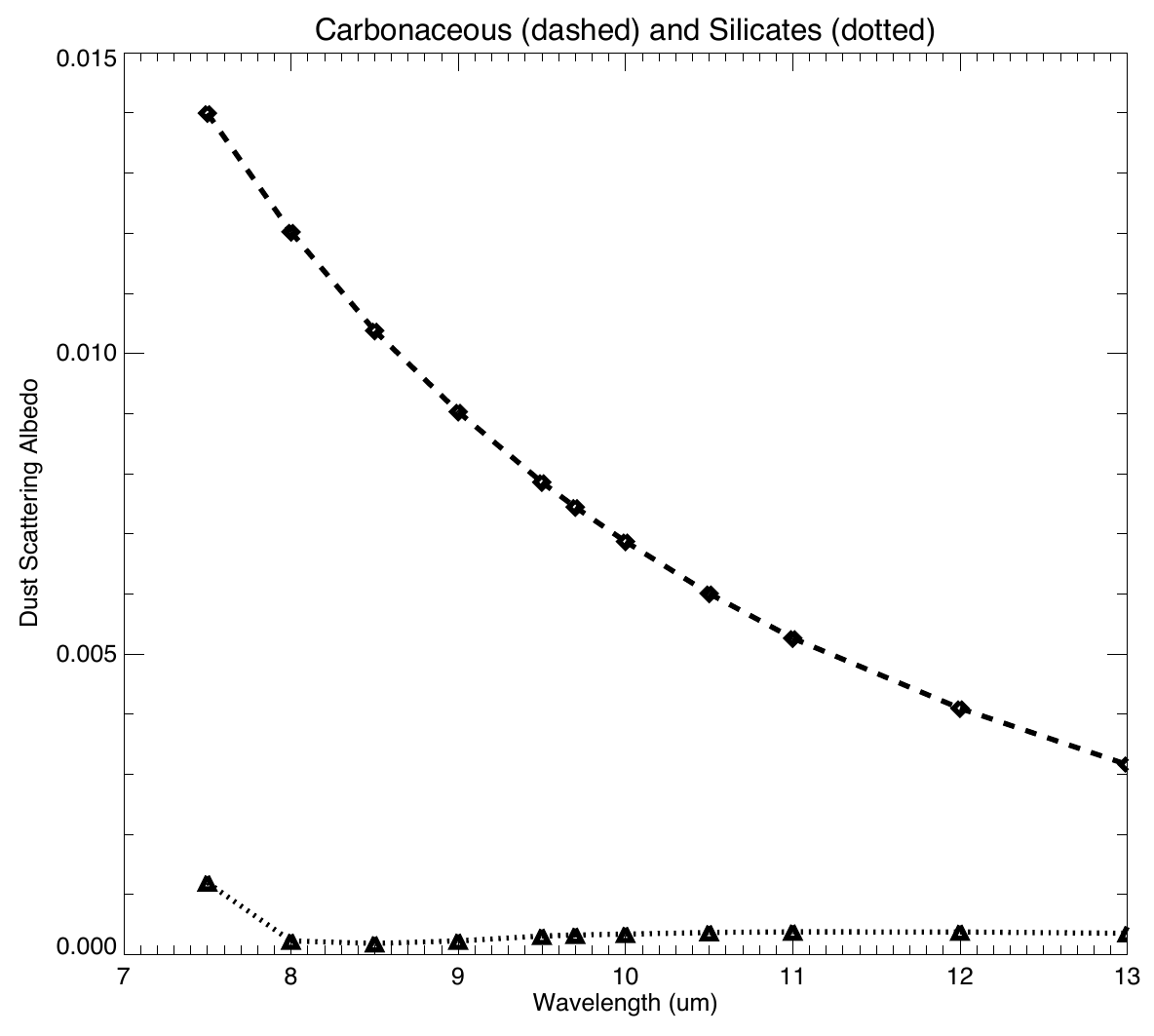}
\caption{Scattering albedo of dust with MRN distribution having maximum grain size = 0.25 $\mu$m. Albedo of silicates is plotted with dotted curve and triangles. Albedo of graphite is plotted with dashed curve and diamonds. The graphite albedo is used to create profile \emph{f}$_S$($\lambda$).}
\label{DustAlbedo}
\end{figure}

\begin{figure}
\includegraphics[width=\columnwidth]{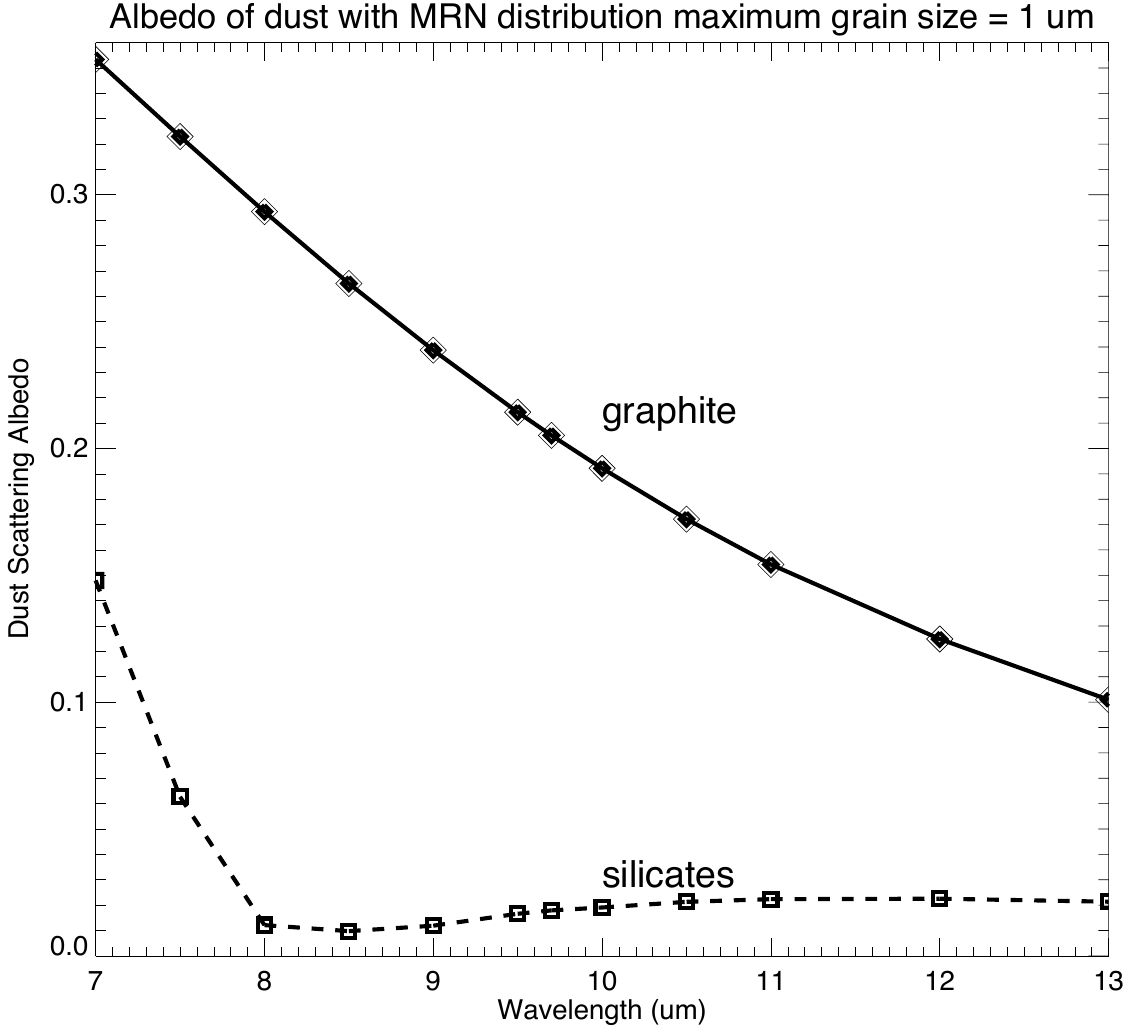}
\caption{Scattering albedo of dust with MRN distribution having maximum grain size = 1 $\mu$m. Albedo of silicates is plotted with dashed curve and squares. Albedo of graphite is plotted with solid curve and diamonds. This larger grains graphite albedo when normalized is basically the same as profile \emph{f}$_S$($\lambda$) derived from smaller grains albedo in Figure \ref{DustAlbedo}, as shown in Figure \ref{DustAlbedo-graphite-maxgs=1um+025um}.}
\label{DustAlbedo-maxgs=1um}
\end{figure}

\begin{figure}
\includegraphics[width=\columnwidth]{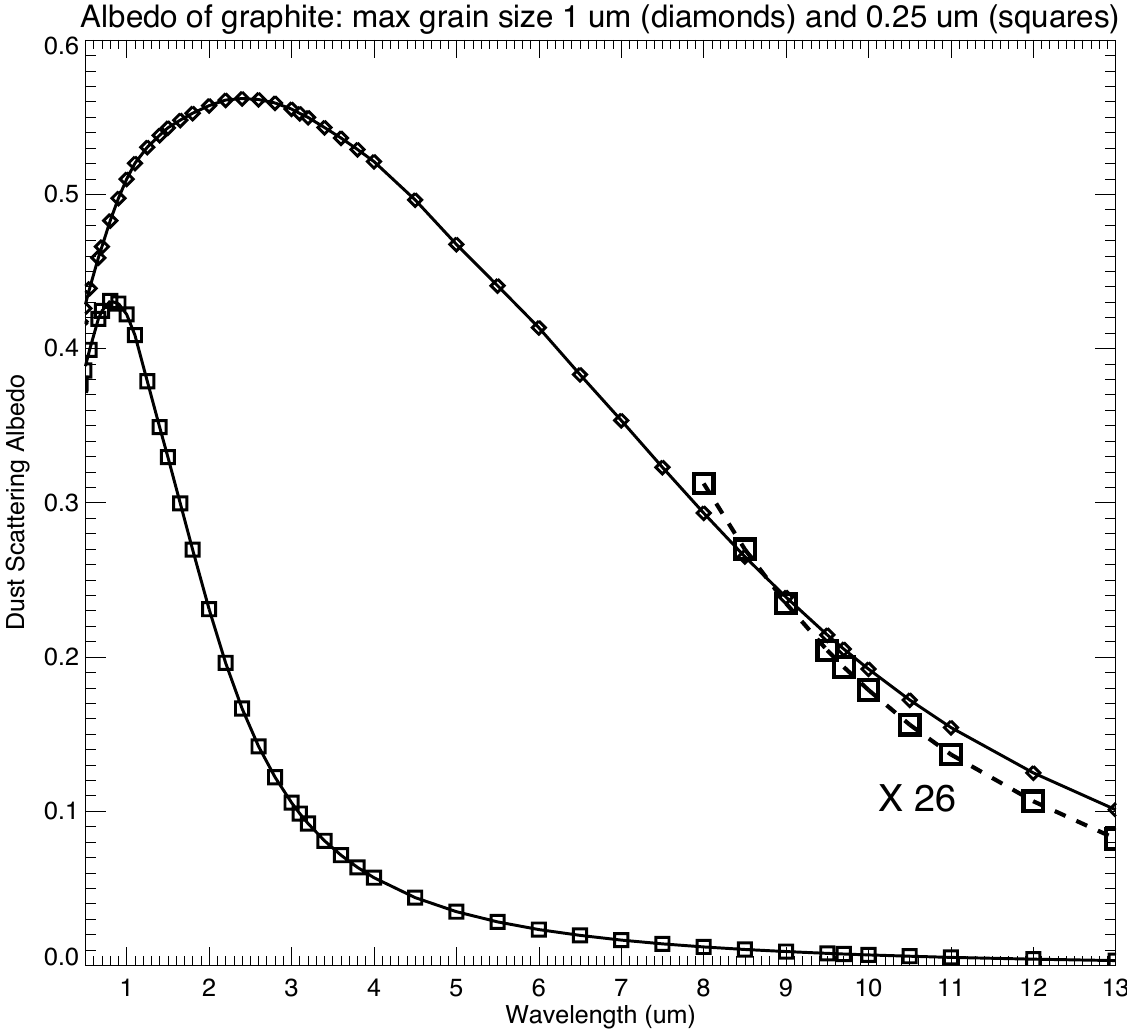}
\caption{Albedo of graphite dust with MRN distribution having maximum grain size = 1 $\mu$m is plotted with diamonds, and maximum grain size = 0.25 $\mu$m is plotted with squares. The dashed curve with squares is the 0.25 $\mu$m max. grain size case times 26, showing nearly identical variation for $\lambda > 8.5 \mu$m. Variation of the albedo in the near-IR is very different for the two cases.}
\label{DustAlbedo-graphite-maxgs=1um+025um}
\end{figure}

Figures \ref{DustAlbedo} and \ref{DustAlbedo-maxgs=1um} show the albedo of graphite and silicate dust in the mid-IR. The albedo curves were computed from mass absorption and scattering coefficients of \citet{DL1984}, assuming the MRN $a^{-3.5}$ grain size distribution \citep{MRN1977}. Albedos computed from grain size range 0.001 $\mu$m $< a <$ 0.25 $\mu$m are shown in Figure \ref{DustAlbedo}, whereas albedos from grain size range 0.001 $\mu$m $< a <$ 1 $\mu$m are shown in Figure \ref{DustAlbedo-maxgs=1um}. Most of scattering in the mid-IR would be produced by graphite since the albedo of silicates is more than a factor of 10 times less than the graphite albedo for wavelengths longer than 8 $\mu$m, in both cases of maximum grain size 0.25 $\mu$m or 1 $\mu$m. The case of maximum grain size 1 $\mu$m shown in Figure \ref{DustAlbedo-maxgs=1um} produces a graphite albedo that is about a factor of 26 times larger in the mid-IR, but when normalized results in a template profile that is almost same as the case of maximum grain size 0.25 $\mu$m, for $\lambda > 8.5 \mu$m, as shown in Figure \ref{DustAlbedo-graphite-maxgs=1um+025um}. We used the graphite albedo in Figure \ref{DustAlbedo} as the template for polarization due to scattering, but could use the albedo for larger grain sizes in Figure \ref{DustAlbedo-maxgs=1um} as the template profile to obtain similar results. Figure \ref{DustAlbedo-graphite-maxgs=1um+025um} also compares the graphite albedo for the two cases of maximum grain size for a wider range of wavelengths. The variation of graphite albedo in the mid-IR is similar for the two cases, whereas the differing variations of albedo in the near-IR could be used to estimate the maximum grain size.

The fraction of scattered photons that are polarized (polarization efficiency factor) has been computed using Mie theory by \citet{Matsu+Seki1986}, and found to have values greater than 0.7 at scattering angles between 50\degr and 130\degr for wavelengths that are longer than graphite grain sizes, peaking near unity at 90\degr scattering angle. The exact value of scattering polarization efficiency is not critical since we are fitting Stokes spectra with the normalized albedo profile in linear combination with emissive and absorptive templates.

\subsection{Accuracy of Aitken Approximation}

\begin{figure}
\includegraphics[width=\columnwidth]{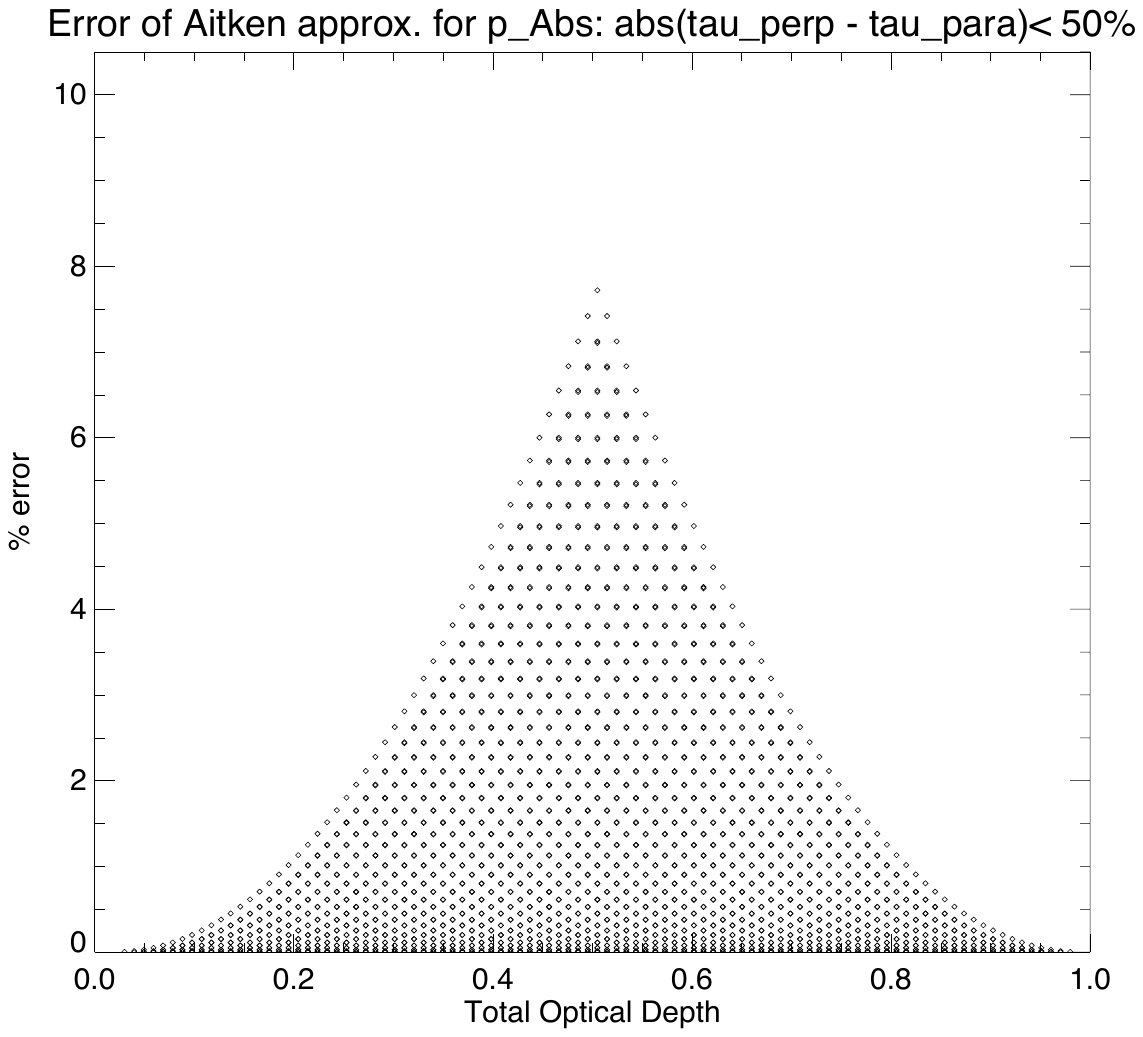}
\caption{Percent error of approximation in Eq.(\ref{Aitken-pA}) for $p_A$ as a function of total optical depth $\tau = \tau_{\parallel}+\tau_{\perp}$.}
\label{Aitken-Approx-pA}
\end{figure}

\begin{figure}
\includegraphics[width=\columnwidth]{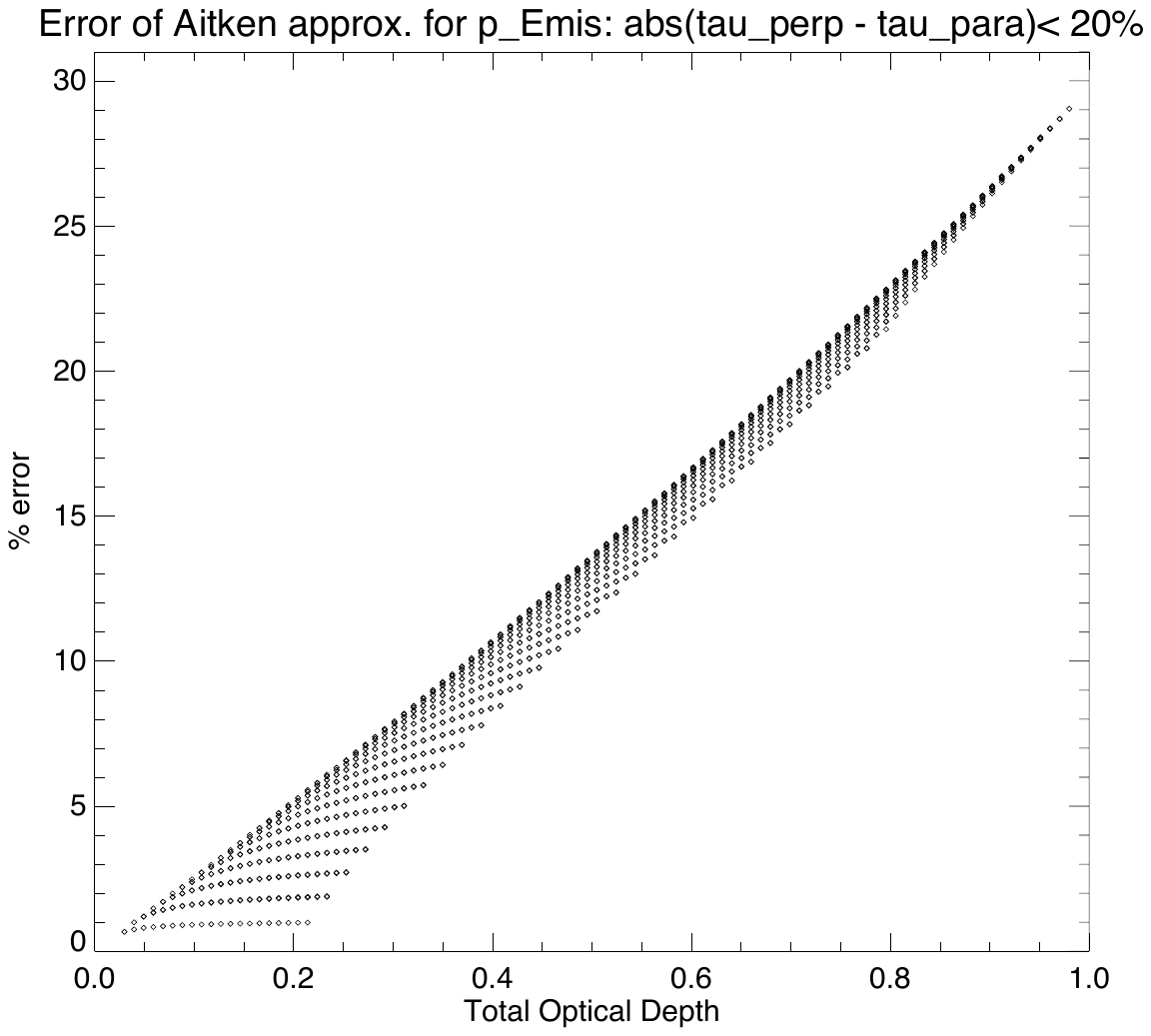}
\caption{Percent error of approximation in Eq.(\ref{Aitken-pE}) for $p_E$ as a function of total optical depth $\tau = \tau_{\parallel}+\tau_{\perp}$.}
\label{Aitken-Approx-pE}
\end{figure}

We next present graphs of the percent error for the approximations in Equations (\ref{Aitken-pA}) and (\ref{Aitken-pE}), which are used to derive the proportional relation in Equation (\ref{Aitken-pE-pA}), then providing means to derive the emissive template from the absorptive template of the Aitken method. Figure \ref{Aitken-Approx-pA} shows that when the absorptive polarization $p_A < 50\%$ the percent error of the approximation in Equation (\ref{Aitken-pA}) is less than 8\% for all optical depths. Figure \ref{Aitken-Approx-pE} shows that when the emissive polarization $p_E < 20\%$ the error of the approximation in Equation (\ref{Aitken-pE}) is less than 10\% when the total optical depth is less than 0.4, which is a good approximation since the optical depth of warm ($\approx 300$ K) dust is usually less than 0.4. Please refer to \citet{Ait2004} for explanation of the second part of the Aitken approximation for emitted photons being polarized by an absorptive layer of dust, arriving at a linear functional decomposition of polarized components.

\begin{figure*}
\centering
\includegraphics[width=240pt, keepaspectratio=true]{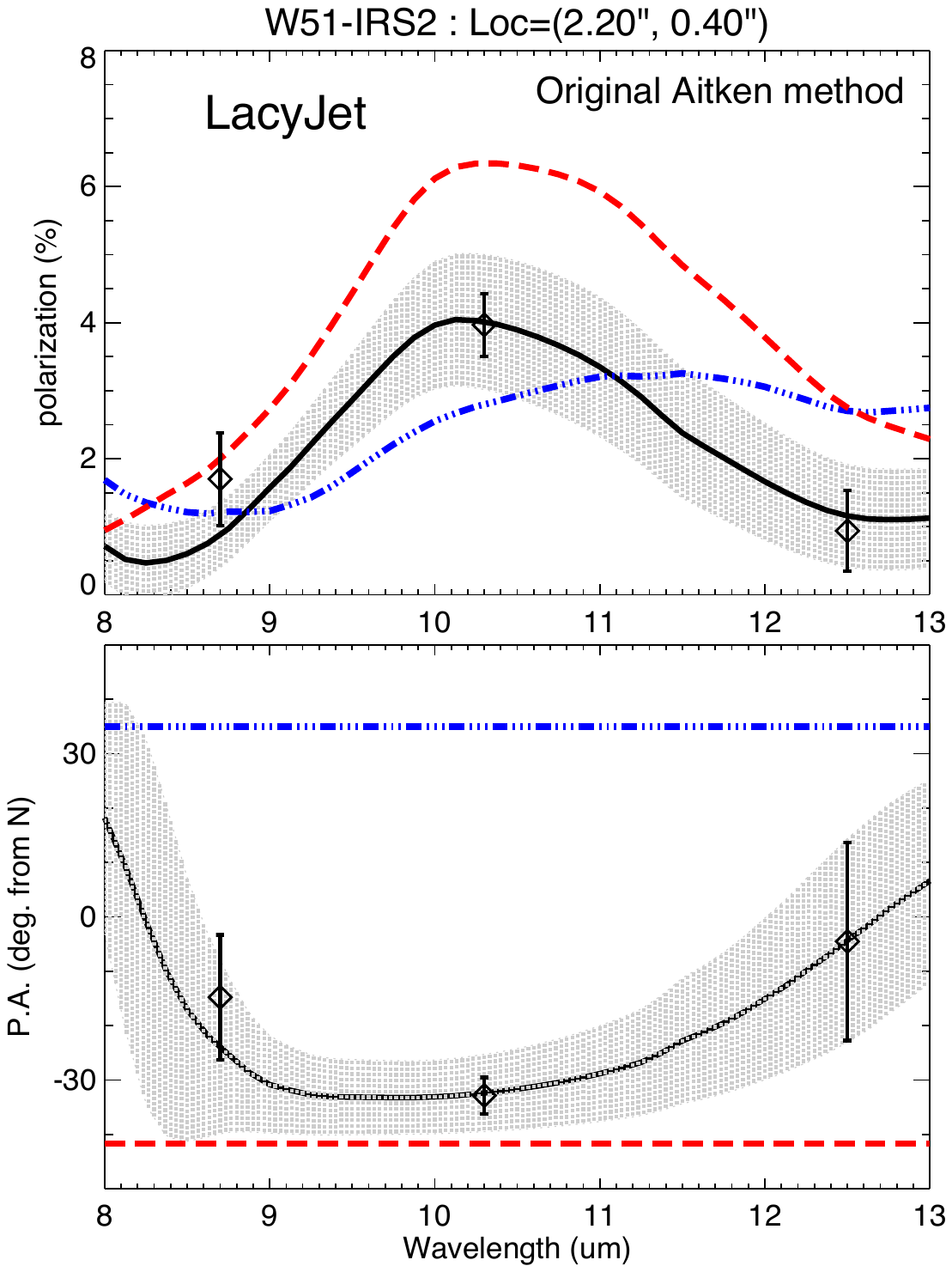}
\includegraphics[width=240pt, keepaspectratio=true]{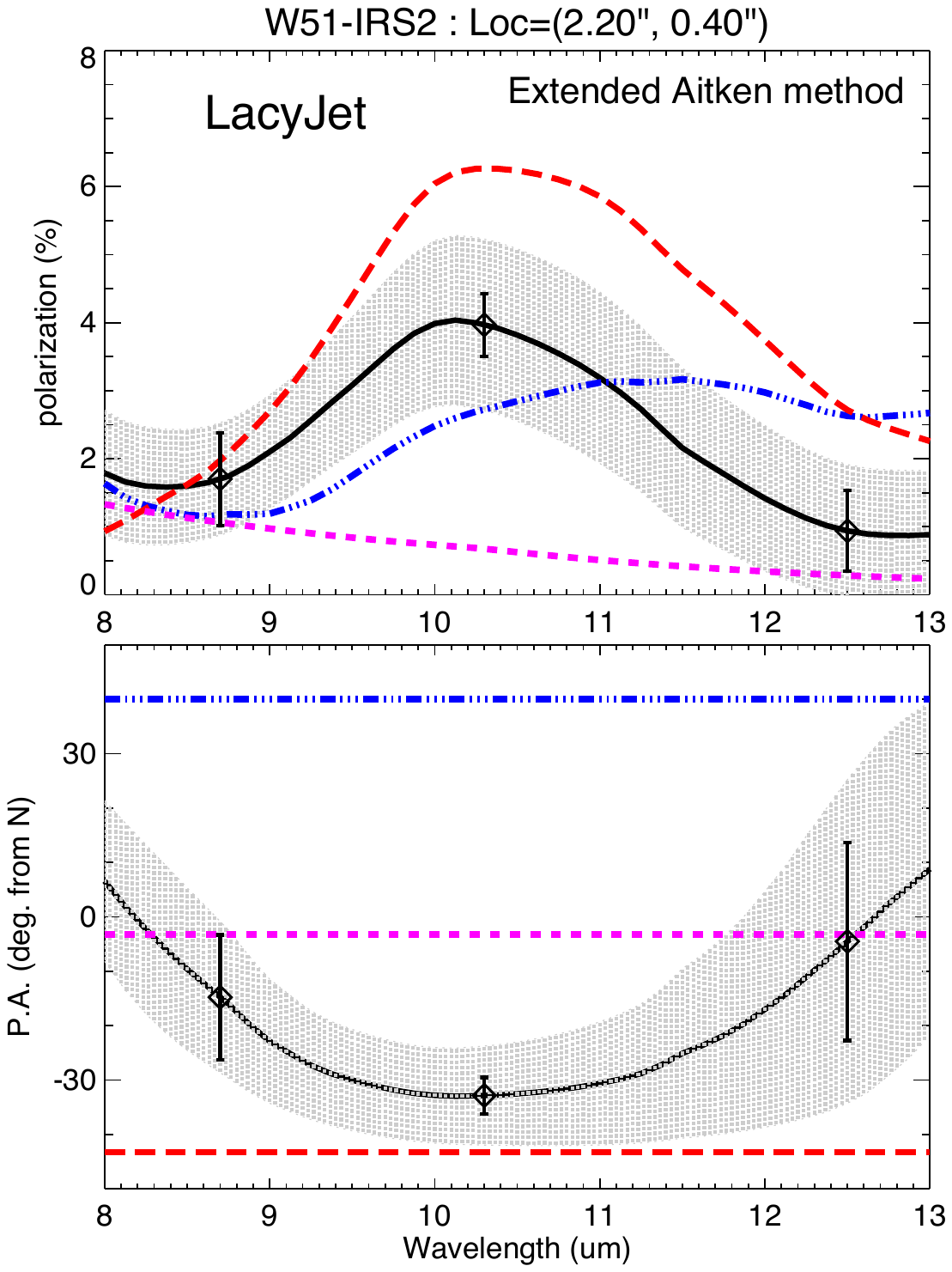}
\caption{Aitken method fits to multi-wavelength imaging polarimetry of W51 IRS2, at the sight-line of the source commonly called the LacyJet (basically in the center of nebula, and "Loc=" offset is from IRS2E). Observations are plotted as diamonds with error bars. Plots show polarization (\%) and PA in degrees computed from Aitken method fits to Stokes $q$ and $u$ spectra. Left plot shows the fit with just two components: absorptive (red long-dashed curves) and emissive (blue dot-dashed curves). Right plot shows the extended Aitken method fits with three components: absorptive, emissive, and scattering (violet short-dashed curves). Solid black curves are the sum of components, with gray shading indicating the 1-sigma range of uncertainties.}
\label{W51-AE+EAS-LacyJet}
\end{figure*}

\subsection{W51 IRS2}

Figure \ref{W51-AE+EAS-LacyJet} shows polarization spectra at another sightline of W51 IRS2 called the "Lacy jet", basically in the center of the nebula. Plots compare the polarization resulting from the original (left plot) and extended (right plot) Aitken method fits to Stokes spectra of imaging polarimetry observations (diamonds with error bars). the elevated 8.7 $\mu$m polarization can be fit better with the extended Aitken method (right plot), but the original Aitken method still produces an adequate fit with just two components. However, this shows again that the extended Aitken method can be always used to give either the same or slightly better results than the original Aitken method when three or more wavelengths are measured.
Clearly it is the 8.7 $\mu$m polarization that determines how much of the scattering component is needed to fit the observations. Note that the polarization and PA of emissive (blue) and absorptive (red) components are about the same result from either the original or extended Aitken method fits. The solid black curves are the sum of components, with gray shading indicating the 1-sigma range of uncertainty of the fits.

\subsection{W3 IRS5}

\begin{figure*}
\centering
\includegraphics[width=235pt, keepaspectratio=true]{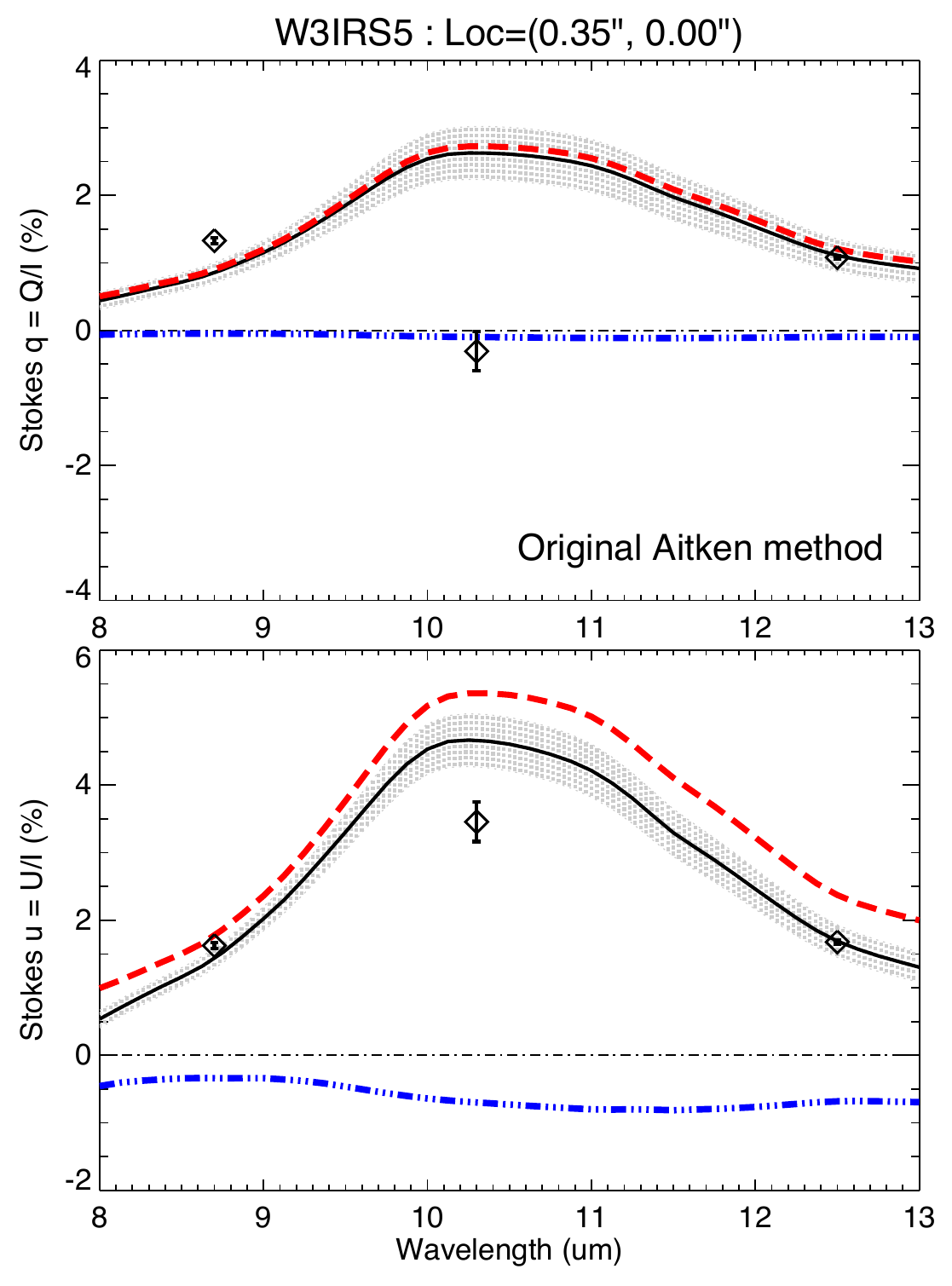}
\includegraphics[width=235pt, keepaspectratio=true]{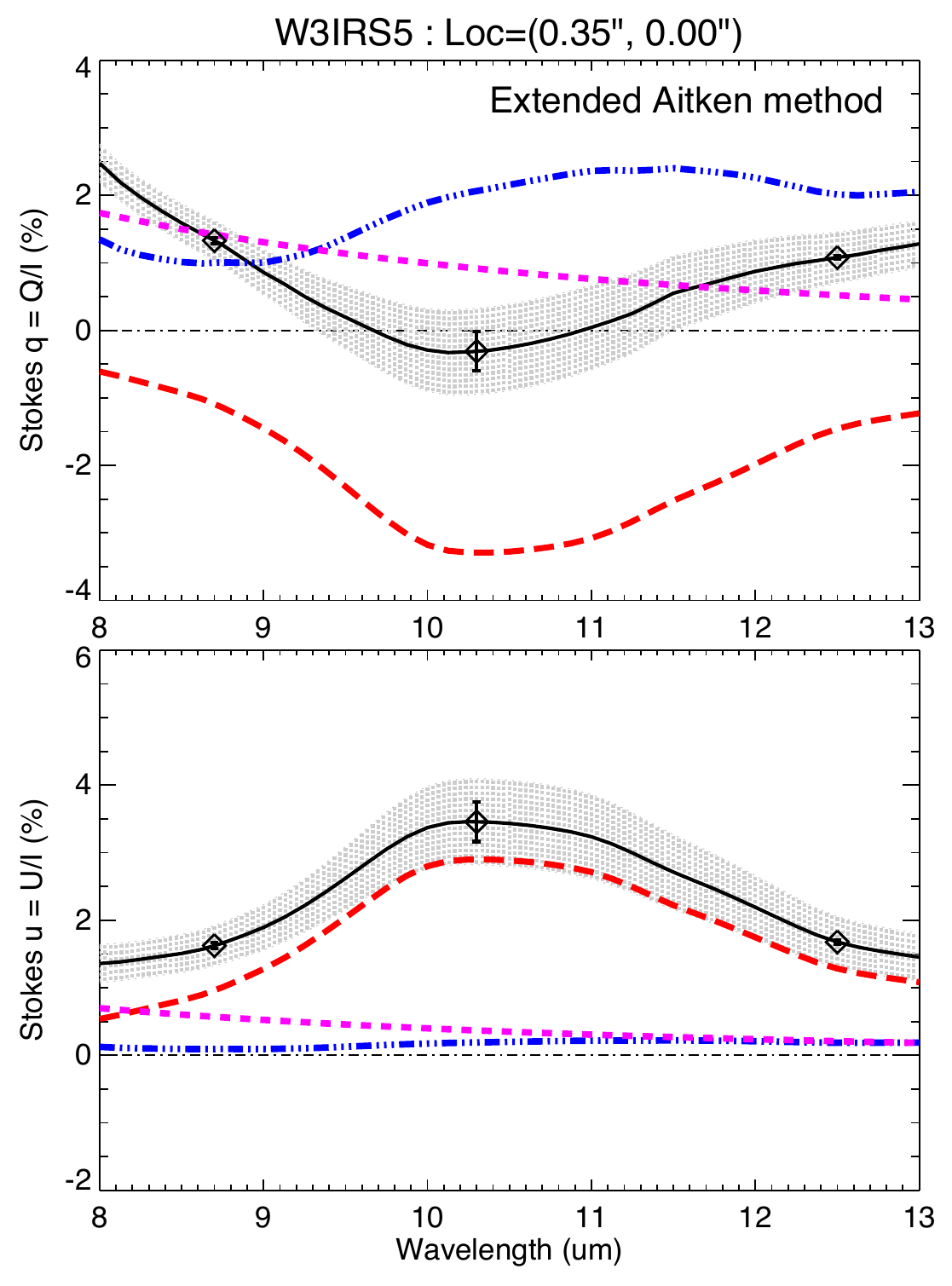}
\includegraphics[width=235pt, keepaspectratio=true]{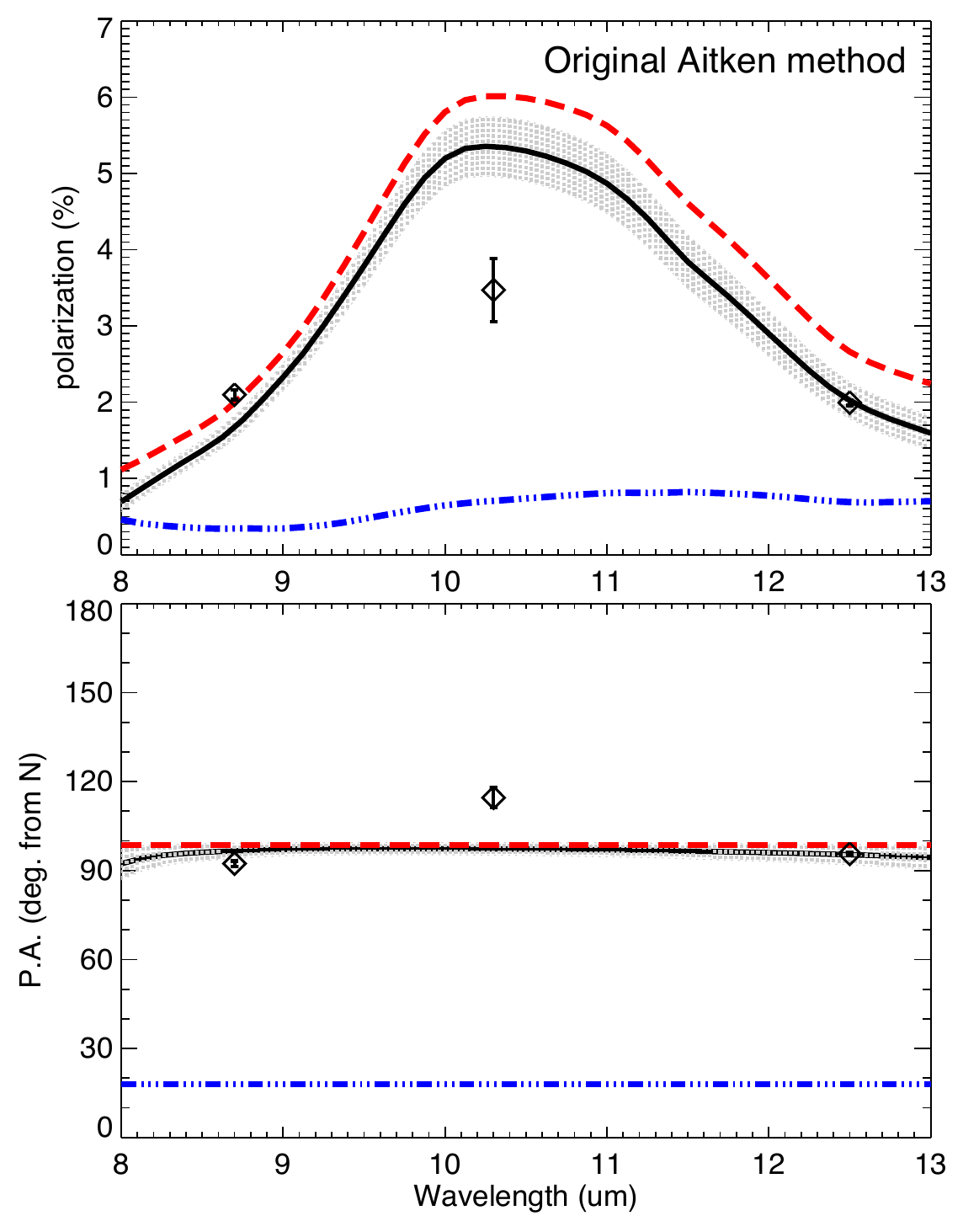}
\includegraphics[width=235pt, keepaspectratio=true]{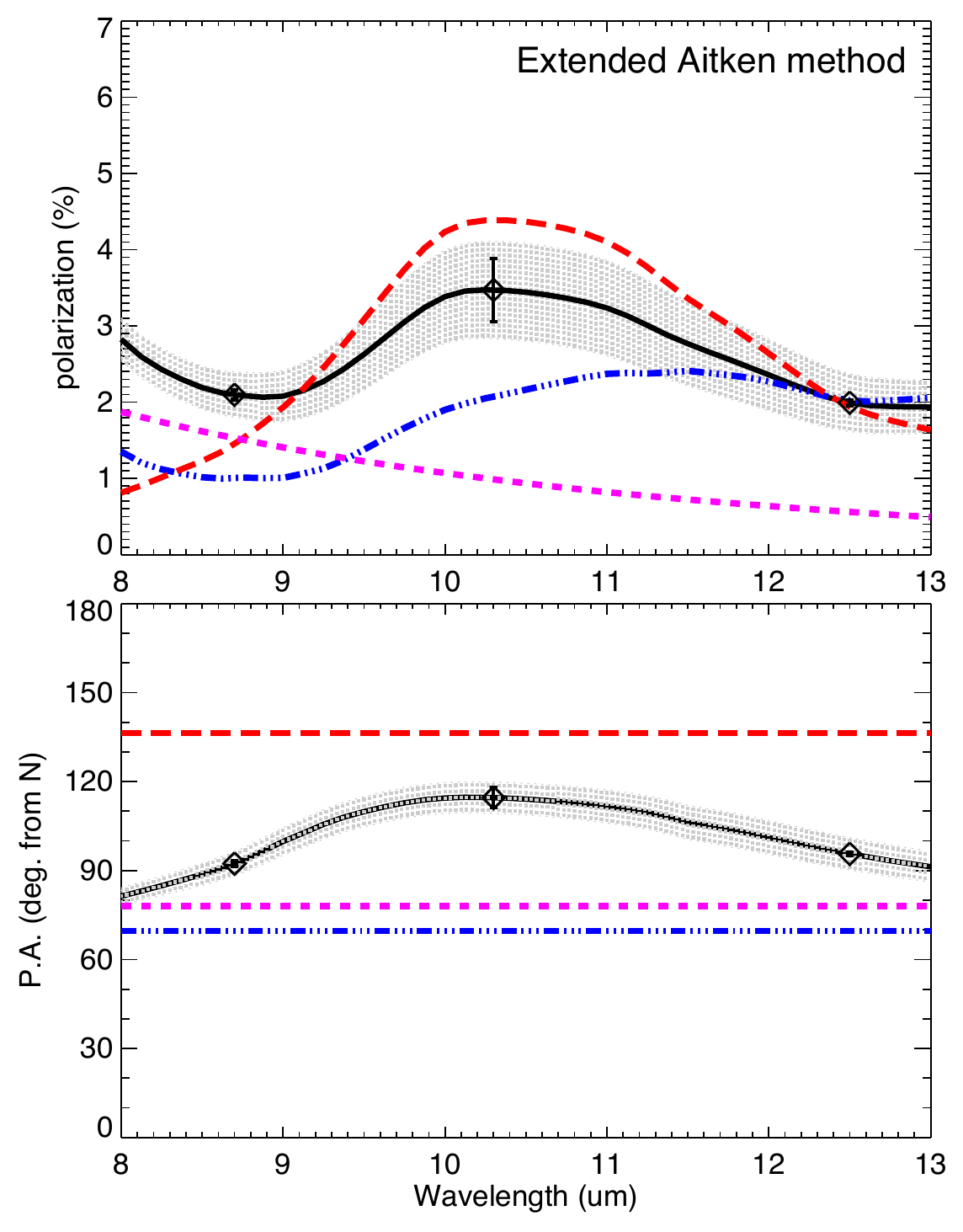}
\caption{Aitken method fits to multi-wavelength imaging polarimetry of W3 IRS5, at sight-line with offset (+0\farcs35, 0\farcs0) to the right of origin in the images of Figure \ref{W3-pvecsPA-EAO+EAS}. This location is marked with a cross inside diamond symbol at -0\farcs3 offset NE from the secondary peak, and was discussed in Section \ref{sec:W3-imp}. Stokes $q$ and $u$ from observations are plotted as diamonds with error bars in top rows. Bottom row of plots show \% polarization and PA computed from Stokes spectra. Left column of plots show the fits with just two components: absorptive (red long-dashed curves) and emissive (blue dot-dashed curves). Right column of plots show the extended Aitken method fits including scattering (violet short-dashed curves). Solid black curves are the sum of components, with gray shading indicating the 1-sigma range of uncertainties.}
\label{W3-AE+EAS-SPR}
\end{figure*}

Next Figures show additional comparisons of the original and extended Aitken method fits to Stokes spectra at other sightlines in images of the W3 IRS5 as discussed in main text. Figure \ref{W3-AE+EAS-SPR} shows plots comparing the original and extended Aitken method fits to Stokes spectra of W3 IRS5 at the scattering polarization peak sightline with offset (+0\farcs35, 0\farcs0) to the right of the origin, marked with a cross inside diamond symbol on the images in Figure \ref{W3-pvecsPA-EAO+EAS}. This sightline is offset -0\farcs3 NE from the secondary mid-IR source in W3 IRS5 (marked by a circled cross). As before, the average Stokes $q$ and $u$ spectra observed in 0\farcs24$\times$0\farcs24 box apertures are plotted as diamonds with error bars in top rows. The plots on side left side present the original Aitken method fits to Stokes spectra (two components), and the plots on the right side present the extended Aitken method fits, using three components. The resulting absorptive components are plotted as long-dashed red curves, the emissive components are plotted as dot-dashed blue curves, and the scattering component is plotted as short-dashed violet curves (right column of plots only). The solid black curves are the sum of components, with gray shading showing the 1-sigma range of uncertainties of the fits. Again, the original Aitken method with just two components has some difficulty fitting the Stokes spectra at shorter wavelengths. The bottom row of plots show the polarization (\%) and PA computed from the Stokes spectra. Clearly the extended Aitken method is able to fit the Stokes spectra exactly with three components, resulting in a more reasonable amount of absorptive polarization.

\begin{figure*}
\centering
\includegraphics[width=235pt, keepaspectratio=true]{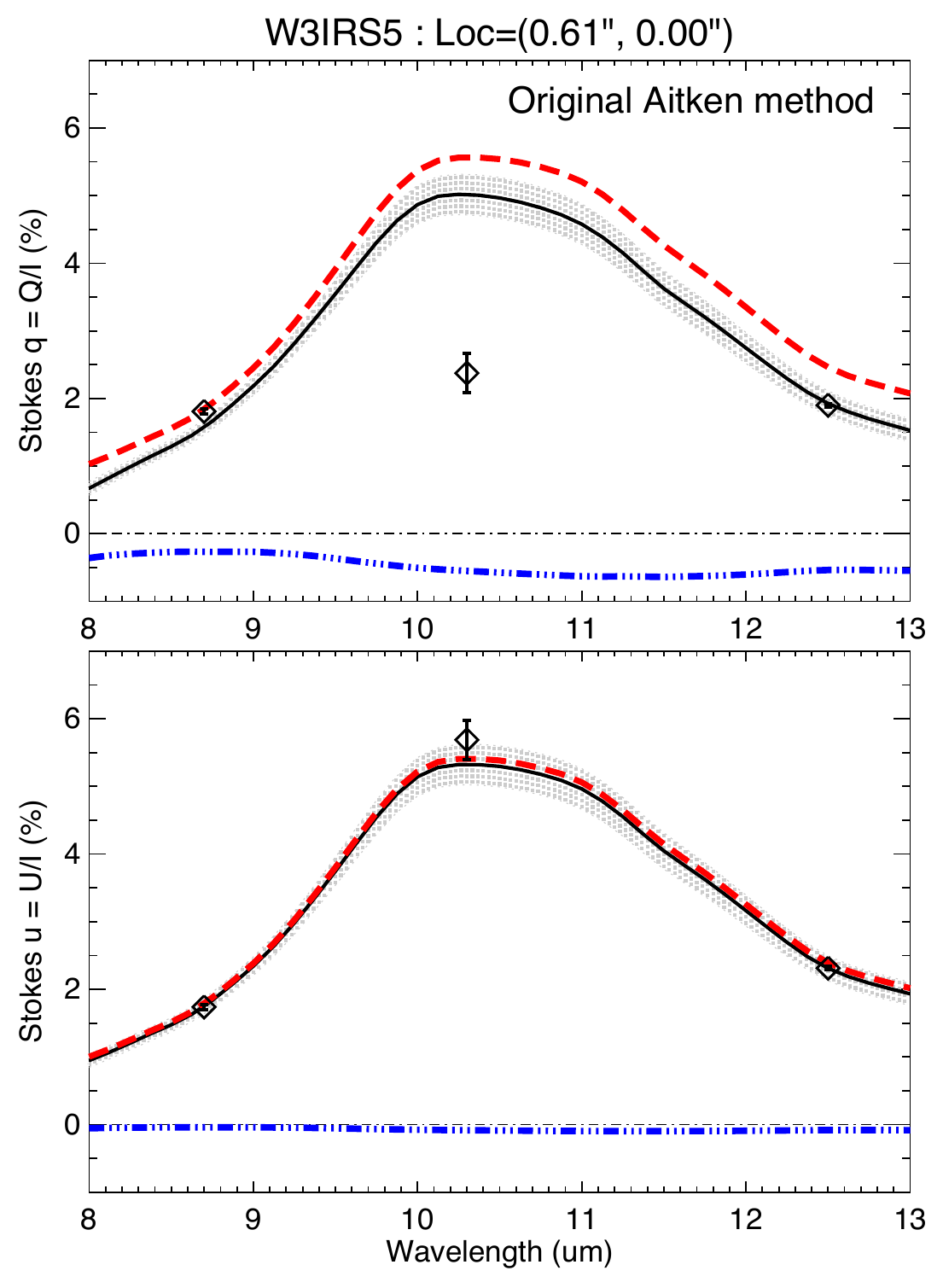}
\includegraphics[width=235pt, keepaspectratio=true]{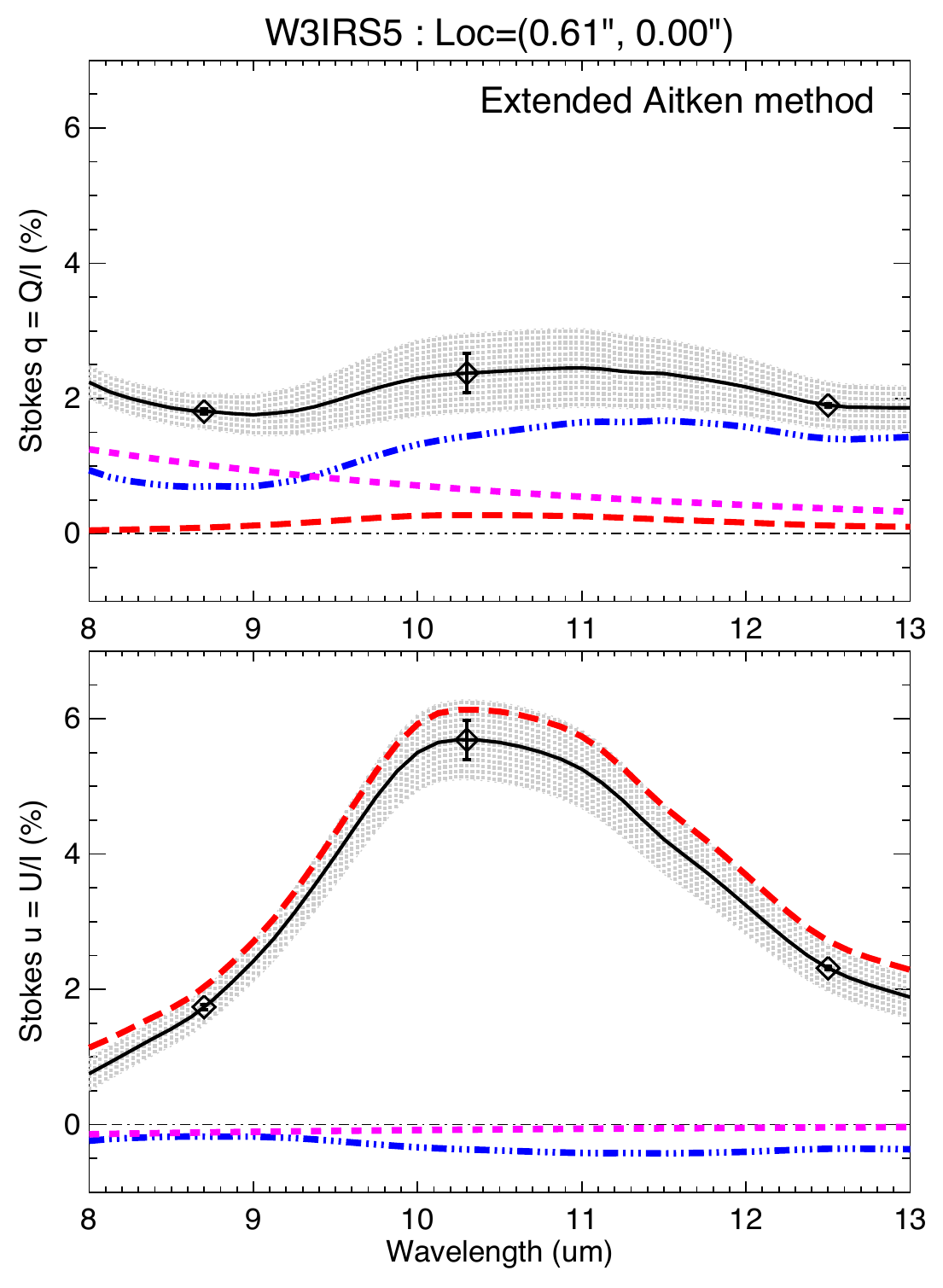}
\includegraphics[width=235pt, keepaspectratio=true]{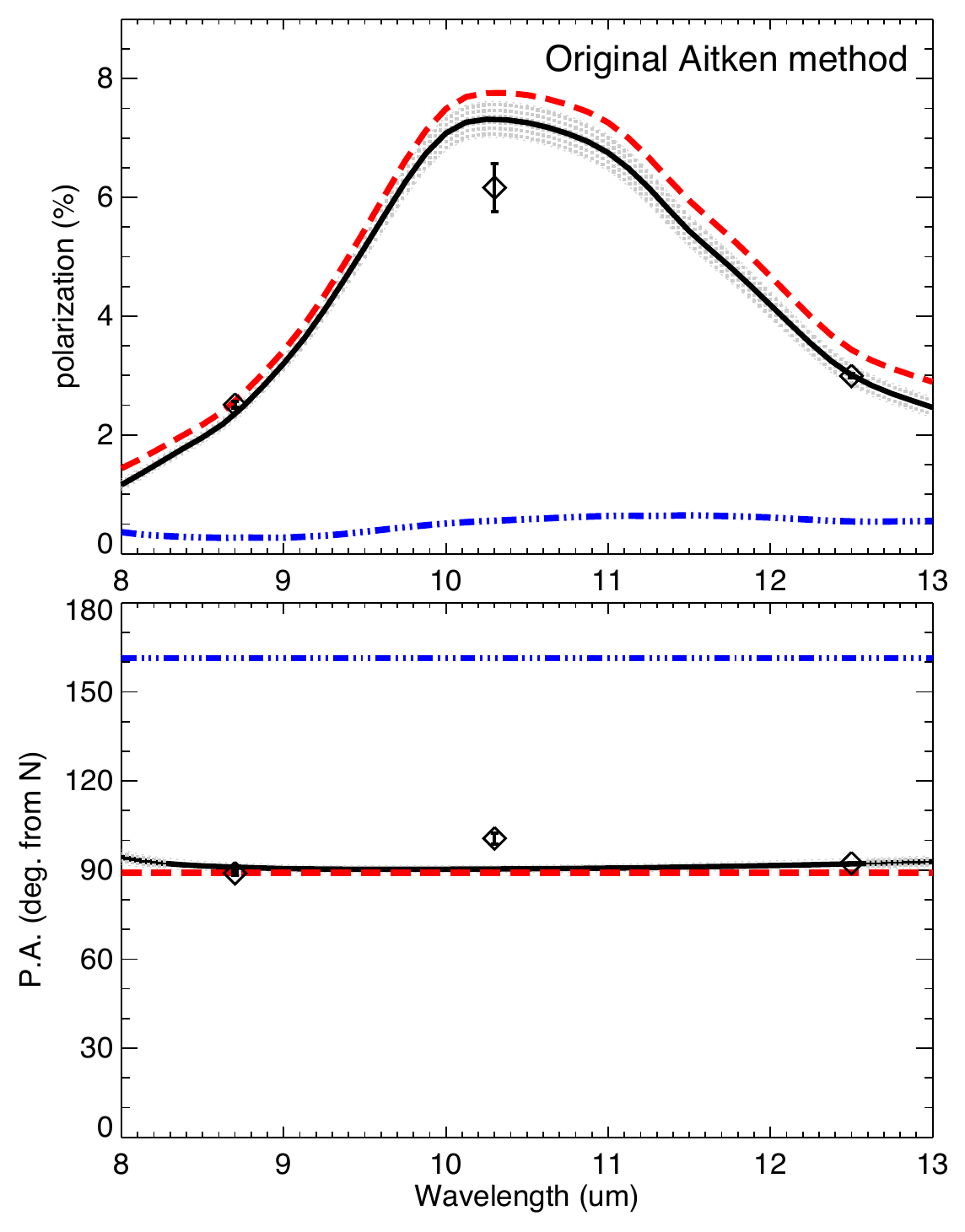}
\includegraphics[width=235pt, keepaspectratio=true]{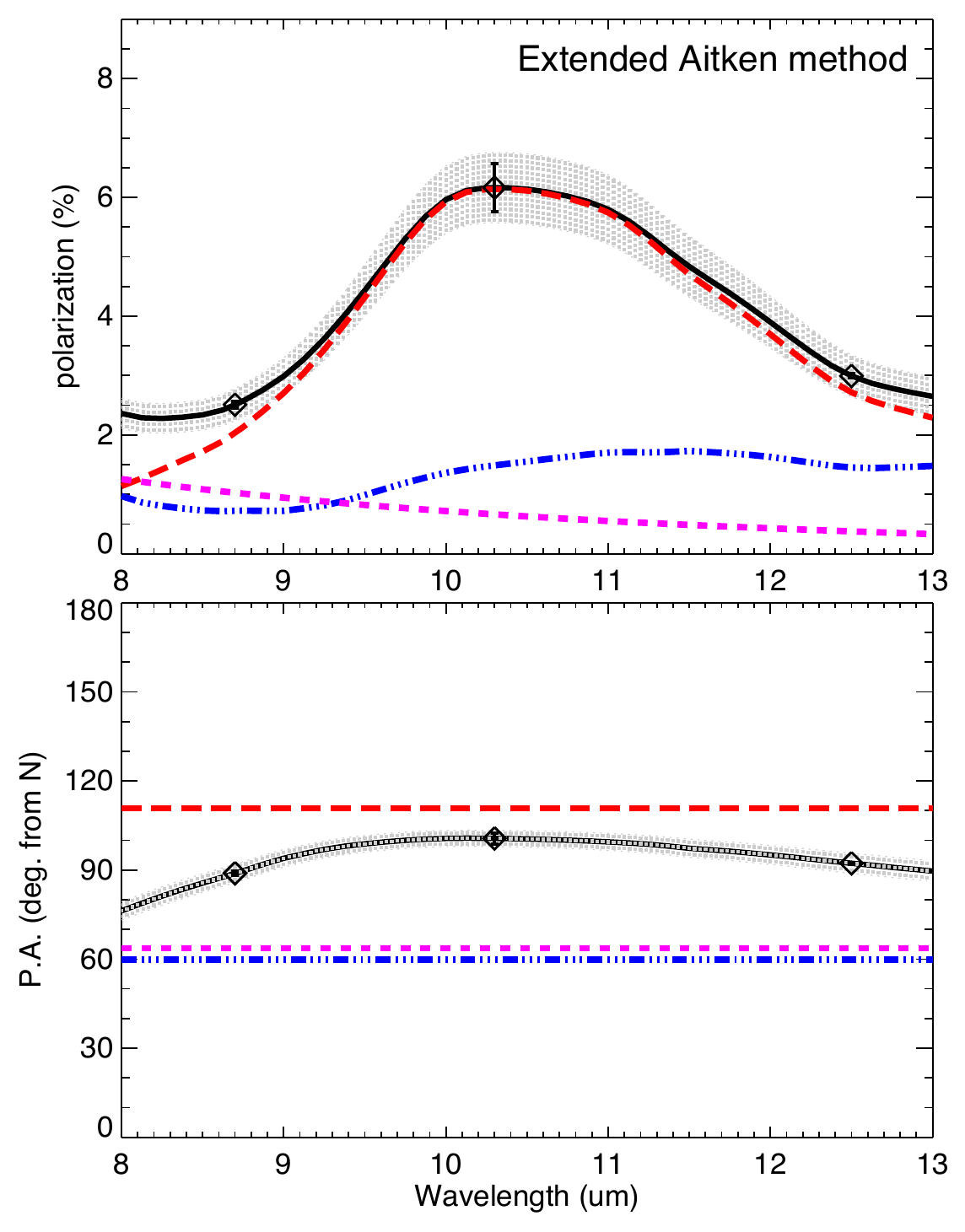}
\caption{Fits to multi-wavelength imaging polarimetry at the secondary mid-IR peak of W3 IRS5, with offset (+0\farcs61, 0\farcs0) SW from the origin in Figure \ref{W3-pvecsPA-EAO+EAS}, and marked with circled cross. This sight-line is $1\farcs2$ SW from main peak. Observations are plotted as diamonds with error bars, with percent polarization in top panels and PA in lower panels (degrees east from north). Left side shows the decomposition of polarization with the original Aitken method, and plots on right shows the extended Aitken method fits with three components. Absorptive components are the red dashed curves, emissive components are the blue dot-dashed curves, and scattering components are the violet short-dashed curve. The solid black curves are the sum of components, with gray shading indicating 1-sigma uncertainty.}
\label{W3-Decomp-AE+EAS-SW}
\end{figure*}

Plots of Aitken method fits of imaging polarimetry at the sightline of the secondary SW peak of W3 IRS5 are shown in Figure \ref{W3-Decomp-AE+EAS-SW}. Stokes spectra are plotted as diamonds with error bars, which are respectively the averages and uncertainties in 0\farcs24$\times$0\farcs24 box apertures centered on the star. Percent polarization and PA computed from the Stokes spectra are shown in the bottom rows of plots. The solid black curves are the sum of components, with shading indicating 1-sigma range of uncertainties. Again the extended Aitken method (right column of plots) is able to exactly fit the Stokes spectra, whereas the original method overestimates the absorptive component due to incorrect fit. The third component may not always be due to scattering. Since the absorption cross-section of graphite has similar functional behavior as the albedo, the extended Aitken method could be fitting absorptive or emissive component from graphite, or similar type of dust grains. The scattering template profile performs the task of fitting spectra that cannot be fit by silicates alone. The interpretation of the third component will be dependent on the source of photons, or the type of dust grains, but utilizing the third component improves the fits.

\begin{figure*}
\centering
\includegraphics[width=237pt, keepaspectratio=true]{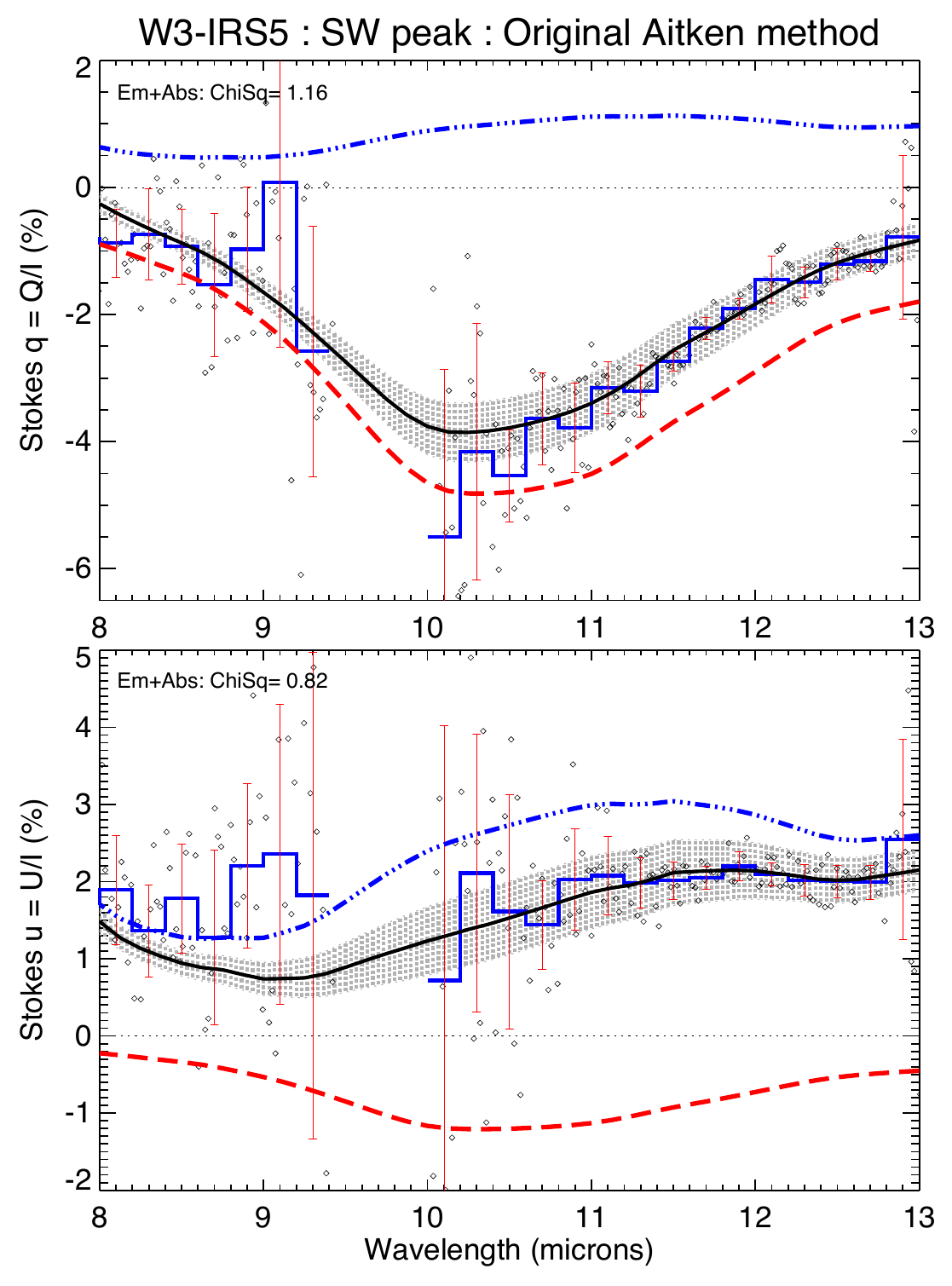}
\includegraphics[width=237pt, keepaspectratio=true]{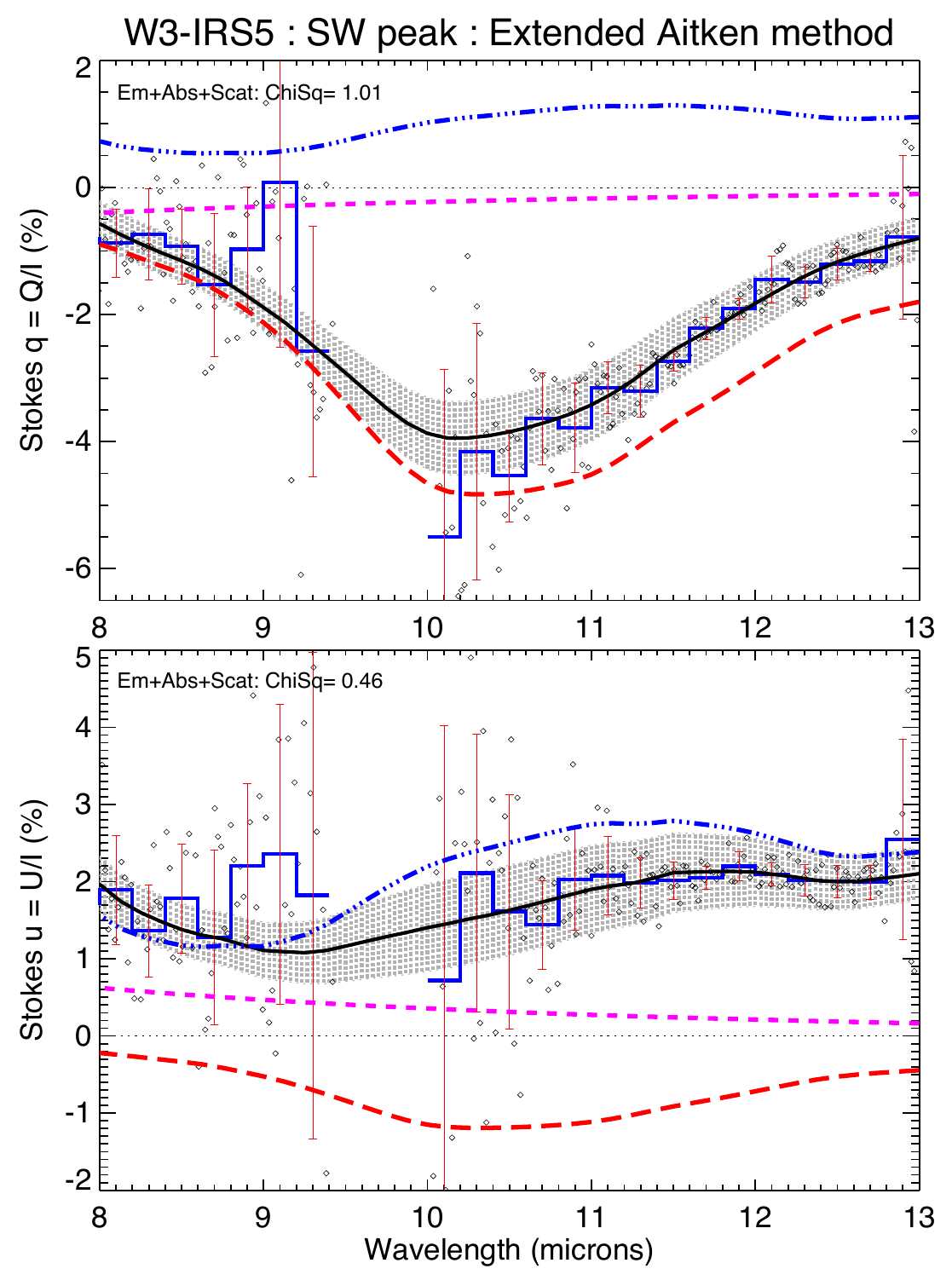}
\includegraphics[width=240pt, keepaspectratio=true]{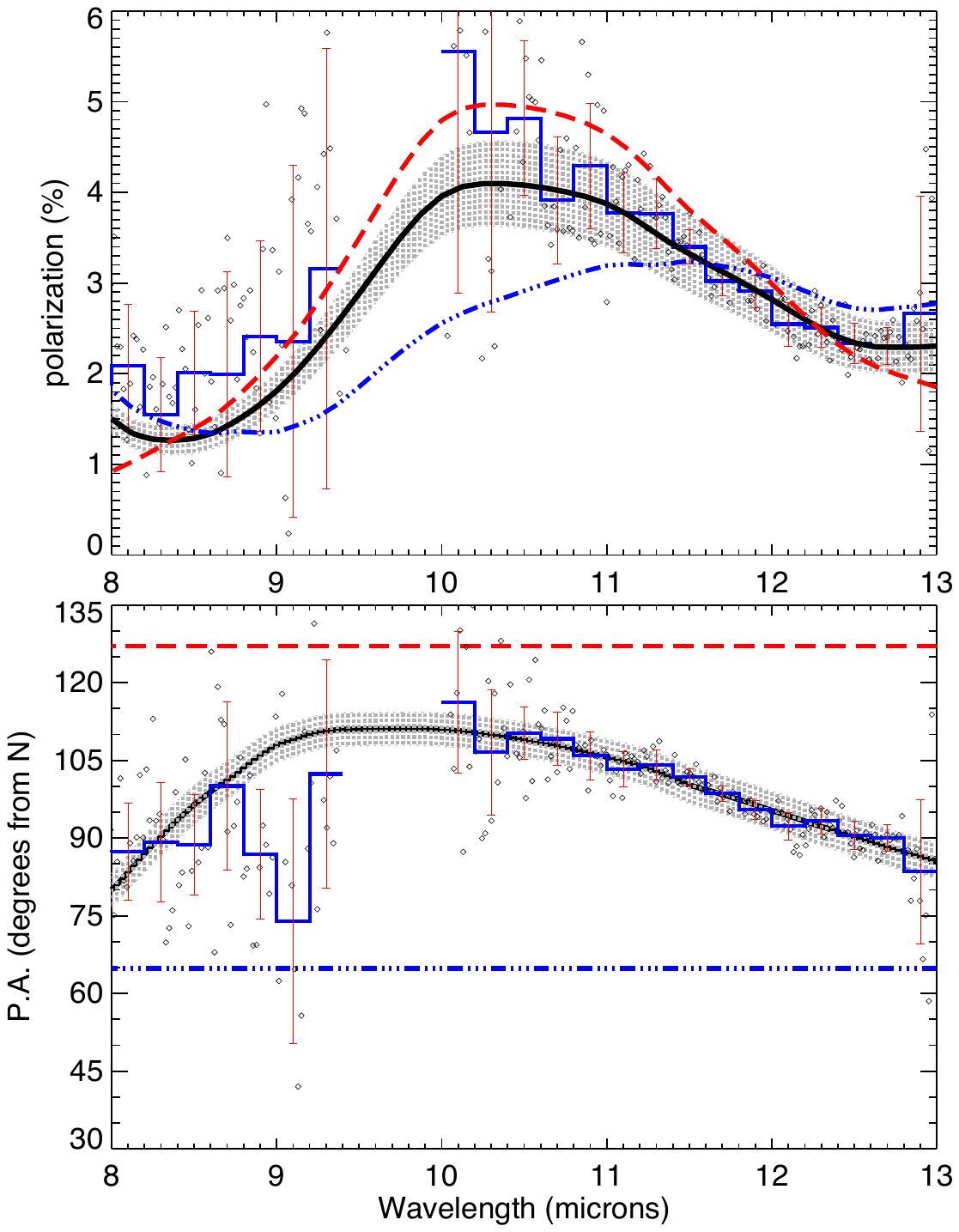}
\includegraphics[width=240pt, keepaspectratio=true]{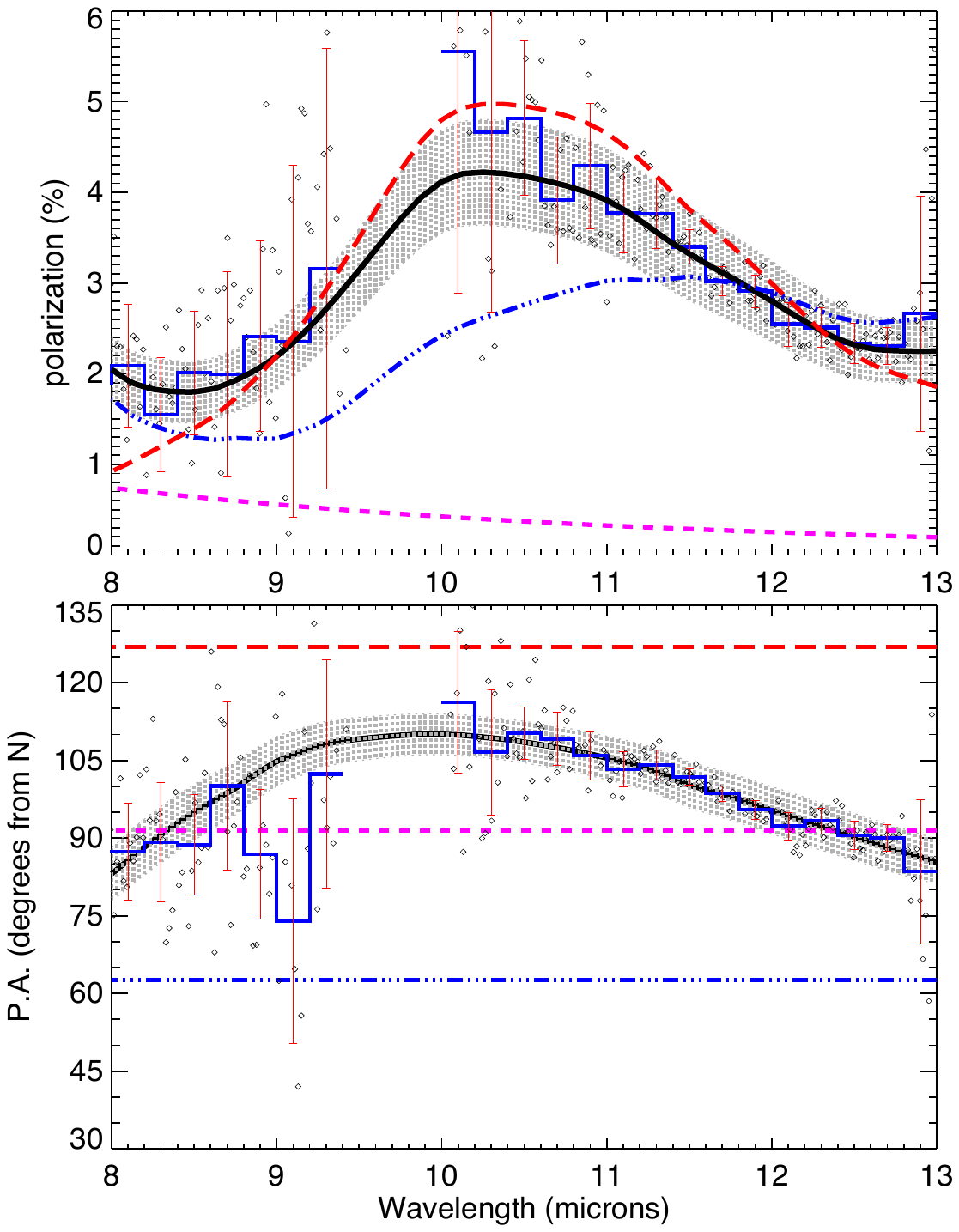}
\caption{Aitken method fits to the spectropolarimetry of the secondary peak of W3 IRS5, marked SW of the origin in Figure \ref{W3-pvecsPA-Obs-SNR+p}. Observations used the 0\farcs52 width slit mask, and resulting polarization spectra are plotted with small diamonds. Binned values are shown with blue histogram style lines, with error bars showing the standard deviation of values in each 0.2 $\mu$m bin. Left column of plots show fits using the original Aitken method, with red dashed curves the absorptive component and blue dot-dash curves the emissive component. Plots on the right side show fits using the extended Aitken method, with scattering component in violet short-dashed curves. Solid black curves are the sum of components, with gray shading indicating the 3-sigma range of uncertainties.} 
\label{W3-SpecPol-AE+EAS-SWpeak}
\end{figure*}

Figure \ref{W3-SpecPol-AE+EAS-SWpeak} shows plots comparing the Aitken method fits to spectropolarimetry of the secondary SW peak of W3 IRS5. The original Aitken method fits are shown in the left column of plots a shown in right column of plots (indicated with "Em+Abs+Scat" for 3 components). For all plots, the raw Stokes spectra (at all wavelengths) are plotted with small diamonds and the blue histogram style lines are the average of values in 0.2 $\mu$m bins, with error bars showing the standard deviation of the values at 11 wavelengths in each bin. Only data with SNR $>$ 30 are analyzed, thus leaving gap in data between 9 and 10 $\mu$m of the plots, The derived absorptive components are plotted as red dashed curves, emissive components are blue dot-dashed curves, and the scattering component is plotted with violet short-dashed curves (right column of plots only). The solid black curves are the sum of components, with shading indicating the 3-sigma range of uncertainties. Note that these Stokes spectra (top rows) are relative to the coordinate system on the detector itself, and so do not agree with Figure \ref{W3-Decomp-AE+EAS-SW} because the orientation of CanariCam on the sky was not the same. The original Aitken method has difficulty fitting the Stokes spectra at shorter wavelengths, so including the scattering component (violet dotted curves) with the extended Aitken method produces a better fit to the Stokes spectra. The $\chi^2$ goodness of fits, measured by the square root of differences between data and fit, is improved by about 15\% to 40\%, as indicated in the plots. This result is further illustrated in the bottom rows of plots showing percentage of polarization and PA: the plots in the bottom-right quadrant shows that the extended Aitken method can fit the polarization spectrum better by including a small amount of scattering component with about 1\% polarization at 8 $\mu$m, and about 0.5\% at 10.3 $\mu$m, in approximate agreement with the polarization images of Figure \ref{W3-pvecsPA-EAO+EAS}. Note that the plot of PA in Figure \ref{W3-SpecPol-AE+EAS-SWpeak} is in reasonable agreement with right-bottom plot in Figure \ref{W3-Decomp-AE+EAS-SW}, since PA is now relative to North on sky, but average values of $p(\%)$ are lower because of the larger aperture of spectropolarimetry (slit width $\times$ polarimetry mask = 0\farcs52 $\times$ 2\farcs1), whereas the smaller 0\farcs24 $\times$ 0\farcs24 aperture used for plots of imaging polarimetry gives higher average values.

The following tables compare the coefficients and uncertainties (in parentheses) resulting from fits of the original and extended Aitken methods to the spectropolarimetry observations of W3 IRS5. Table \ref{Table-SP-W3-NE} compares the coefficients, uncertainties, and $\chi^2$ of fits to spectropolorimetry of the NE peak (Figure \ref{W3-SpecPol-AE+EAS-NEpeak}), and Table \ref{Table-SP-W3-SW} compares results from the original and extended Aitken method fits to the SW peak spectropolarimetry (Figure \ref{W3-SpecPol-AE+EAS-SWpeak}). The tables present the coefficients from linear least-squares fitting of Equations (\ref{qfit}) and (\ref{ufit}) to the Stokes spectra, with uncertainties given in parentheses. The original Aitken method fits use just two coefficients ($A$ and $B$), whereas the extended Aitken method fits use the full three coefficients of the equations, resulting in better fits with essentially the same parameter uncertainties.

\begin{table}
\caption{Comparison of Original and Extended Aitken Method Fits to Spectropolarimetry of NE peak of W3 IRS5 (Figure \ref{W3-SpecPol-AE+EAS-NEpeak})}
\label{Table-SP-W3-NE}
\begin{tabular}{lrrrrr}
\hline
Aitken & \\
Method & Absorptive & Emissive & Scattering &\\
\hline
$q(\lambda)$    & $100 \times A_q$ & $100 \times B_q$ & $100 \times C_q$ & $\chi^2$ &\\
\hline
Original  & -5.39 (0.13) & 1.21 (0.07) & N/A & 1.70 & \\
\hline
Extended  & -5.66 (0.14) & 1.71 (0.07) & -0.97 (0.07) & 0.89  & \\
\hline
$u(\lambda)$     & $100 \times A_u$ & $100 \times B_u$ & $100 \times C_u$ & $\chi^2$ & \\
\hline
Original  &  4.02 (0.13) & 1.15 (0.07) & N/A & 0.84 & \\
\hline
Extended  &  4.23 (0.14) & 0.75 (0.08) & 0.79 (0.07) & 0.31 & \\
\hline
\end{tabular}
\end{table}
\begin{table}
\caption{Comparison of Original and Extended Aitken Method Fits to Spectropolarimetry of SW peak of W3 IRS5 (Figure \ref{W3-SpecPol-AE+EAS-SWpeak})}
\label{Table-SP-W3-SW}
\begin{tabular}{lrrrrr}
\hline
Aitken & \\
Method & Absorptive & Emissive & Scattering &\\
\hline
$q(\lambda)$    & $100 \times A_q$ & $100 \times B_q$ & $100 \times C_q$ & $\chi^2$ &\\
\hline
Original  & -4.81 (0.10) & 1.12 (0.05) & N/A & 1.16 & \\
\hline
Extended  & -4.82 (0.10) & 1.29 (0.06) & -0.46 (0.08) & 1.01 & \\
\hline
$u(\lambda)$     & $100 \times A_u$ & $100 \times B_u$ & $100 \times C_u$ & $\chi^2$ & \\
\hline
Original  & -1.21 (0.10) & 3.04 (0.05) & N/A & 0.82 & \\
\hline
Extended  & -1.19 (0.10) & 2.79 (0.06) &  0.72 (0.08) & 0.46 & \\
\hline
\end{tabular}
\end{table}
\begin{figure*}
\centering
\includegraphics[width=240pt, keepaspectratio=true]{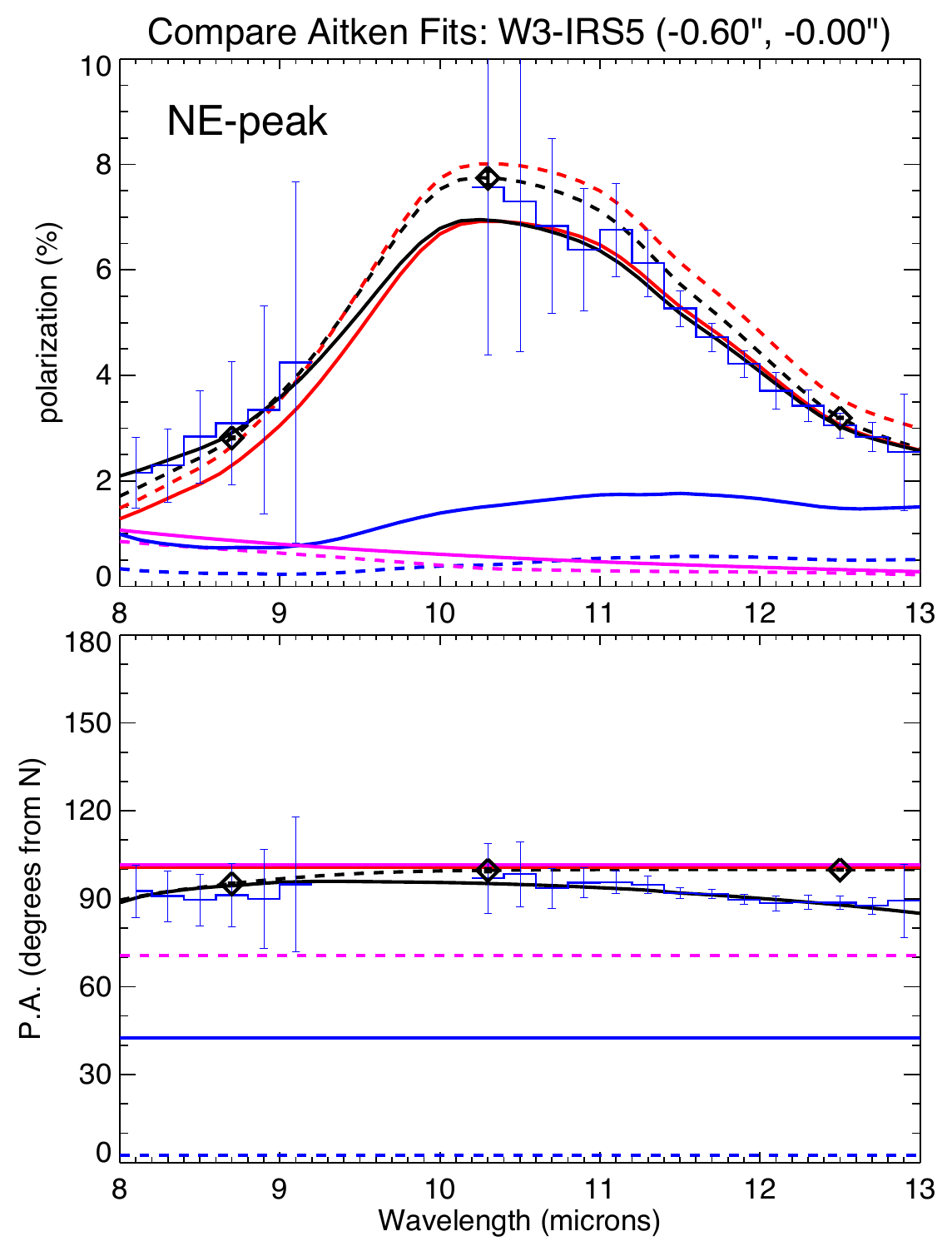}
\includegraphics[width=240pt, keepaspectratio=true]{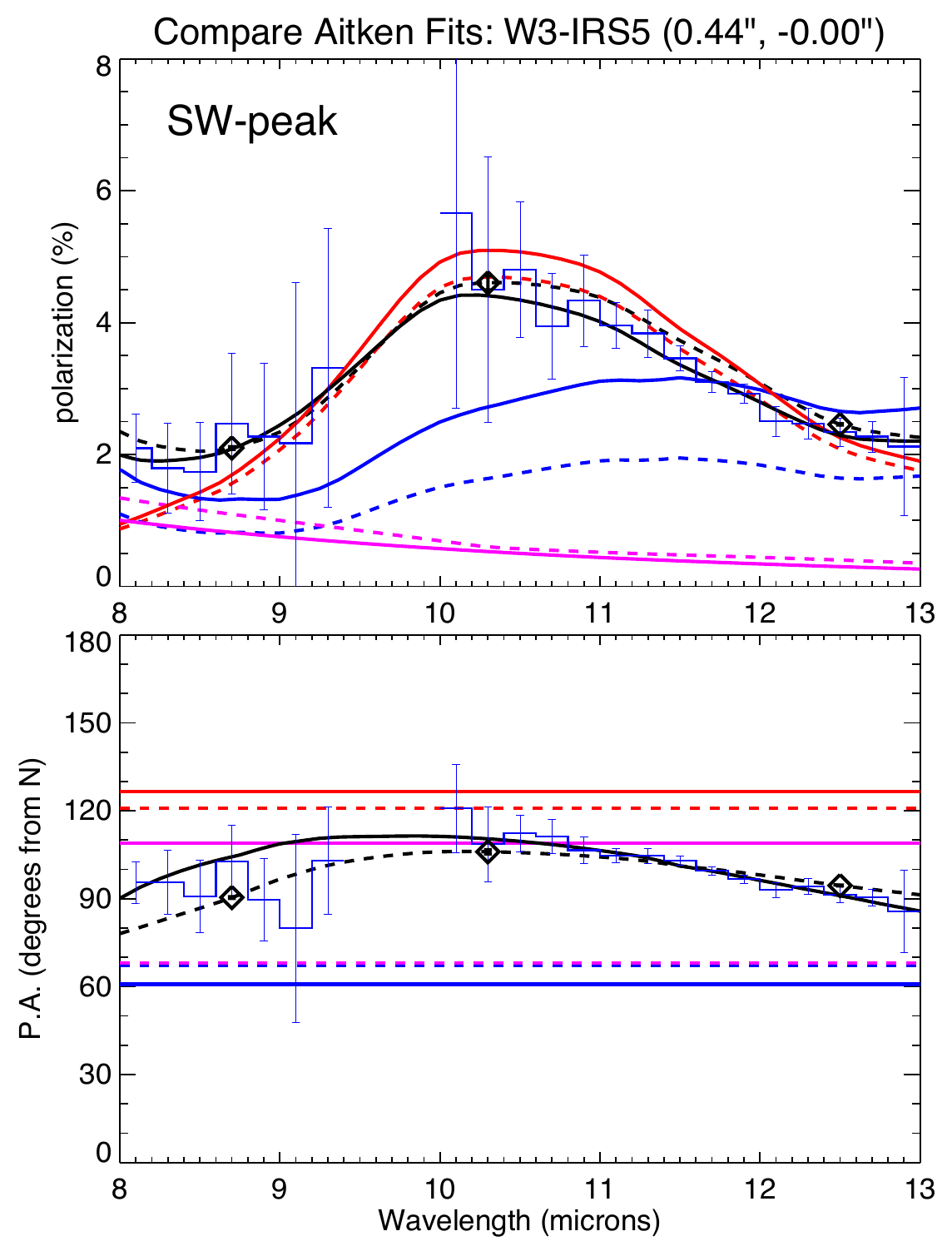}
\caption{Comparison of extended Aitken method fits to multi-wavelength imaging polarimetry and spectropolarimetry of W3 IRS5 at each of the two source peaks: NE of origin in plots on the left side, and SW of origin in plots on the right side. The percent polarization and PA at three wavelengths of imaging polarimetry observations are plotted as diamonds with error bars (uncertainty is negligible).
The polarization and PA of 0.2 $\mu$m binned spectropolarimetry observations are plotted as histograms with error bars (note much more uncertainty). Curves show polarization and PA that result from extended Aitken method fits to Stokes $q$ and $u$ spectra. Both plots show decomposition into three components, indicated by colors, but the results from fitting imaging polarimetry are indicated by dashed curves and results from fitting spectropolarimetry are indicated by solid curves. As before the absorptive components are plotted in red, emissive components are plotted in blue, scattering components are plotted in violet, and the sums of components are plotted in black.}
\label{W3-CompAitkenFits}
\end{figure*}

\subsection{Comparison of Fitting Imaging Polarimetry with Fits of Spectropolarimetry Observations}

Of course three values can always be fit exactly by three functions, therefore to further investigate the question of whether fitting three wavelengths suffices to achieve the same result as fitting a full mid-IR spectrum, we present comparisons of the extended Aitken methods fits of imaging polarimetry to the results from fits to spectropolarimetry of W3 IRS5. These results were presented separately in Figures \ref{W3-AE+EAS-Main-NE}, \ref{W3-Decomp-AE+EAS-SW}, \ref{W3-SpecPol-AE+EAS-NEpeak}, and \ref{W3-SpecPol-AE+EAS-SWpeak}, but we compare them directly here. Figure \ref{W3-CompAitkenFits} shows plots comparing extended Aitken method fits to Stokes spectra derived from observations of the NE peak and SW peak of W3 IRS5. Diamonds show the \% polarization and PA from imaging polarimetry and the histograms with error bars show polarization and PA from 0.2 $\mu$m binned spectropolarimetry observations. The curves are results from extended Aitken method fits, and the  linetypes denote which type of polarimetry is fit:  all dashed curves indicate results from fits to imaging polarimetry whereas solid curves indicate results from fits to the spectropolarimetry observations. As before colors of curves indicate which of the three components resulting from extended Aitken method fits are plotted:  red shows the absorptive components, blue shows the emissive components, violet shows the scattering components, and black curves are the sums of components. There is reasonable agreement between the absorptive components (red) derived from imaging (dashed curves) and spectropolarimetry (solid curves), but the emissive components (blue) are quite different. The magnitude of scattering polarization (violet) and in approximate agreement but the PA values are quite different. Note that polarization from observations of imaging and spectropolarimetry are not in exact agreement, and therefore the resulting Aiken method fits and decomposition are not exactly the same. However, the resulting components are providing approximately the same information. The spectropolarimetry observations were performed with telescope pointing alternating between the two sources peaks, and upon examining the data in more detail it was found that the slit was not always in exactly the same location. Furthermore the SNR of spectropolarimetry is worse than SNR of imaging polarimetry. Therefore it was not expected that the imaging polarimetry observations can match the spectropolarimetry exactly in Figure \ref{W3-CompAitkenFits}. For this reason we procceded to perform another test of the methods.

\subsection{Comparison of Fitting Spectropolarimetry Observations with Fits of Imaging Polarimetry Synthesized from Spectropolarimetry}

\begin{figure*}
\centering
\includegraphics[width=234pt, keepaspectratio=true]{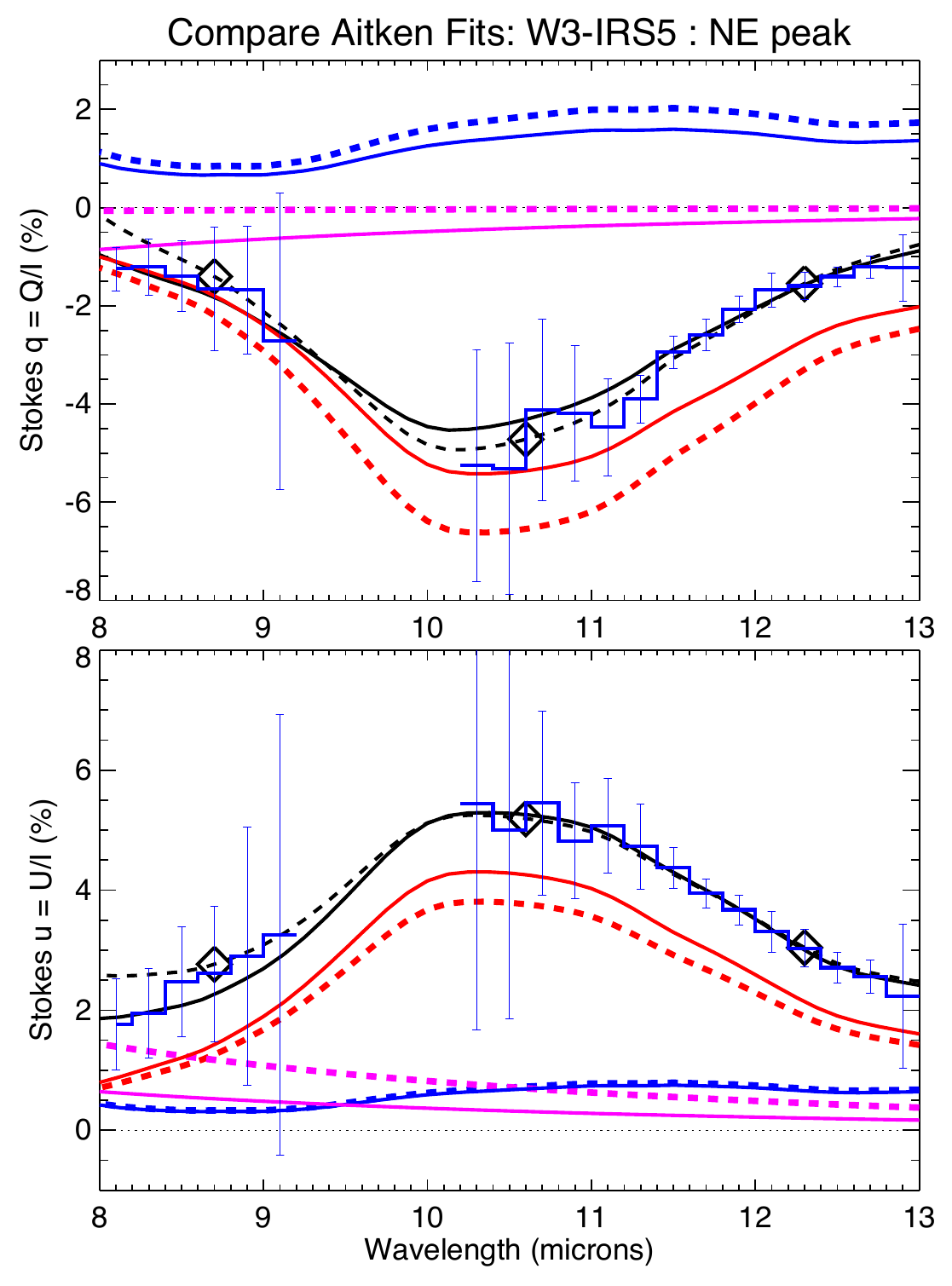}
\includegraphics[width=234pt, keepaspectratio=true]{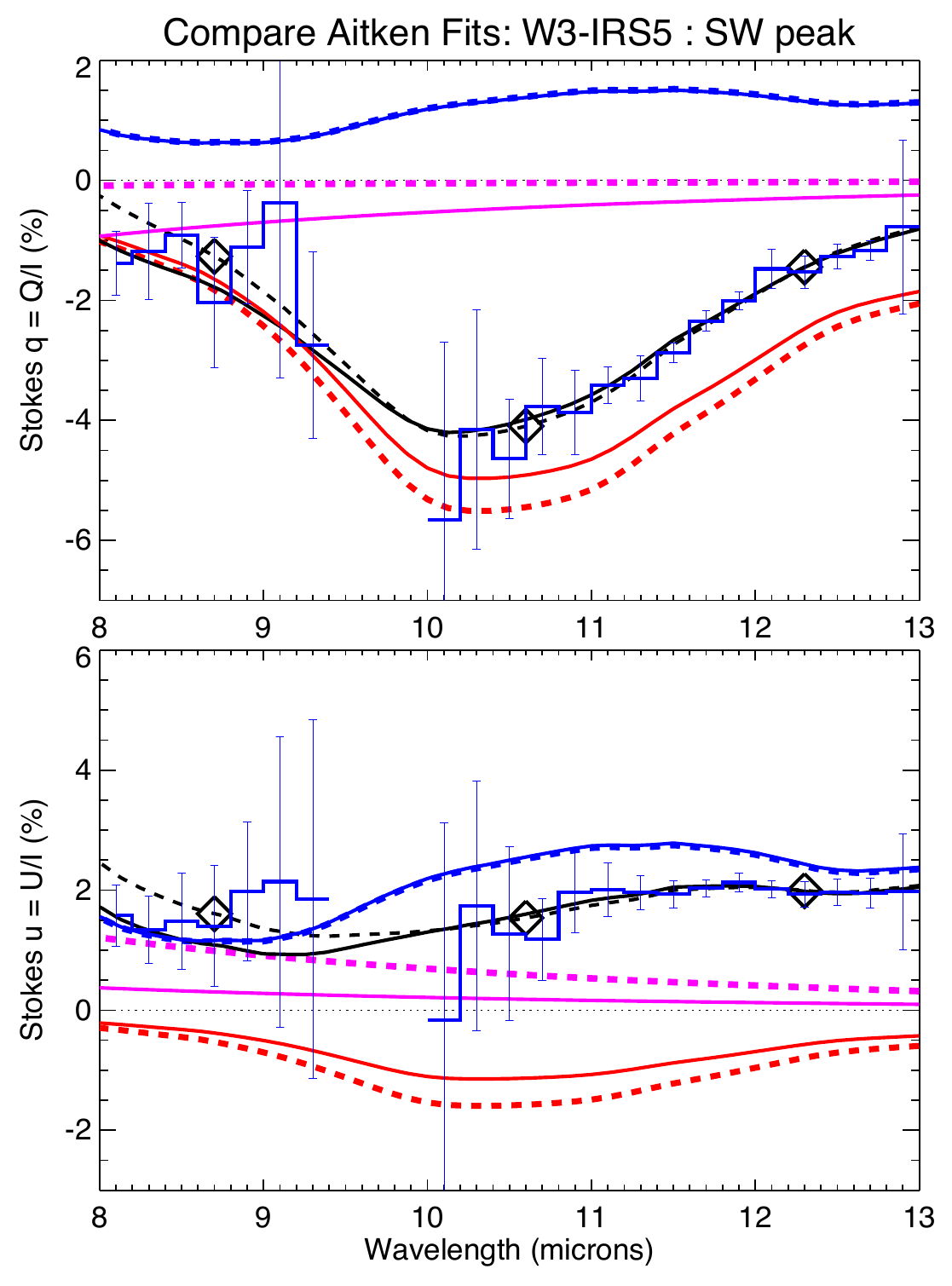}
\includegraphics[width=234pt, keepaspectratio=true]{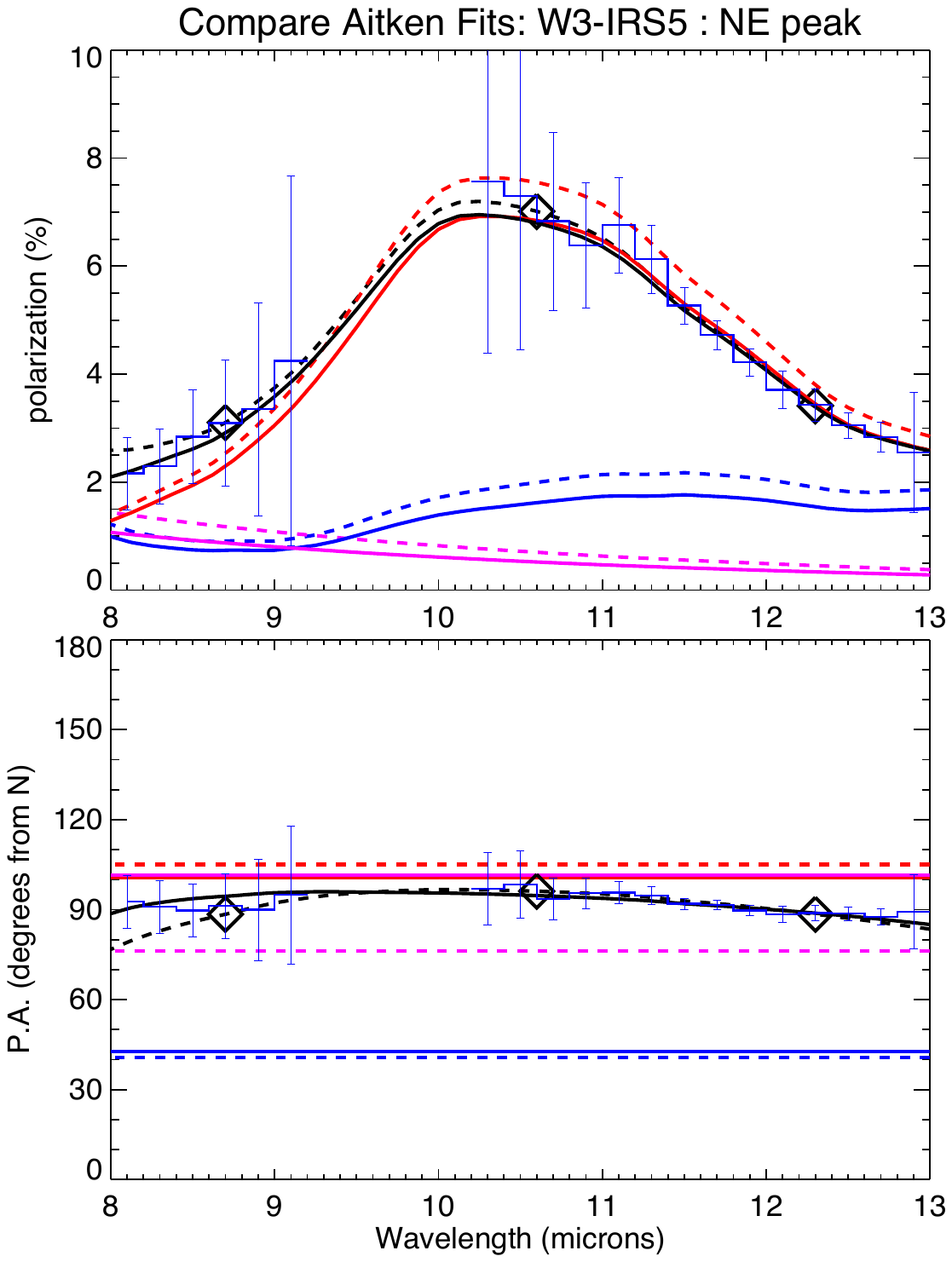}
\includegraphics[width=234pt, keepaspectratio=true]{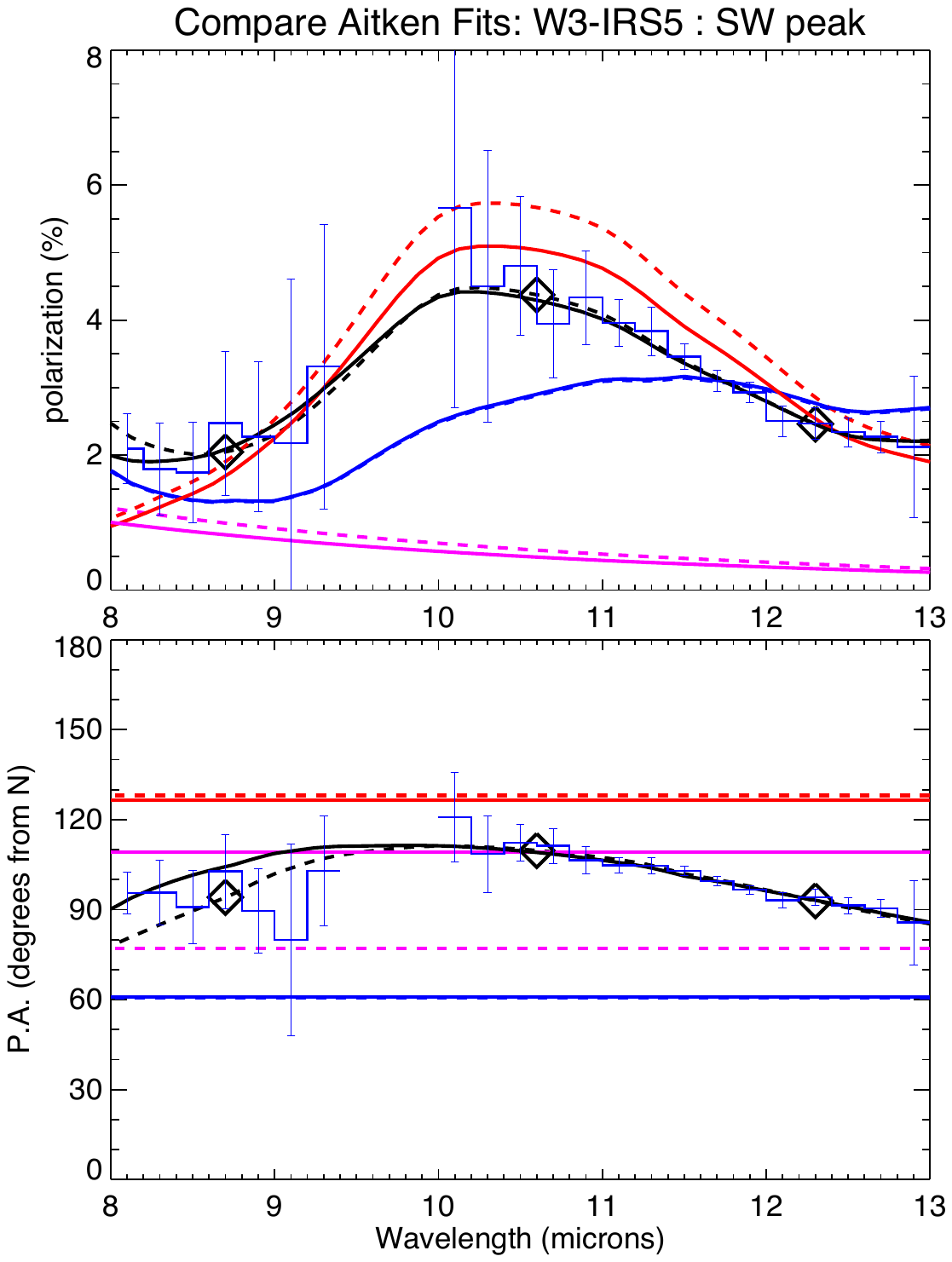}
\caption{Comparison of extended Aitken method fits of synthetic imaging polarimetry and observed spectropolarimetry of W3 IRS5 at the two source peaks: NE of origin plots left side, and SW of origin on right side. Stokes $q$ and $u$ at just three 0.8 $\mu$m wavelength bins extracted from spectropolarimetry are plotted as diamonds. The Stokes spectra of 0.2 $\mu$m binned spectropolarimetry are plotted as histograms with error bars. Curves show results from fitting Stokes $q$ and $u$ spectra with the extended Aitken method. All plots show decomposition into three components, indicated by colors, but results from fitting just three wavelengths are plotted with dashed curves whereas results of fitting spectropolarimetry are plotted with solid curves. As before, the absorptive components are plotted in red, emissive components are blue, scattering components are violet, and the sums of components are black.}
\label{W3-CompAitkenSimoFits}
\end{figure*}

We can simulate imaging polarimetry observations at three wavelengths by binning the spectropolarimetry at three chosen wavelengths, creating sythesized observations. We use bin sizes of 0.8 $\mu$m, but at 10.3 $\mu$m the uncertainty of spectropolarimety was found to be too high because of low SNR, so we use a bin centered at 10.6 $\mu$m instead, whereas the bins at 8.7 and 12.5 $\mu$m have acceptable uncertainty. Figure \ref{W3-CompAitkenSimoFits} shows the results of extended Aitken method fits to both the full spectropolarimety, shown with solid curves, and to the synthetically constructed imaging polarimetry at three wavelengths, shown with dashed curves. As before, red shows the absorptive components, blue shows the emissive components, violet shows the scattering components, and black curves are the sums of components. The left column of plots shows fitting of Stokes spectra with slit approximately positioned on the NE peak of W3, whereas the right column of plots shows Stokes spectra from the SW peak of W3. The top two rows of plots show the Stokes $q$ and $u$ spectra and the bottom two rows of plots show the resulting \% polarization and PA degrees from North, computed from the Stokes spectra. Clearly the resulting emissive (blue) and absorptive (red) components are approximately the same whether fitting just three wavelengths (dashed curves) or the full spectrum (solid curves). The scattering component (violet) has more disagreement between fitting three wavelengths (dashed) or all wavelengths (solid) in the Stokes spectra plots. However, the resulting polarization magnitude of scattering is essentially the same (third row of plots), and just the PA of scattering is in disagreement. Note that for this object, W3 IRS5, we do not know if actual scattering occurs or if the third component is simply fitting for emission/absorption by another dust type such as graphite. However, by including the third independent component in the extended Aitken method fits the other two components, emissive \& absorptive, are more accurately derived. Conclusion from this synthetic data test is that fitting three wavelengths with the extended Aitken method does suffice to decompose the polarization into approximately the same values as fitting a full spectrum, but of course having more wavelengths is usually better if the SNR of the data is good.

\bsp	
\label{lastpage}
\end{document}